\documentclass[aoas,preprint]{imsart}

\usepackage{amsthm,amsmath,natbib}
\usepackage{epsfig,amssymb,latexsym,graphicx,algorithm, multirow}
\usepackage{array,enumitem,url,booktabs}
\RequirePackage[OT1]{fontenc}
\RequirePackage[colorlinks,citecolor=blue,urlcolor=blue]{hyperref}
\arxiv{arXiv:1603.063580}

\startlocaldefs
\numberwithin{equation}{section}

\newcommand{\ra}[1]{\renewcommand{\arraystretch}{#1}}
\endlocaldefs

\graphicspath{{images/}}

\begin{document}
\sloppy

\begin{frontmatter}

\title{Bayesian inference for multiple Gaussian graphical models with application to metabolic association networks}
\runtitle{Multiple Gaussian graphical models}




\begin{aug}
  \author{\fnms{Linda S. L.}  \snm{Tan}\thanksref{m1}\ead[label=e1]{statsll@nus.edu.sg}},
  \author{\fnms{Ajay} \snm{Jasra}\thanksref{m1}\ead[label=e2]{staja@nus.edu.sg}}
  \author{\fnms{Maria}
  \snm{De Iorio}\thanksref{m2}\ead[label=e3]{m.deiorio@ucl.ac.uk}}
\and
  \author{\fnms{Timothy M. D.}  \snm{Ebbels}\thanksref{m3}
  \ead[label=e4]{t.ebbels@imperial.ac.uk}}

  \runauthor{Tan, Jasra, De Iorio and Ebbels}

  \affiliation{National University of Singapore\thanksmark{m1}, University College London\thanksmark{m2} and Imperial College London\thanksmark{m3}}
           
\address{Department of Statistics \& Applied Probability \\
Block S16, Level 7, 6 Science Drive 2 \\
Faculty of Science \\
National University of Singapore \\
Singapore 117546 \\
          \printead{e1}\\
          \printead{e2}}

  \address{Department of Statistical Science \\
  	University College \\
  	Gower Street \\
  	London WC1E 6BT \\
  	United Kingdom \\
          \printead{e3}}
        
  \address{Department of Surgery and Cancer \\
    Imperial College London  \\
    South Kensington Campus  \\
    London SW7 2AZ
  	United Kingdom \\
  	\printead{e4}}

\end{aug}

\begin{abstract}
We investigate the effect of cadmium (a toxic environmental pollutant) on the correlation structure of a number of urinary metabolites using Gaussian graphical models (GGMs). The inferred metabolic associations can provide important information on the physiological state of a metabolic system and insights on complex metabolic relationships. Using the fitted GGMs, we construct differential networks, which highlight significant changes in metabolite interactions under different experimental conditions. The analysis of such metabolic association networks can reveal differences in the underlying biological reactions caused by cadmium exposure. We consider Bayesian inference and propose using the multiplicative (or Chung-Lu random graph) model as a prior on the graphical space. In the multiplicative model, each edge is chosen independently with probability equal to the product of the connectivities of the end nodes. This class of prior is parsimonious yet highly flexible; it can be used to encourage sparsity or graphs with a pre-specified degree distribution when such prior knowledge is available. We extend the multiplicative model to multiple GGMs linking the probability of edge inclusion through logistic regression and demonstrate how this leads to joint inference for multiple GGMs. A sequential Monte Carlo (SMC) algorithm is developed for estimating the posterior distribution of the graphs.
\end{abstract}


\end{frontmatter}

\section{Introduction}
Technological advances have enabled quantitative measurements and profiling of metabolites (products of metabolic reactions), which is important to the understanding of complex biological systems as well as the diagnosis and monitoring of disease states. A key feature of such data is that a significant number of metabolite levels are often highly interrelated. Analysis of these associations may provide further information about the physiological state of a system and lend insights on complex metabolic relationships \citep{Steuer2006}. In this article, we analyze urinary metabolic data acquired using $^1$H NMR spectroscopy for 127 individuals. These subjects live close to a lead and zinc smelter at Avonmouth in Bristol, UK, that produces large quantities of airborne cadmium \citep{Ellis2012}. An extremely toxic metal, cadmium is commonly released through industrial processes and acute exposure poses numerous health risks. Here, we use Gaussian graphical models \citep[GGMs,][]{Dempster1972} to investigate the correlation structure of 22 urinary metabolites for each individual in response to cadmium exposure. Differential networks \citep{Valcarcel2011}, which highlight significant changes in metabolite interactions under different experimental conditions, are inferred jointly with the GGM characterizing different levels of cadmium exposure. This is a strength of our modelling framework  as it allows borrowing of strength across different biological conditions. Analysis of such metabolic association networks can point to differences in the underlying biological reactions caused by cadmium exposure.

Gaussian graphical models \citep[GGMs,][]{Dempster1972} provide an important tool for studying the dependence structure among a set of random variables. Under the assumption that the variables have a joint Gaussian distribution, a zero in the precision matrix indicates conditional independence between the associated variables. This corresponds to the absence of an edge in the underlying graph, where nodes denote variables and edges represent conditional dependencies \citep{Lauritzen1996}. GGMs are widely used, for instance, in biological networks to study the dependence structure among genes from expression data \citep[e.g.][]{Dobra2004, Chun2014} and financial time series for forecasting and predictive portfolio analysis \citep[e.g.][]{Carvalho2007, Wang2011}. In applications where the effect of different experimental conditions on the dependence relationships among variables is of interest, multiple GGMs (one for each condition) have to be estimated. Under such circumstances, joint inference can encourage sharing of information across graphs and allow for common structure where appropriate \cite[e.g.][]{Guo2011, Peterson2015}.

We focus on Bayesian inference for GGMs using the G-Wishart prior \citep{Roverato2002, Atay2005}. The G-Wishart is the family of conjugate distributions for the precision matrix, where entries corresponding to missing edges in the underlying graph are constrained to be zero. The normalizing constant of the G-Wishart can only be computed in closed form for decomposable graphs. In this work, we consider the unrestricted graph space where non-decomposable graphs are allowed. Where necessary, we use the Monte Carlo method of \cite{Atay2005} and the Laplace approximation of \cite{Lenkoski2011} to estimate the normalizing constant efficiently. 

The main idea of this paper is to propose a prior for $p_{ij}$, the probability of a link between nodes $i$ and $j$, that is grounded in the network literature. We start with one of the simplest random graph models; the multiplicative model where
\begin{equation*}
p_{ij}= \pi_i \pi_j.
\end{equation*}
This model is additive on a log scale: $\log p_{ij} = \alpha_i + \alpha_j$ where $\alpha_i = \log \pi_i$. Alternatively, and without substantial difference in performance, we could have assumed a logistic/probit model. Incorporating interaction can be achieved by including an extra term. Extension to include more complex structures is in principle straightforward. For example, scale-free models can be achieved by placing a discrete prior on $\pi_i$ such as the Barab\'{a}si-Albert model so that the probability that node $i$ has $k$ connections is proportional to $A+k^{\alpha}$. To incorporate a community structure, we could assume
\begin{equation*}
\text{logit}(p_{ij})= \alpha_i + \alpha_j+\theta_{g_ig_j};
\end{equation*}
where $g_i$ denotes the community $i$ belongs to and $\theta_{g_ig_j}$ denotes an offset for   node $i$ and $j$ belonging to the same community, with $\theta_{g_ig_j}$ equal to 0 otherwise. 

Here, we propose using the multiplicative model of \cite{Chung2002} as a prior on graphs for estimating GGMs. This prior is further extended to multiple GGMs via logistic regression. To obtain joint posterior inference for multiple GGMs, we develop a novel sequential Monte Carlo (SMC) algorithm \citep{Moral2006} which uses tempering techniques. We apply proposed methods to a simulated dataset in addition to the urinary metabolic dataset.

The rest of the paper is organized as follows. Section \ref{S-bg} provides the background and review of existing methods. In Section \ref{S-multmodel}, we introduce the multiplicative model and discuss its degree and clustering properties. Section \ref{S-GGM} specifies the model setup for multiple GGMs. Section \ref{S-postdistn} describes posterior inference and a Laplace approximation for the prior probabilities of graphs. The SMC algorithm is outlined in Section 5. Proposed methods are illustrated using simulations and an application to urinary metabolic data in Section 6. Section 7 concludes.

\section{Background} \label{S-bg}

\begin{figure}
	\centering
	\includegraphics[width=0.55\textwidth]{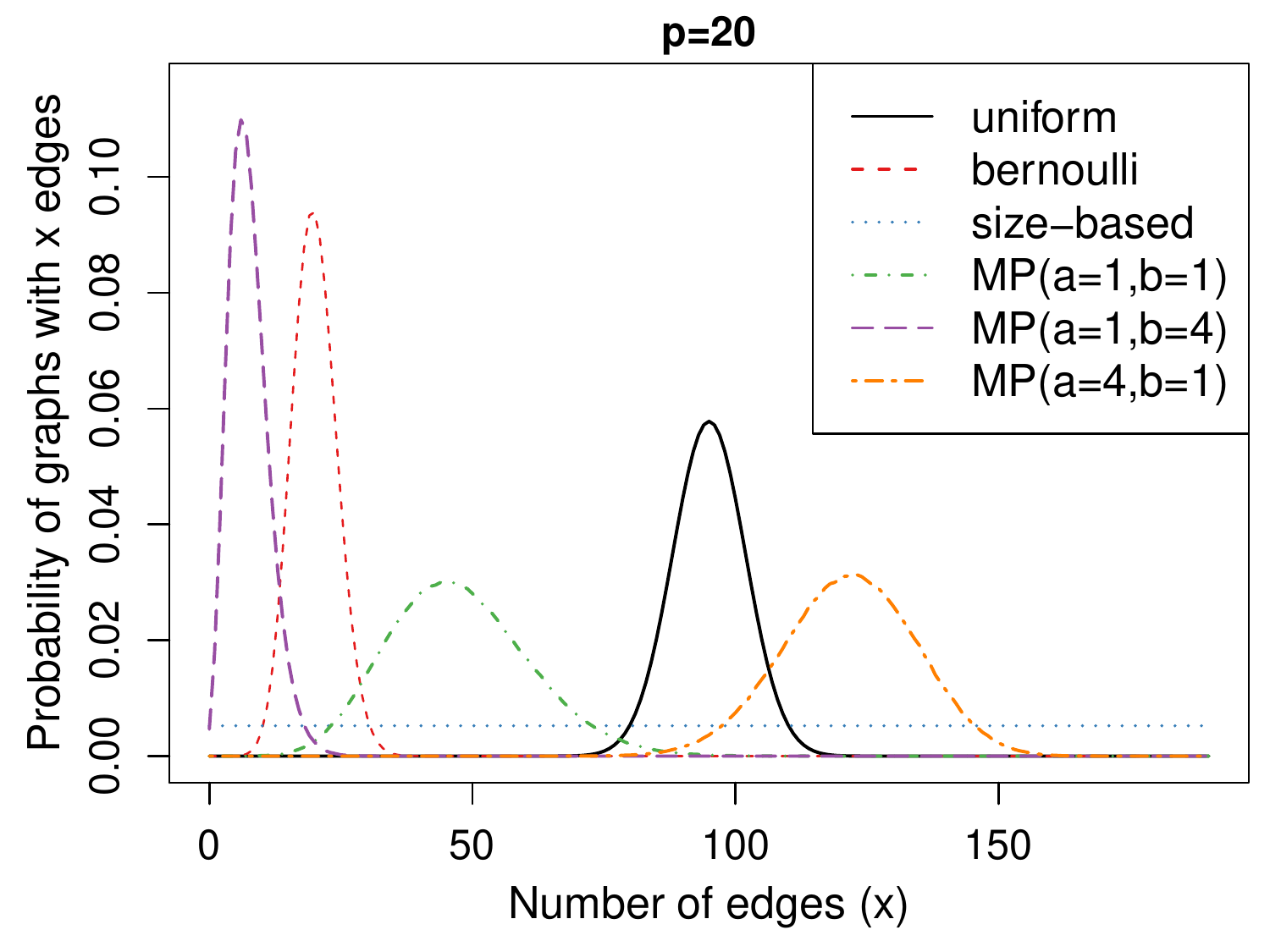}
	\caption{Plot shows the probability allocated to graphs with $x$ edges by the uniform prior, the Bernoulli prior \citep{Jones2005} with probability of inclusion of each edge: $\alpha =2/(p-1)$, the size-based prior \citep{Armstrong2009} or equivalently the Bernoulli prior with $\alpha \sim \text{Uniform}[0,1]$ integrated out \citep{Carvalho2009} and the multiplicative prior (MP) for different values of $a$ and $b$. \label{pxedges}}
\end{figure}

In the absence of any prior belief on the graphical structure, a uniform prior over all graphs is often used in estimating GGMs \citep[e.g.][]{Lenkoski2011, Wang2012}. That is, given $p$ nodes, it is assumed that each of the $2^r$ possible graphs, where $r= p(p-1)/2$, has equal probability of arising. This prior concentrates its mass on graphs with moderately large number of edges and the expected number of edges as well as the mode is $r/2$ (see Figure \ref{pxedges}). Thus, this prior may not be appropriate when sparse graphs are desired. Several alternatives have been developed. To encourage sparse graphs, \cite{Dobra2004} and \cite{Jones2005} propose a prior where every edge is included independently with a small probability $\alpha$ so that a graph with $x$ edges has prior probability $\alpha^{x}(1-\alpha)^{r-x}$. This prior is known as the Erd\H{o}s-R\'{e}nyi model in random graph theory and it reduces to the uniform prior when $\alpha = 0.5$. \cite{Jones2005} recommend taking $\alpha =2/(p-1)$ so that the expected number of edges is $p$. \cite{Carvalho2009} treat $\alpha$ as a model parameter rather than a fixed tuning constant. They place a Beta$(a,b)$ prior on $\alpha$ so that a graph with $x$ edges has prior probability $B(a+x,r+b-x)/B(a,b)$, where $B(a,b)$ denotes the Beta function. When $a=b=1$, this probability simplifies to $\frac{1}{r+1}{r \choose x}^{-1}$. This prior is equivalent to the size-based prior \citep{Armstrong2009} when the graph space is unrestricted. Under the size-based prior, every size $0, \dots, r,$ has equal probability and every graph of the same size has equal probability. \cite{Carvalho2009} demonstrate that their proposed prior has strong control over the number of spurious edges and corrects for multiple hypothesis testing automatically, where each null hypothesis corresponds to the exclusion of one edge. 

We propose using the multiplicative or Chung-Lu random graph model as a prior on the graphical space of GGMs. Given a desired or expected degree sequence $\{d_1, \dots, d_p\}$, where $d_i$ denotes the degree (number of neighbours) of node $i$, the multiplicative model \citep{Chung2002} assumes that the edge between each pair of nodes $i$ and $j$ is formed independently with probability $p_{ij}$ proportional to the product $d_id_j$. Allowing self-loops and provided $(\operatorname*{max}_i d_i)^2 < \sum_i d_i$, the expected degree of node $i$ is exactly $d_i$. The multiplicative model is able to capture degree distributions which are more diverse (e.g. right-skewed, U-shaped) and closer to that of real-world networks than the Erd\H{o}s-R\'{e}nyi model. Notably, the Erd\H{o}s-R\'{e}nyi model has a degree distribution that is binomial and can be viewed as a special case of the multiplicative model with a constant expected degree sequence. We consider an alternative parametrization of the multiplicative model introduced by \cite{Olhede2013}, which dispenses with self-loops and the normalization constraint by taking $p_{ij} = \pi_i \pi_j$ and $0 < \pi_i < 1$ for each $i$. They derive degree characteristics and large-sample approximations of this model, which lends insight on the variation attainable in degree structure. Adopting a Bayesian approach, we treat each $\pi_i$ as a variable with a Beta$(a,b)$ prior. We present degree and clustering properties of the multiplicative model, showing how they depend on choices of $a$ and $b$. In the context of GGMs, we show that the multiplicative model provides an avenue to encourage sparsity or graphs that exhibit particular degree patterns based on prior knowledge obtained through expert opinion or past data. We further demonstrate how the multiplicative model can be extended to include covariates and become a prior on joint graphs for multiple GGMs. 

Several approaches for joint inference of multiple GGMs have been developed recently. \cite{Guo2011} estimate precision matrices for different groups jointly by parameterizing each entry as a product of two factors: one factor is common across all groups while the other is group specific. A hierarchical $l_1$ penalty is imposed and optimization is performed using graphical lasso \citep{Friedman2008}. \cite{Danaher2014} formulate a more general framework called joint graphical lasso and introduce two convex penalty functions; the fused graphical lasso which encourages edge value on top of structural similarity and the group lasso which only encourages a shared sparsity pattern. \cite{Chun2014} extend the approach of \cite{Guo2011} to a wider class of nonconvex penalty functions. \cite{Mohan2014} consider a node-based approach where multiple GGMs are estimated using a convex regularizer by assuming that the similarities between networks are due to the shared presence of certain highly-connected nodes which serve as hubs and the differences are due to some nodes whose connectivity changes across conditions. \cite{Yajima2015} compare the Gaussian directed acyclic graphs of two subgroups using Bayesian inference via Gibbs sampling. The strength of association between two variables in the differential group is modeled as the strength in the baseline group plus an edge-specific parameter controlling the difference in association between the two subgroups. They define a prior on the graphical space by centering on a prior graph constructed from a database \citep{Telescar2012}. \cite{Peterson2015} consider an alternative Bayesian approach which links graphs from different groups using a Markov Random Field. The probability of inclusion of each edge in graphs $1, \dots, K,$ is parameterized in terms of a $K \times K$ symmetric matrix which measures the pairwise similarity of groups and is common across all edges, and an edge-specific $K \times 1$ vector which controls the inclusion probability in each group independently of group relatedness. Priors are further placed on these parameters and a block Gibbs sampler is used for inference.

The approach that we use to link multiple graphs is based on the multiplicative model by expressing the connectivity of each node as a logistic regression function of graph specific covariates. As the multiplicative model decouples the inclusion probability of each edge into the product of the connectivities of the end nodes, the resulting model is parsimonious and scales linearly in the number of variables and graphs. For inference, we develop an SMC sampler for estimating the joint posterior distribution of the graphs. Using tempering techniques \citep[see e.g.][and the references therein]{Moral2006}, we create a sequence of probability distributions from which to sample, moving gradually from a distribution that is easy to sample from, through artificial intermediate distributions towards the posterior distribution of interest.

\section{Multiplicative model} \label{S-multmodel}
Here we define our notation and present the multiplicative model, followed by a study of its properties. These properties lend insight on the structure and range of networks that can be generated from the multiplicative model. They are also useful in the determination of suitable hyperparameters based on prior understanding of data obtained through expert opinion or a database. 

Let $G = (V,E)$ be a simple graph with vertex set $V=\{1, 2, \dots, p\}$ and edge set $E \subseteq \{(i,j) \in V \times V:i<j\}$. A simple graph is undirected and does not contain self-loops or multiple edges. The adjacency matrix $A = [A_{ij}]$ of $G$ is a $p \times p$ binary matrix where $A_{ij}$ is 1 if an edge is present between nodes $i$ and $j$, and 0 otherwise for $i,j \in V$. As $G$ is simple, $A$ is symmetric and has zeros on its diagonal.

In the multiplicative model, each edge is modeled independently as
\begin{gather} \label{E-multmodel}
A_{ij} \sim \text{Bernoulli}(p_{ij}) \text{ for } 1 \leq i < j \leq p, \nonumber \\ 
p_{ij} = \pi_i \pi_j, \text{ where }  0 \leq \pi_i \leq 1 \text{ for } i=1, \dots,p. 
\end{gather}
Thus, every edge $A_{ij}$ is formed independently with probability $p_{ij}$, where $p_{ij}$ is a product of the tendencies of nodes $i$ and $j$ to form edges with other nodes. The parameter $\pi_i$ is characteristic of node $i$ and reflects its activity level. We refer to $\pi_i$ as the \textit{connectivity} of $i$. The Erd\H{o}s-R\'{e}nyi random graph model arises as a special case when $\pi_i$ is constant across all $i$, that is, every link is formed independently with equal probability. 

We adopt a Bayesian approach and place an independent Beta prior on each $\pi_i$. Let
\begin{equation} \label{E-betaprior}
\pi_i \sim \text{Beta}(a,b) \text{ for } i=1, \dots, p,
\end{equation}
where $a, b>0$. We have $p(\pi_i) = \pi_i^{a-1}(1-\pi_i)^{b-1}/ B(a,b)$, where the Beta function $B(a,b) =\frac{\Gamma(a+b)}{\Gamma(a)\Gamma(b)}$. Let $\pi=(\pi_1, \dots, \pi_p)^T$ and $p(\pi) = \prod_{i=1}^p p(\pi_i)$. Networks of highly varying densities and structures can be formed by choosing different hyperparameters $a$ and $b$. 

\subsection{Degree and clustering properties} \label{S-Properties}
The degree $D_i$ of a node $i$ is the number of links that involve $i$ or the number of neighbours of $i$, and is given by $D_i = \sum_{j \neq i} A_{ij}$. The properties below describe the degree distribution and cohesiveness of networks generated from the multiplicative model. Their implications are discussed later in the section. We follow the framework in \cite{Rastelli2015}, which is based on probability generating functions \cite[see][]{Newman2001}. Proofs are given in the Supplementary Material. We note that some of these results have been discussed in \cite{Olhede2013} but not with regards to the Beta$(a,b)$ prior. In the following, let $\mu = \frac{a}{a+b}$ and $\sigma^2 = \frac{ab}{(a+b)^2 (a+b+1)}$ denote the mean and variance of a Beta$(a,b)$ distribution respectively. 

\begin{enumerate}[label=\textbf{P\arabic*:},ref=P\arabic*]
\item \label{prob_given_pii} 
The probability that a randomly chosen node is a neighbour of a node with connectivity $\pi_i$ is $\mu \pi_i$. 
	
\item \label{av_deg_given_pii} 
The degree of a node with connectivity $\pi_i$ is distributed as $\text{Binomial}(p-1,\mu \pi_i)$. Hence its average degree is $(p-1)\mu \pi_i$, which is proportional to $\pi_i$.
	
\item \label{pgf} 
The probability generating function of the degree of a randomly chosen node is given by 
\begin{equation}
G_{D_i}(z) =\sum_{d=1}^{p-1} \text{P}(D_i=d)z^d= \int_0^1 (1 - \mu \pi_i + \mu \pi_i z )^{p-1} p(\pi_i) \, \text{d} \pi_i.
\end{equation}
The $k$th factorial moment of $D_i$, $\text{E}\{D_i(D_i-1) \dots (D_i - k+1)\}$, is given by 
\begin{equation*}
\text{E}\left( \frac{D_i !}{(D_i-k)!} \right) = G_{D_i}^{(k)}(1) = \frac{(p-1)!\, B(a+k,b)}{(p-1-k)! \, B(a,b)} \mu^k 
\end{equation*}
for any positive integer $k$.
	
\item \label{av_deg} 
The average degree of a randomly chosen node is $\text{E}(D_i) = (p-1)\mu^2$ and the variance is $\text{Var}(D_i) = (p-1)\mu^2 \{1-\mu^2 +(p-2) \sigma^2\}$.
	
\item \label{Rdegdistn}
The degree distribution of a randomly chosen node is given by
\begin{equation*}
P(D_i = d) = {p-1 \choose d}  \frac{\mu^d}{ B(a,b)} \int_0^1 \pi_i^{a+d-1} (1-\mu \pi_i)^{p-1-d} (1-\pi_i)^{b-1} \, \text{d}\pi_i
\end{equation*}	
for $d \in \{0, \dots, p-1\}$. When $b=1$,  $P(D_i = d) = {p-1 \choose d} a B(\mu, a+d,p-d)/\mu^a$, where $B(x; a, b) = \int_0^x t^{a-1} (1-t)^{b-1} \; dt$ is the incomplete Beta function.
	
\item \label{Rdisp}
The dispersion index of the degree distribution is given by 
\begin{equation*}
1 -\frac{a\{a^2 + (b+1)a -(p-2)b\}}{(a+b)^2(a+b+1)}.
\end{equation*}
\begin{itemize}[noitemsep, leftmargin=1em, labelsep=5pt]
\item When $0 < a < \{ \sqrt{b^2+(4p-6)p+1} -b-1\}/2$, the distribution has dispersion index greater than 1 and is over-dispersed.
\item When $a = \{ \sqrt{b^2+(4p-6)p+1} -b-1\}/2$, the distribution has dispersion index 1 (equal to that of a Poisson distribution).
\item When $a > \{ \sqrt{b^2+(4p-6)p+1} -b-1\}/2$, the distribution has dispersion index less than 1 (similar to that of a Binomial distribution) and is under-dispersed.
\end{itemize}
	
\item \label{Rskew}
The skewness index or Pearson's moment coefficient of skewness of the degree distribution can be computed as $\{ \text{E}(D_i^3) - 3\text{E}(D_i)\text{Var}(D_i) -  \text{E}(D_i)^3 \} / \text{Var}(D_i)^{1.5} $, where $\text{E}(D_i^3) = (p-1)\mu^2 \{1+3(p-2)( \mu^2+\sigma^2)+(p-2)(p-3)\mu E(\pi_i^3) \} $ and $\text{E}(\pi_i^3) = \frac{(a+2)(a+1)a}{(a+b+2)(a+b+1)(a+b)}$.
	
\item \label{Rdegneighbour_connect}
The average degree of the neighbours of a node is independent of the connectivity or degree of that node, and is given by $1 +(p-2)(\mu^2 + \sigma^2)$.
	
\item \label{PClust}The global clustering coefficient, which measures the probability that nodes $j$ and $k$ are linked given that both nodes are linked to $i$,  is given by $\frac{a+1}{a+b+1}$.
\end{enumerate}

In the Erd\H{o}s-R\'{e}nyi model, $D_i$ is distributed as Binomial$(p-1,\alpha)$, where $\alpha$ is the probability of inclusion of each edge. When $\alpha = \mu^2$, the mean degree of a randomly chosen node in the multiplicative model is equal to that in the Erd\H{o}s-R\'{e}nyi model from \ref{av_deg}. However, the variance of the degree distribution in the multiplicative model is greater than that in the Erd\H{o}s-R\'{e}nyi model by $(p-1)(p-2)\sigma^2$. Thus, as the number of nodes increases, the multiplicative model can accommodate greater variation in the degree distribution than in the Erd\H{o}s-R\'{e}nyi model by $\mathcal{O}(p^2)$ given the same mean degree. 

\begin{figure}
	\centering \includegraphics[width=0.88\linewidth]{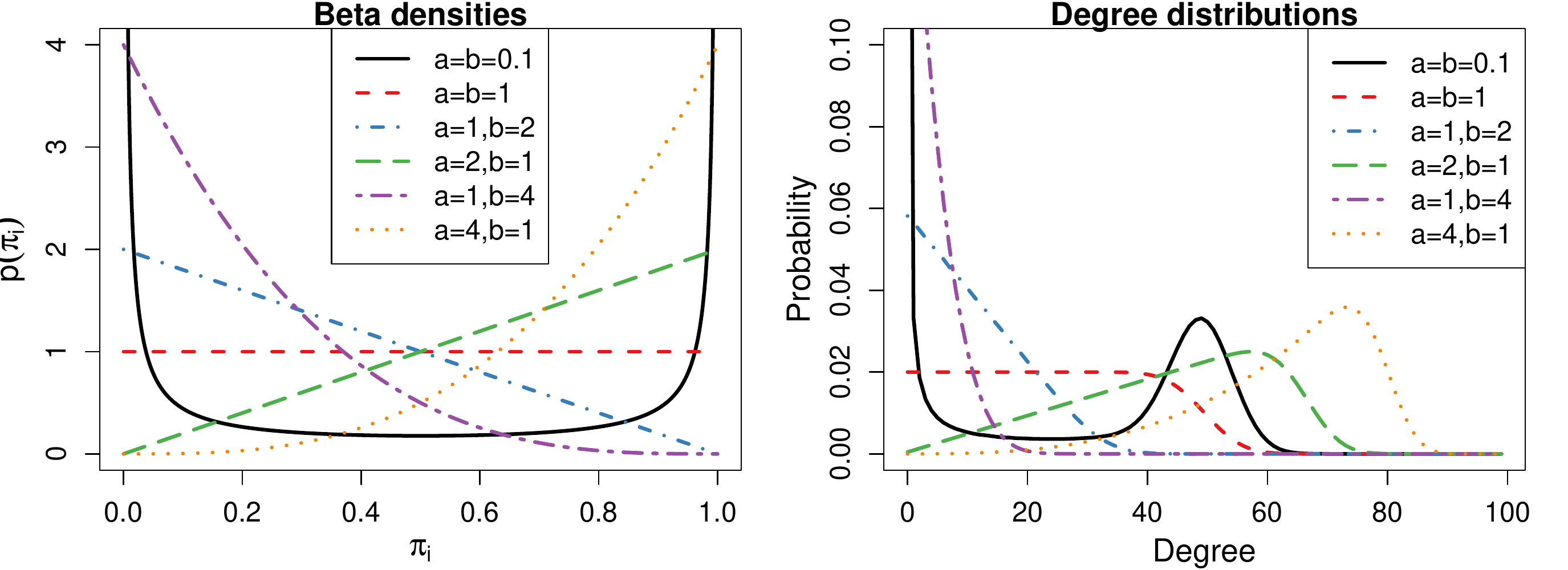}
	\caption{\label{F-degdistn} Beta densities (left) and degree distributions of the multiplicative model (right) corresponding to different hyperparameter settings when $p=100$.}
\end{figure}

Figure \ref{F-degdistn} shows the degree distributions of the multiplicative model for graphs with $p=100$ nodes under different hyperparameter settings. When the degree distributions cannot be computed directly using \ref{Rdegdistn}, they are estimated via simulation using $10^5$ graphs. Degree distributions of a variety of shapes (e.g. right-skewed, U-shaped) can be obtained from the multiplicative model by varying $a$ and $b$.

The dispersion index measures how clustered a distribution is compared to standard statistical models. From \ref{Rdisp}, the degree distribution is over-dispersed when $a$ is small and under-dispersed when $a$ is large. In fact, as $a  \rightarrow \infty$ (and/or $b \rightarrow \infty$), $\sigma^2 \rightarrow 0$ and each $\pi_i$ reduces to a point mass. The multiplicative model thus degenerates and reduces to the Erd\H{o}s-R\'{e}nyi model with constant probability of inclusion for every edge. As the degree distribution is over-dispersed for a wide range of hyperparameter values, the multiplicative model is able to represent well heterogeneity in degree sequences. 

The skewness index in \ref{Rskew} is useful for identifying asymmetries in degree distributions. Generally, the degree distribution is positively skewed when $a$ is small and $b$ is large and negatively skewed vice versa (plots of the dispersion index and skewness as a function of $a$ and $b$ can be found in the Supplementary Material Figures \ref{disp} and \ref{skew}). 

\begin{figure}[tb!]
\centering \includegraphics[width=0.9\linewidth]{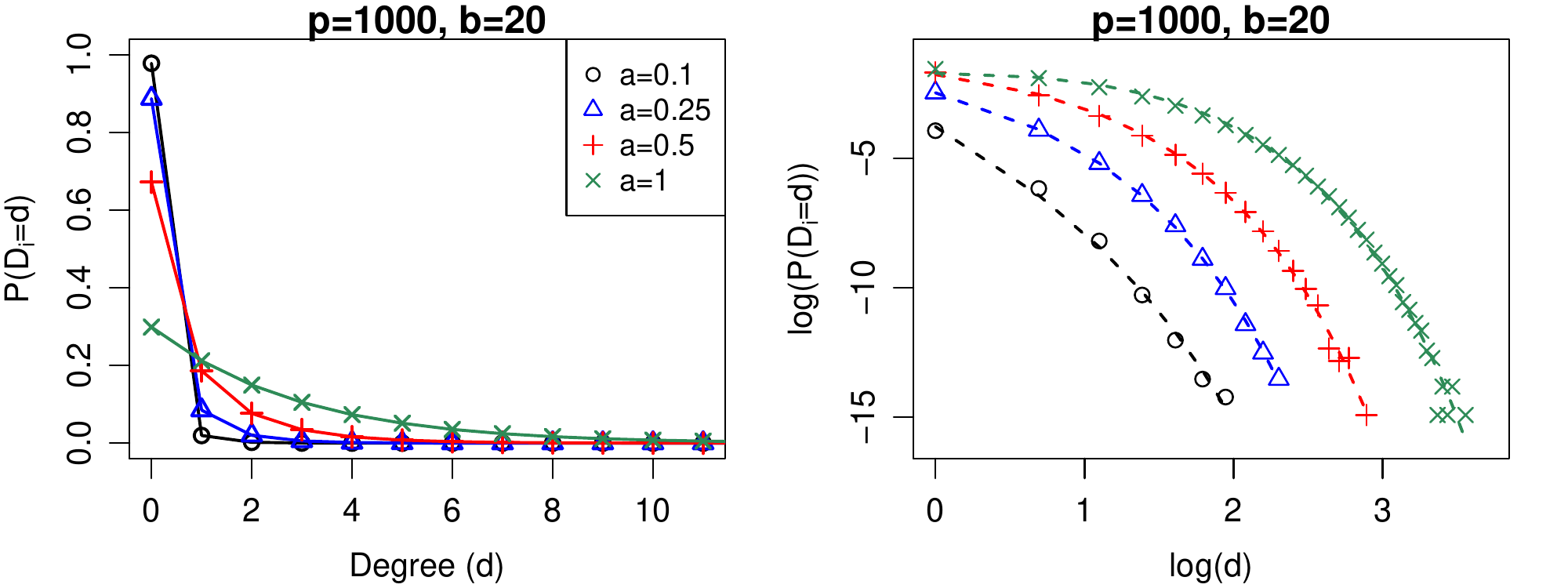}
\caption{\label{powerlaw} Degree distributions for investigating power-law. Right plot is in log-scale.}
\end{figure}

Of particular interest are scale-free networks whose degree distributions follow a power law ($P(D_i=d) \propto d^{-\gamma}$ where $\gamma$ is a positive constant). \cite{Olhede2013} show that the multiplicative model can lead to networks with power law degree distributions when $p$ is large, the elements of $\pi$ are ordered such that $\pi_1 \geq \pi_2 \geq \dots \geq \pi_p$ and a polynomial decay of $\pi_i$ with $i$ is assumed; $\pi_i \propto i^{-\gamma}$ for $0 < \gamma < 1$. We investigate via simulations the behavior of the degree distributions when $\{\pi_i\}$ is modeled instead as a random sample from a Beta$(a,b)$ prior. As scale-free networks tend to have large positive values for the skewness index, we consider a large $b=20$ and some small values of $a \in \{0.1,0.25,0.5,1\}$. The left plot in Figure \ref{powerlaw} shows the degree distributions obtained via simulation and the right plot shows the relationship between $\log P(D_i=d)$ and $\log d$, which should be a straight line if the power-law is satisfied. We observe that the multiplicative model (with a Beta prior) comes close to but does not quite induce power law networks as the right tail is not sufficiently heavy. However, we find that these points are well fitted by a power law with an exponential cutoff \citep[$P(D_i=d) \propto d^{-\gamma} \exp(-\tau d)$,][]{Newman2001a}. Fits of these form are shown as dotted lines in the right plot of Figure \ref{powerlaw}. In such networks, the power law dominates for small $d$ but turns into an exponential decay for large $d$. A broad range of empirical data such as protein interaction networks \citep{Jeong2001, Giot2003} and scientific collaboration networks \citep{Fenner2007} have been found to exhibit power-laws with exponential cutoffs instead of pure power laws due to finite-size effects such as the physical limitation of the binding sites in the protein structure and the finite working lifetime of a scientist. \cite{DSouza2007} provides further examples.

\section{Gaussian Graphical Models} \label{S-GGM}
Suppose we have a dataset with $p$ variables and $K$ groups or classes. Let $y_h=(y_{h1}, \dots, y_{hp})$ denote the $h^{\text{th}}$ observation of the $p$ variables for $h=1, \dots, H,$ and $S_k$ be an index set containing the indices of observations which belong to group $k$ for $k=1, \dots, K$. The number of observations in $S_k$ is denoted by $|S_k|$ and $H= \sum_{k=1}^K |S_k|$. Without loss of generality, we assume that the observations in each group are centered at 0 along each variable. We consider
\begin{equation} \label{E-GGM}
y_h|\Omega_k \sim N(0, \Omega_k^{-1}) \;\; \text{for} \;\; h \in S_k,
\end{equation}
where $\Omega_k$ is a $p \times p$ precision matrix and $k \in \{1,\dots,K\}$. 

Let $G_k=(V,E_k)$ be a simple graph with vertex set $V=\{1, 2, \dots, p\}$ and edge set $E_k \subseteq \{(i,j) \in V \times V:i<j\}$ for $k=1,\dots,K$. Node $i \in V$ represents the $i$th variable and each edge $(i,j) \in E_k$ corresponds to $\Omega_{k,ij} \neq 0$. That is, $y_{hi}$ and $y_{hj}$ are conditionally independent  (in $G_k$) given the rest of the elements in $y_h$ if and only if $\Omega_{k,ij}=0$, or equivalently $(i,j) \notin E_k$. The conjugate prior for $\Omega_k$ is the $G$-Wishart distribution \citep{Atay2005}, $W_{G_k}(\delta_k, D_k)$, which has density
\begin{equation*}
p(\Omega_k|G_k) = \frac{1}{I_{G_k}(\delta_k, D_k)} |\Omega_k|^{(\delta_k-2)/2} \exp\left\{ -\frac{1}{2} \text{tr}(\Omega_k D_k)  \right\}.
\end{equation*}
Here, $\Omega_k$ is constrained to the cone $P_{G_k}$ of positive definite matrices with entries equal to zero for all $(i,j) \notin E_k$ and $I_{G_k}(\delta_k, D_k)$ is a normalizing constant such that
\begin{equation*}
I_{G_k}(\delta_k, D_k) = \int_{\Omega_k \in P_{G_k}} |\Omega_k|^{(\delta_k-2)/2} \exp\left\{ -\frac{1}{2} \text{tr}(\Omega_k D_k)\right\} \; \text{d}K.
\end{equation*} 
This normalizing constant is guaranteed to be finite if $\delta_k >2$ and $D_k^{-1} \in P_{G_k}$ \citep{Diaconis1979}. The $G$-Wishart distribution reduces to the Wishart distribution when $G_k$ is complete, and the normalizing constant can then be evaluated in closed form as 
\begin{equation}\label{normal_const_closed_form}
I_{G_k}(\delta_k,D_k) = 2^{(\delta_k+p-1)p/2} \; \Gamma_p\{(\delta_k+p-1)/2\} |D_k|^{-(\delta_k+p-1)/2},
\end{equation}
where $\Gamma_p(a) = \pi^{p(p-1)/4} \prod_{i=0}^{p-1} \Gamma(a-\frac{1}{2})$ for $a>(p-1)/2$.

\subsection{Priors over graphs} 
We use the multiplicative model to assign prior probabilities to graphs. Let $A_{k}=[A_{k,ij}]$ be the adjacency matrix of $G_k$ for $k=1,\dots,K$. Consider
\begin{equation} \label{bernoulli_distn}
A_{k,ij}|\pi_{k,i} \pi_{k,j} \sim \text{Bernoulli} (\pi_{k,i} \pi_{k,j}),
\end{equation}
where $0 \leq \pi_{k,i} \leq 1$ for $i=1, \dots,p$ and $k=1,\dots,K$. As in Section \ref{S-multmodel}, the probability that an edge $(i,j)$ is present in $E_k$ is given by $\pi_{k,i} \pi_{k,j}$, the product of the propensities of nodes $i$ and $j$ to form edges with other nodes in $G_k$. Priors are further placed on each $\pi_{k,i}$. We consider the cases $K=1$ and $K>1$ separately.

\subsubsection{When $K=1$}
When $K=1$, there is only one group and the subscript $k$ indicating different groups may be dropped so that $G_1 = G $ and $\pi_{1,i}= \pi_i $ for $i=1, \dots, p$. We place a Beta$(a,b)$ prior on each $\pi_i$ as in \eqref{E-betaprior}. The prior probability of $G$ with adjacency matrix $A$ is then given by 
\begin{equation}\label{E-Gprior}
\begin{aligned}
p(G|a,b) &= \int p(G|\pi) p(\pi|a,b) \; d\pi \\
&= \frac{1}{B(a,b)^p} \int  \prod_{i,j: \,i<j} \; (1-\pi_i \pi_j )^{(1-A_{ij})} \prod_{i=1}^p \pi_i^{(a+d_i-1)} (1-\pi_i)^{(b-1)} \, d\pi,
\end{aligned}
\end{equation} 
where $d_i$ denotes the degree of node $i$.

\subsubsection{When $K>1$} \label{sec: mp_K_big}
We propose a joint prior for $G_1, \dots, G_K,$ which is allowed to depend on covariates specific to each graph. First, we express $\pi_{k,i} $ in terms of a logistic regression as
\begin{equation*}
\pi_{k,i}  =  \frac{\exp(\beta_i^T x_k)}{1+\exp(\beta_i^Tx_k)}
\end{equation*}
for $i=1, \dots, p,$ and $k=1,\dots,K$, where $x_k = (x_{k1}, \dots, x_{kQ})^T$ is a vector of covariates for $G_k$ and $\beta_i = (\beta_{i1}, \dots, \beta_{iQ})$ is a vector of coefficients specific to node $i$. Let $x = (x_1, \dots,x_K)$ and $\beta=(\beta_1^T, \dots, \beta_p^T)^T$. We consider a normal prior for each $\beta_{iq}$ such that
\begin{equation*}
\beta_{iq}|\sigma_q^2 \sim N(0, \sigma_q^2)
\end{equation*}
for $i=1, \dots,p$ and $q=1,\dots,Q$. Let $\sigma^2=(\sigma_1^2, \dots, \sigma_Q^2)$ be a hyperparameter assumed to be known. The joint prior probability of $G_1, \dots, G_K,$ is then given by 
\begin{multline} \label{E-jointprior}
p(G_1, \dots, G_K|x,\sigma^2) = \int p(\beta|\sigma^2)  \prod_{k=1}^K p(G_k|x_k,\beta)  d\beta \\
= \int  \prod_{i=1}^p \prod_{q=1}^Q \left\{ \frac{\exp (-\frac{\beta_{iq}^2}{2\sigma_q^2})}{\sqrt{(2\pi \sigma_q^2)}}   \right\}   \prod_{k=1}^K \left\{ \prod_{i=1}^p \pi_{k,i}^{d_{k,i}} \prod_{i <j} (1-\pi_{k,i} \pi_{k,j})^{1-A_{k,ij}} \right\}  d\beta,
\end{multline}
where $d_{k,i}$ denotes the degree of node $i$ in $G_k$ for $k=1,\dots,K$.

As an example, in the application on urinary metabolic data, we consider $K=2$ and the covariates $x_k$ to include an intercept and an indicator for level of exposure to cadmium (1 if above the median and 0 otherwise). We take $x_1 = (1,0)$ and $x_2=(1,1)$ so that $G_1$ and $G_2$ represent the correlation structure of the groups with exposure to cadmium below or equal to the median, and above the median respectively. The connectivity of node $i$ is $\pi_{1,i} = \{1+\exp(-\beta_{i1})\}^{-1}$ in $G_1$ and $\pi_{2,i} = \{1+\exp(-\beta_{i1} - \beta_{i2})\}^{-1}$ in $G_2$. Thus $\beta_{i1}$ determines the popularity of node $i$ in $G_1$ while $\beta_{i2}$ is a differential parameter which controls the difference in popularity of node $i$ between $G_1$ and $G_2$. If $\beta_{i2}$ is close to zero, the connectivity of node $i$ in $G_1$ and $G_2$ is similar. As the magnitude of $\beta_{i2}$ increases, the difference in connectivity of node $i$ between $G_1$ and $G_2$ becomes greater. See illustration in Supplementary Material Figure \ref{F-betas}. 

Figure \ref{F-degpair} shows the prior degree distributions of $G_1$ and $G_2$ for $p=50$ and different values of $\sigma^2$. These plots are obtained via simulation of $10^5$ joint pairs of graphs in each case. When $\sigma_1^2=\sigma_2^2 = 0.1$, both $\beta_{i1}$ and $\beta_{i2}$ are close to zero, and $\pi_{1,i}$ and $\pi_{2,i}$ are close to 0.5. Thus the degree distribution is shaped like a Binomial curve, resembling the Erd\H{o}s-R\'{e}nyi model where each edge is formed independently with constant probability 0.25. As $\sigma_1^2$ increases, there is greater variation in the degree sequence of $G_1$. When $\sigma_1^2$ is large, the connectivity of each node tends to the extremes of 0 and 1 (each node has a high probability of being either very connected or isolated). Thus the degree distribution resembles the case where each $\pi_i$ is allocated a U-shaped Beta(0.1,0.1) prior as shown in Figure \ref{F-degdistn}. The distinction between the degree distribution of $G_1$ and $G_2$ becomes greater as $\sigma_2^2$ increases. 
\begin{figure}[htb!]
	\centering
	\includegraphics[width=\linewidth]{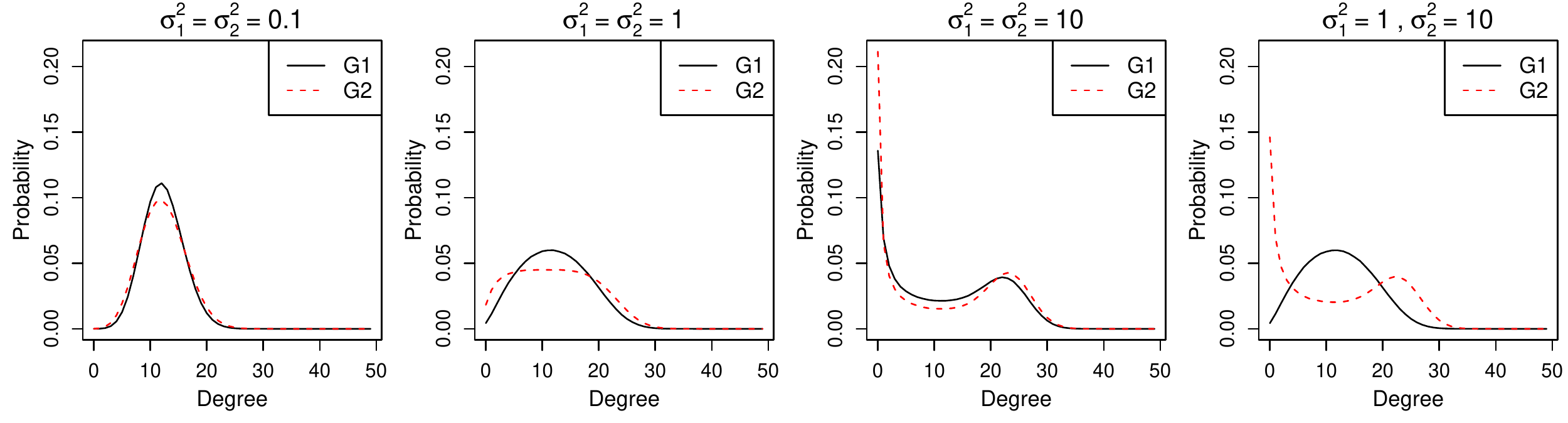}
	\caption{\label{F-degpair} Prior degree distributions of $G_1$ and $G_2$ for $p=50$ and different combinations of $\sigma_1^2$ and $\sigma_2^2$. Covariates for $G_1$ and $G_2$ are (1,0) and (1,1)  respectively.}
\end{figure}

We can also add a third covariate say for gender (1 if male and 0 for female) so that $K=4$ and take $x_1 = (1,0,0)$, $x_2=(1,0,1)$, $x_3 = (1,1,0)$ and $x_4=(1,1,1)$. Then $G_1$, for instance, will represent the correlation structure for the group of females with exposure to cadmium below or equal to the median level.

\section{Posterior distribution} \label{S-postdistn}
Let $y=(y_1, \dots, y_H)$. For $K>1$, the joint distribution of the model can be written as
\begin{multline*}
p(y,\Omega_1, \dots,\Omega_K,G_1,\dots,G_K,\beta|x,\sigma^2) \\
= p(\beta|\sigma^2)  \prod_{k=1}^K\left\{ p(G_k|x_k,\beta) p(\Omega_k|G_k) \prod_{h \in S_k} p(y_h|\Omega_k) \right\}.
\end{multline*}
Integrating out $\Omega_k$, the marginal likelihood $p(\{y_h|h \in S_k\}|G_k)$ can be shown \citep[see, e.g.][]{Atay2005} to be given by
\begin{equation*}
p(\{y_h|h \in S_k\}|G_k) = (2\pi)^{-p|S_k|/2} I_{G_k}(\delta_k+|S_k|,D_k+ \sum_{h \in S_k} y_h y_h^T)/I_{G_k}(\delta_k,D_k).
\end{equation*}
Integrating out $\beta$ as well, we have
\begin{multline} \label{E-targetjoint}
p(G_1, \dots, G_K|y, x, \sigma^2) \\
 \propto p(G_1, \dots, G_K|x,\sigma^2)  \prod_{k=1}^K I_{G_k}(\delta_k+|S_k|,D_k+ \sum_{h \in S_k} y_h y_h^T) / I_{G_k}(\delta_k,D_k).
\end{multline}
When $K=1$, the posterior distribution can be derived similarly. The only difference is that the dependence on $x$ and $\sigma^2$ is replaced by the Beta prior hyperparameters $a$ and $b$. We have
\begin{equation} \label{E-targetG}
p(G|y, a,b) \propto p(G|a,b) I_{G}(\delta+H,D_k+ \sum_{h=1}^H y_h y_h^T) / I_G(\delta,D).
\end{equation}

For posterior inference, we propose a SMC algorithm to obtain samples from the posterior distribution. To compute the right-hand side of \ref{E-targetjoint} and \ref{E-targetG}, we note that for any graph $G$ (not necessarily decomposable), normalizing constants of the form $I_G(\delta,D)$ can be evaluated by first factorizing $G$ into its prime components and their separators \citep[see, e.g.][]{Lauritzen1996}. Suppose $(\mathcal{P}_1, \dots, \mathcal{P}_L)$ is a perfect sequence of the prime components of $G$ and $(\mathcal{S}_2, \dots, \mathcal{S}_L)$ is the corresponding set of separators. Then
$I_G(\delta,D) = \prod_{l=1}^L I_{G_{\mathcal{P}_l}} (\delta,D) / \prod_{l=2}^L I_{G_{\mathcal{S}_l}} (\delta,D)$, where $G_{\mathcal{P}_l}$ and $G_{\mathcal{S}_l}$ denote the subgraphs induced by $\mathcal{P}_l$ and $\mathcal{S}_l$ respectively. As the separators are complete, $I_{G_{\mathcal{S}_l}} (\delta,D)$ can be evaluated as in \eqref{normal_const_closed_form}. The same applies to $I_{G_{\mathcal{P}_l}} (\delta,D)$ for any prime component $\mathcal{P}_l$ which is complete. Otherwise, we estimate $I_{G_{\mathcal{P}_l}} (\delta,D)$ using the Monte Carlo method of \cite{Atay2005} when $\delta$ is small and the Laplace approximation of \cite{Lenkoski2011} when $\delta$ is large. This combination of using Laplace approximation and Monte Carlo integration to evaluate the normalizing constants is feasible as the size of the graphs considered in this paper is moderately small ($p \leq22$). When $p$ is large, the size of the Monte Carlo sample has to be increased dramatically in order for the variance to be controlled and Monte Carlo integration becomes a computational bottleneck \citep[see][]{Jones2005, Wang2012}. At this point, techniques that avoid evaluation of prior normalizing constants (and that explore the space of graphs and precision matrices jointly) based on for instance, the exchange algorithm \citep{Murray2006} have to be integrated with SMC sampler. The priors on graphs are estimated using Laplace approximation, which is described next.

\subsection{Laplace approximation for prior on graphs}
Evaluating $p(G|a,b)$ or $p(G_1, \dots, G_K|x,\sigma^2)$ via Monte Carlo becomes more computationally intensive as $p$ increases and we estimate these quantities efficiently using Laplace approximation instead. We consider 
\begin{equation} \label{E-Laplaceapprox}
\int \exp \{f(u) \} \;\text{d}u  \approx (2\pi)^{\frac{n}{2}} |-H(u_0)|^{-\frac{1}{2}} \exp\{f(u_0)\},
\end{equation}
where $u = (u_1, \dots, u_d)^T$, $u_0$ is the mode of $f(u)$ and $H(u_0)$ denotes the Hessian of $f$ evaluated at $u_0$. The mode $u_0$ can be found using numerical methods.

For $K=1$, we estimate $p(G|a,b)$ in \eqref{E-Gprior} using \eqref{E-Laplaceapprox} by first making a change of variable and letting $\pi_i=\frac{\exp(u_i)}{1+\exp(u_i)}$. For $K>1$, we estimate $p(G_1, \dots, G_K|x,\sigma^2)$ in \eqref{E-jointprior} using \eqref{E-Laplaceapprox} by taking $u = \beta$. Detailed functional and Hessian expressions are given in the Supplementary Material.

\section{Sequential Monte Carlo sampler} \label{S-SMC}
We use SMC samplers for posterior inference. Suppose we are interested in sampling from a complex target $\lambda(x)$. The idea is to start with some distribution $\lambda_1$ that is easy to sample from and move via a sequence of intermediate distributions, $\lambda_2, \dots, \lambda_{T-1}$, towards the distribution of interest $\lambda_T = \lambda$. At any time $t$, a large collection of weighted samples $\{W^{(n)}_t, X^{(n)}_t | n=1, \dots N\}$ is maintained, and these particles are used to generate samples from the target distribution at the next time point using sequential importance sampling (SIS) and resampling methods. The motivation is that it would be easier to move the particles from one target to the next if $\lambda_t$ is close to $\lambda_{t+1}$. 

\subsection{Review of methodology}
We first review SIS and SMC briefly. Let $\lambda_1, \dots, \lambda_T,$ be the target densities, $\gamma_1, \dots, \gamma_T,$ be unnormalized densities such that $\lambda_t(x_{1:t}) \propto \gamma_t(x_{1:t})$ and $\eta_t$ be an arbitrary proposal density for $t=1, \dots,T$. In importance sampling, the unnormalized weights are given by 
\begin{equation}\label{E-ISweights}
w_t(x_{1:t}) = \gamma_t(x_{1:t}) / \eta_t(x_{1:t}).
\end{equation}
Let $\{X_{1:t}^{(n)}|n=1,\dots,N\}$ be a sample from $\eta_t(x_{1:t})$ and $w^{(n)}_t = w_t(X^{(n)}_{1:t})$. Then
\begin{equation} \label{E-normedw}
W^{(n)}_t = w^{(n)}_t/ \sum\nolimits_{n=1}^N w^{(n)}_t
\end{equation}
are the normalized weights. Given $\{W^{(n)}_{1:t}, X^{(n)}_{1:t} | n=1, \dots N \}$  approximating $\lambda_t(x_{1:t})$ at time $t$, samples $\{ X^{(n)}_{1:t+1} \}$ approximating $\lambda_{t+1}$ at time $t+1$ are obtained in SIS by sampling from $\{ X^{(n)}_{1:t} \}$ using a Markov kernel $K_{t+1}(x_t,x_{t+1})$. The proposal density is $\eta_{t+1}(x_{1:t+1}) = \eta_t(x_{1:t})K_{t+1}(x_t,x_{t+1})$. From \eqref{E-ISweights}, the corresponding unnormalized weights can be computed recursively using 
\begin{equation*}
w_{t+1} (x_{1:t+1})=  \frac{\gamma_{t+1}(x_{1:t+1}) }{\eta_{t+1}(x_{1:t+1})} = \frac{\gamma_{t+1}(x_{1:t+1}) }{\gamma_t(x_t)K_{t+1}(x_t,x_{t+1})}w_t(x_{1:t}).
\end{equation*}

In SMC, artificial joint target distributions $\tilde{\lambda}_t(x_{1:t}) \propto \tilde{\gamma}_t(x_{1:t}) $ are introduced, where $\tilde{\gamma}_t(x_{1:t})  = \gamma_t(x_t) \prod_{l=1}^{t-1} L_l(x_{l+1},x_l)$ and $L_l(x_{l+1},x_l)$ is an artificial backward in time Markov kernel. Assume $\{W^{(n)}_{1:t}, X^{(n)}_{1:t} | n=1, \dots N \}$ is a weighted sample approximating $\tilde{\lambda}_t(x_{1:t})$ at time $t$ distributed according to $\eta(x_{1:t})$. Moving the samples to $\{X^{(n)}_{1:{t+1}}\}$ using the Markov kernel $K_{t+1}$, the unnormalized importance weights can be computed as
\begin{equation}\label{E-SMCwts}
w_{t+1}(x_{1:t+1}) = \tilde{\gamma}_{t+1}(x_{1:t+1}) / \eta_{t+1}(x_{1:t+1}) = w_t(x_{1:t}) \tilde{w}_{t+1}(x_t,x_{t+1}) ,
\end{equation}
where $\tilde{w}_{t+1}(x_t,x_{t+1}) = {\gamma_{t+1}(x_{t+1}) L_t(x_{t+1},x_t)}/\{\gamma_t(x_t) K_{t+1}(x_t,x_{t+1})\}$ are unnormalized incremental weights. In the proposed algorithm, we take $K_{t+1}$ to be an MCMC kernel of invariant distribution $\lambda_{t+1}$ and $L_t(x_{t+1},x_t) = {\lambda_{t+1}(x_t) K_{t+1}(x_t,x_{t+1})}/{\lambda_{t+1}(x_{t+1})}$. See \cite{Moral2006} Section 3.3.2.3. The unnormalized incremental weights then simplify to
\begin{equation} \label{E-SMCwts2}
\tilde{w}_{t+1}(x_t,x_{t+1}) = {\gamma_{t+1}(x_t)}/{\gamma_t(x_t)}.
\end{equation}

\subsection{Proposed Algorithm}
Our aim is to sample from $p(G_1, \dots, G_K|y, x, \sigma^2)$ in \eqref{E-targetjoint} when $K>1$ and $p(G|y,a,b)$ in \eqref{E-targetG} when $K=1$. Let $p(G_1, \dots, G_K|y)$ denote the posterior density generally omitting dependence on covariates and hyperparameters. We have $p(G_1, \dots, G_K|y) \propto \gamma(G_1, \dots, G_K|y)$ where  
\begin{multline*}
\gamma(G_1, \dots, G_K) = \\
\begin{cases}
p(G|a,b) I_{G}(\delta+H,D+ \sum_{h=1}^H y_h y_h^T) / I_G(\delta,D) & \text{ if } K=1, \\
p(G_1, \dots, G_K|x,\sigma^2)  \prod_{k=1}^K \frac{I_{G_k}(\delta_k+|S_k|,D_k+ \sum_{h \in S_k} y_h y_h^T)}{I_{G_k}(\delta_k,D_k)}  & \text{ if } K>1.
\end{cases}
\end{multline*}
For simplicity, we do not state the dependence of $\gamma$ on other variables explicitly. To construct the SMC sampler, we devise the following sequence of intermediate target densities,
\begin{equation*}
p(G_1, \dots, G_K|y) ^{\phi_1}, \; p(G_1, \dots, G_K|y) ^{\phi_2},\; \dots,\; p(G_1, \dots, G_K|y) ^{\phi_T},
\end{equation*}
where $0 < \phi_1 < \phi_2 < \dots \phi_T = 1$ is a sequence of user-specified temperatures that can be set adaptively \citep[see][]{Jasra2011}.  For greater stability, we use tempering to bridge the target densities so that they evolve smoothly. At each time $t$, we maintain $N$ weighted samples $\{W_t^{(n)}, (G_1, \dots, G_K)^{(n)}_t|n=1, \dots, N\}$ approximating the target $p(G_1, \dots, G_K|y)^{\phi_t}\propto \gamma(G_1, \dots, G_K)^{\phi_t}$ and the annealing temperature is raised gradually from 0 to 1.

\subsection{Initialization and computation of weights}
To generate $N$ samples from the initial target $p(G_1, \dots, G_K|y) ^{\phi_1}$ at time $t=1$, we sample $(G_1, \dots, G_K)$ uniformly from the joint graphical space. This can be accomplished by sampling each edge in $G_k$ independently with probability 0.5 for each $k=1, \dots, K$. This process is performed $N$ times independently to obtain $\{(G_1, \dots, G_K)^{(n)}_1|n=1, \dots, N \}$. The weight of each sample can be computed using importance sampling. Let $r=p(p-1)/2$. From \eqref{E-ISweights},  
\begin{equation}\label{E-wtsupdate1}
w^{(n)}_1 = {\gamma((G_1, \dots, G_K)^{(n)}_1)^ {\phi_1}}2^{rK}.
\end{equation} 

Suppose we increase the temperature from $\phi_{t-1}$ to $\phi_t$ at time $t \geq 2$. From \eqref{E-SMCwts} and \eqref{E-SMCwts2}, unnormalized weights for the $n$th sample can be computed as
\begin{equation}\label{E-wtsupdate2}
w^{(n)}_t =  w_{(t-1)}^{(n)} \gamma((G_1, \dots, G_K)^{(n)}_{t-1})^ {\phi_t - \phi_{t-1}},
\end{equation} 
Normalized weights may be obtained using \eqref{E-normedw}.

\subsection{Resampling}
To prevent degeneracy of the particle approximation, we measure the effective sample size, ESS $= \{ \sum_{n=1}^N (W^{(n)}_t)^2 \}^{-1}$, at each time $t$ and resample when the ESS falls below a threshold, say $N_{\text{threshold}} = N/3$. Resampling is performed by drawing $N$ new particles from the multinomial distribution with parameters $(W_t^{(1)}, \dots, W_t^{(N)})$. In this way, particles with high weights will be duplicated multiple times while particles with low weights will be eliminated. Resampled particles are then assigned equal weights.

\subsection{MCMC steps} \label{sect_MCMC}
Suppose we have weighted samples $\{W_{t-1}^{(n)}, (G_1, \dots, G_K)^{(n)}_{t-1}|n=1, \dots, N\}$. At time $t$, these samples are moved using an MCMC kernel with invariant distribution $p_t(G_1, \dots, G_K|y) $ by performing a small number of MCMC steps. This improves mixing and helps to restore the heterogeneity lost during resampling. During this step, candidates for each sample are generated by selecting a small number, say $M$, of edges uniformly at random from the set of all possible edges and proposing to flip each edge (a 1 (present) to 0 (absent) and vice versa) in turn in each $G_k$ for $k=1, \dots, K$. For the selected edge, let $(G_1, \dots, G_K)^{(n)}_c$ denote the sample obtained after flipping this edge in one of the $K$ graphs in $(G_1, \dots, G_K)^{(n)}_{t-1}$. As the proposal is symmetric, the candidate is accepted with probability given by
\begin{equation}\label{E-MCMCaccept}
\min \Big[ \left\{{\gamma((G_1, \dots, G_K)^{(n)}_c)}/{\gamma((G_1, \dots, G_K)^{(n)}_t)}\right\}^{\phi_t},1\Big].
\end{equation}
If the candidate is accepted, we update the $n$th sample as $(G_1, \dots, G_K)^{(n)}_t = (G_1, \dots, G_K)^{(n)}_c$, otherwise it remains unchanged. The proposed SMC sampler is summarized in Algorithm \ref{Alg}. Note that Algorithm 1 is easily parallelizable as computation of weights as well as the MCMC steps can be performed independently for the $N$ samples.

\begin{algorithm}	
	\caption{ \label{Alg} SMC Algorithm for multiple GGMs}\vspace{1mm}
	\begin{flushleft}
    At $t=1$, 
	\end{flushleft}
	\begin{itemize}[leftmargin=1.3em,topsep=4pt,itemsep=1pt]
		\item draw $(G_1, \dots, G_K)^{(n)}_1$ at random uniformly from the joint graphical space for $n=1, \dots, N$.
		\item Compute weights $\{w^{(n)}_1\}$ using \eqref{E-wtsupdate1} and obtain normalized weights $\{W^{(n)}_1\}$ using \eqref{E-normedw}.
	\end{itemize}
	\vspace{2mm}
	\begin{flushleft}
	For $t=2, \dots, T$,	
	\end{flushleft}
	\begin{itemize}[leftmargin=1.3em,topsep=4pt,itemsep=1pt]
		\item Update weights $\{w^{(n)}_t\}$ using \eqref{E-wtsupdate2} and obtain normalized weights $\{W^{(n)}_t\}$ using \eqref{E-normedw}.
		\item If $\text{ESS} < N_{\text{threshold}}$, resample the particles and set $W^{(n)}_t = 1/N$ for $n=1, \dots, N$.
		\item  For $n=1, \dots, N$,
		\begin{itemize}[leftmargin=1.3em,topsep=1pt,itemsep=1pt]
			\item Randomly select $M$ edges from the set of all possible edges..
			\item Set $(G_1, \dots, G_K)^{(n)}_t = (G_1, \dots, G_K)^{(n)}_{t-1}$.
			\item For $m=1, \dots, M,$ and $k=1,\dots,K,$ let $(G_1, \dots, G_K)^{(n)}_c$ be the sample candidate obtained from $(G_1, \dots, G_K)^{(n)}_t$ by flipping the $m$th selected edge in $G_k$. Accept sample candidate with probability in \eqref{E-MCMCaccept}. If sample candidate is accepted, set $(G_1, \dots, G_K)^{(n)}_t = (G_1, \dots, G_K)^{(n)}_c$.
		\end{itemize} 
	\end{itemize}
\end{algorithm}

Note that Algorithm 1 can be fully automated to the extent that one only needs to set the first temperature and MCMC proposal. The rest of the algorithm, such as in \citep{DelMoral2012, Jasra2011, Schafer2011} can be made entirely adaptive with stable and mathematically proven convergence \citep{Beskos2016}. 

\subsection{Scalability} \label{S-scalability}
The proposed algorithm scales linearly in $K$ due to the MCMC steps which have to be performed for each graph. The algorithm does not scale well with respect to the number of nodes $p$ as the computation of the normalizing constants $I_{G_k}(\delta_k, D_k)$ using Monte Carlo integration \citep{Atay2005} is computationally expensive when $p$ is large (scales approximately as the cube of $p$).

\section{Results} \label{S-results}
In this section, we discuss the results obtained from simulations and application of the proposed GGM to the urinary metabolic data.

To set the hyperparameters for the multiplicative prior, we suggest using prior data or R packages such as GeneNet \citep{Schaefer2015} or GGMselect \citep{Giraud2012} to obtain a quick preliminary estimate of the degree distribution. The hyperparameters of the multiplicative prior can then be selected so that the shape of the prior degree distribution matches that of the estimated one. For $K=1$, one can compute $(a,b)$ using the formulas in \ref{av_deg} so that the mean and variance of the prior degree distribution matches that of the estimated one. This procedure is implemented in Section \ref{sec:simulations}. Note that the estimated degree distribution may have a variance smaller than that in \ref{av_deg} for any $a>0$, $b>0$. In this case, we set $b$ to be large (e.g. 1000) so that the variance is very small and then find $a$ to match the mean degree.

In the following experiments, we take $\delta_k=3$ and $D_k=I_p$ for $k=1, \dots, K$ in the G-Wishart prior. The number of samples used in SMC is $N=500$. Our code is written in Matlab and is available as part of the Supplementary Material. We run the experiments on HPC (High Performance Computing) where each job is run in parallel across 12 cores. 

\subsection{Simulations} \label{sec:simulations}
In this section, we investigate the performance of the multiplicative model as a prior on graphs for GGMs and compare it with the commonly used uniform prior which assigns equal prior probability to every graph, and the size-based prior \citep{Armstrong2009} or equivalently the prior that corrects for multiple hypothesis testing proposed by \cite{Carvalho2007} (see Section \ref{S-bg} for details). First we consider $K=1$, $p=20$ nodes and generate data from three different types of networks:
\begin{description}
\item[Multiplicative model] : We generate $\pi_i \overset{\text{i.i.d.}}{\sim} \text{Beta}(0.1,0.1)$ and simulate the edges using $A_{i,j} \sim \text{Bernoulli}(\pi_i \pi_j)$ for $i<j$.
\item[Scale-free network] : A scale-free network with $p$ nodes is generated using the Barab\'{a}si-Albert (BA) model. Starting with a connected network with 2 nodes, new nodes are added one at a time to the network. Each new node is connected to 2 existing nodes with a probability proportional to the degree of existing nodes.
\item[Community network] : The $p$ nodes are divided into two communities of equal-size and a network is generated by assuming that the within-community interaction rate is 0.6 and across-community interaction rate to be 0.02.
\end{description}

\begin{figure}[htb!]
\centering
\includegraphics[width=0.32\textwidth]{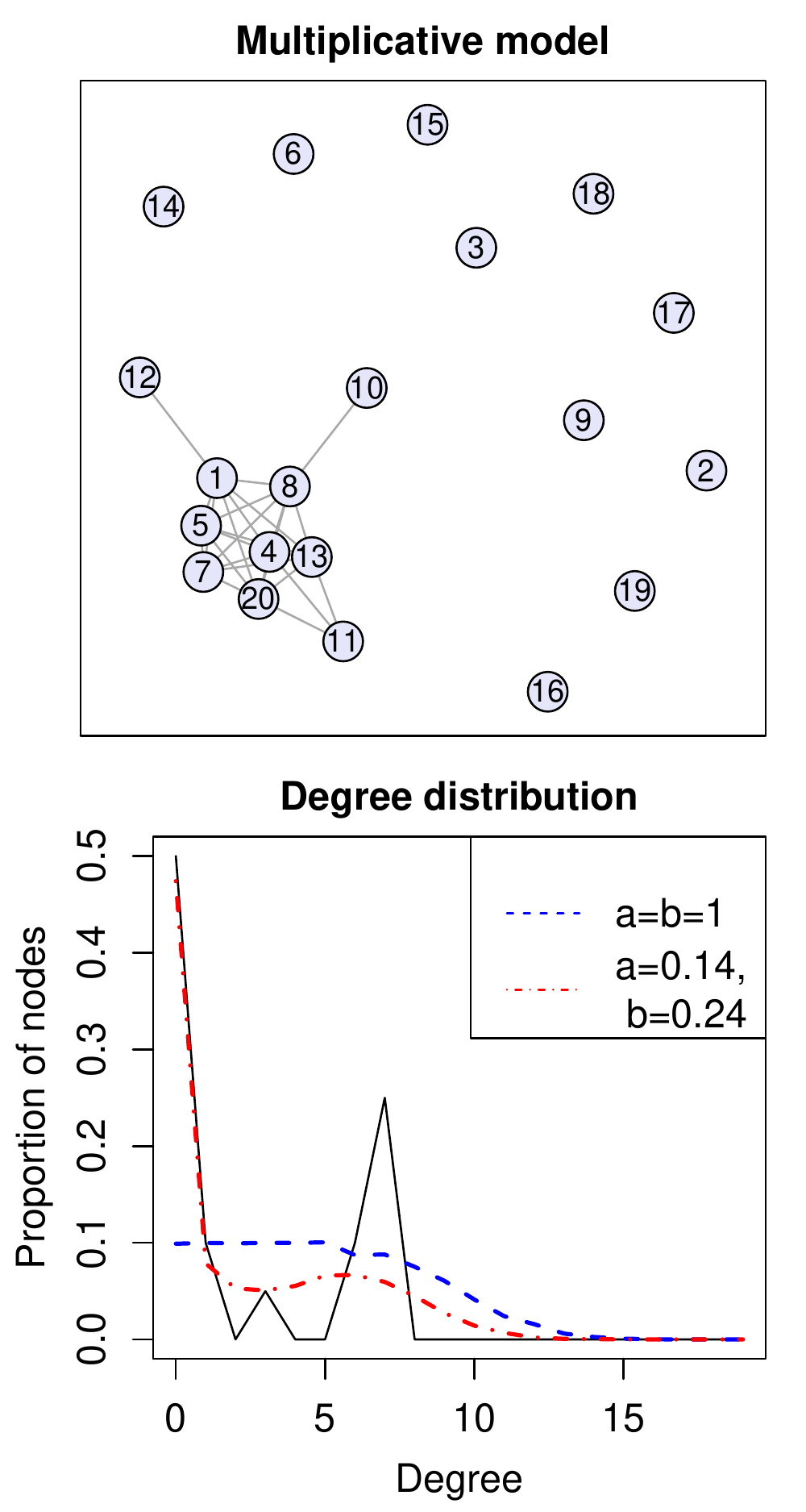}
\includegraphics[width=0.32\textwidth]{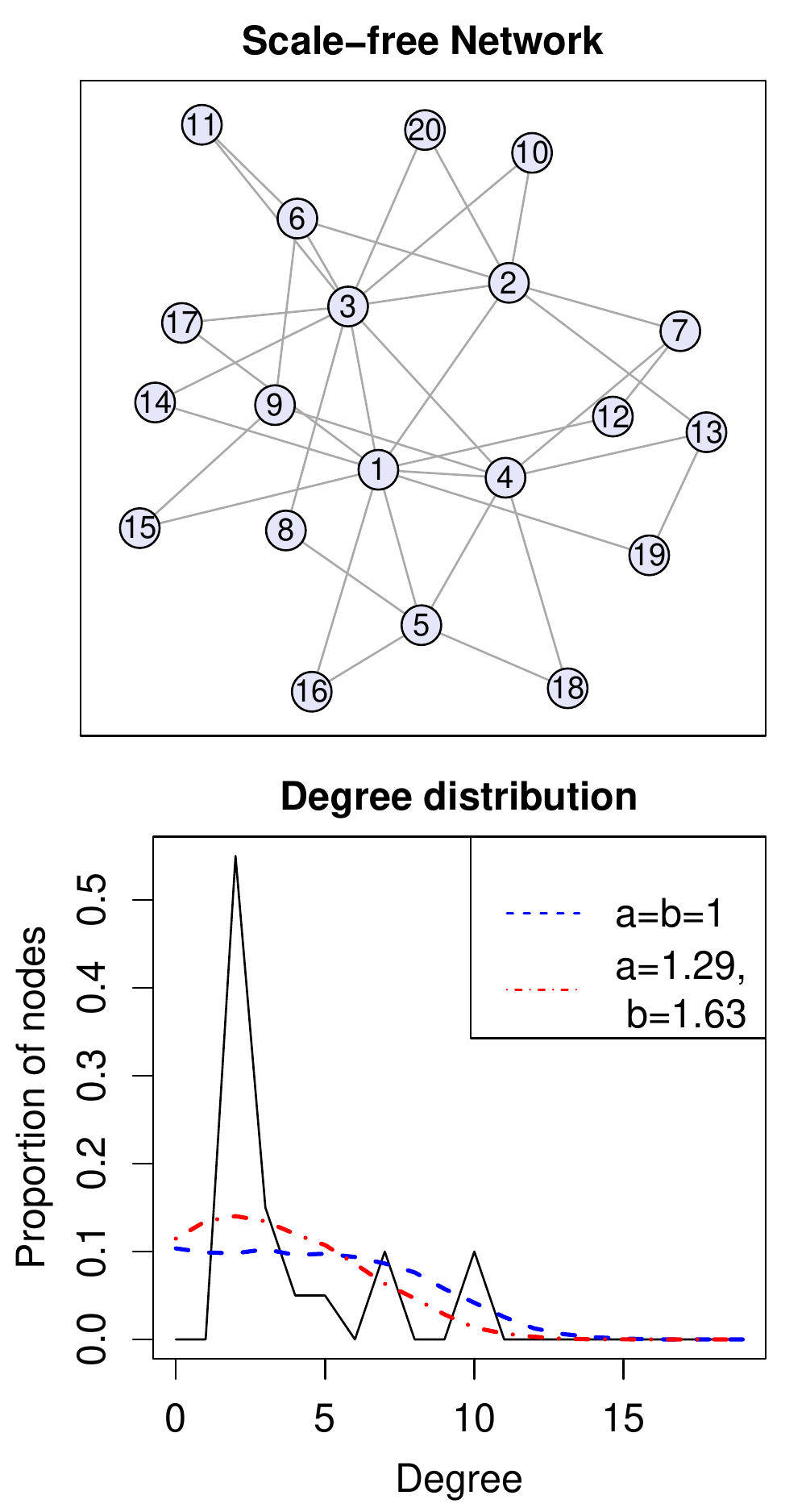}
\includegraphics[width=0.32\textwidth]{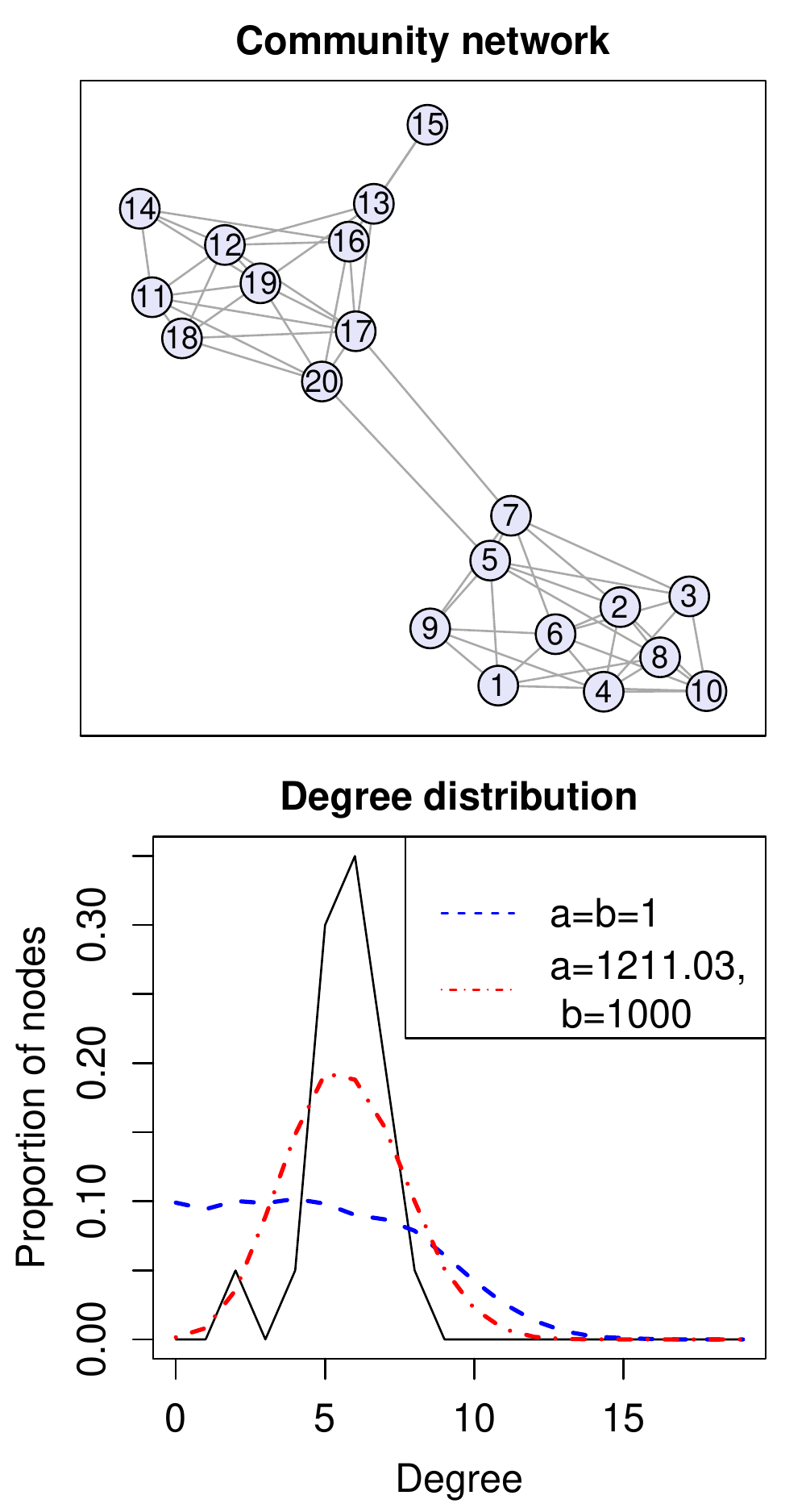}
\caption{Generated networks (top) and corresponding degree distribution (bottom)\label{networks_degdistn}. The degree distribution of the multiplicative model with hyperparameters given by $a$ and $b$ are superimposed.}
\end{figure}

The generated networks and their degree distributions are shown in Figure \ref{networks_degdistn}. 
For each network we created a $p \times p$ symmetric matrix $C$ where entries corresponding to missing edges are set to zero and non-zero entries are simulated randomly from the uniform distribution on ${[-0.6,-0.3] \cup [0.3,0.6]}$. To ensure that the precision matrix $\Omega$ is positive, we let $c$ be the smallest eigenvalue of $C$ and set $\Omega = C + (0.1+|c|)I$, following \cite{Mohan2014}. Ten datasets are then simulated  from the GGM in \eqref{E-GGM} by setting the number of observations $H=100$ and $K=1$. The underlying network is regarded as the ``true" graph.

Using Algorithm 1, weighted samples from the posterior distribution are obtained for each simulated data set under the uniform prior, size-based prior and multiplicative prior respectively. For the multiplicative model, we consider two settings. For one setting, we set the Beta hyperparameters as $a=b=1$. For the second setting we try to find $a$ and $b$ such that the shape of the prior degree distribution resembles that of the true graph. These prior degree distributions are superimposed on the true degree distributions in Figure \ref{networks_degdistn}. For the SMC sampler, we set the number of edges flipped at each iteration in the MCMC step $M=3$. The sequence $\{\phi_t\}$ is set as (0.01, 0.02, \dots, 1) with $T=100$. The CPU time taken on average for each experiment is ($6.4 \pm 0.5$) hours.

\begin{figure}[htb!]
\centering
\includegraphics[width=0.32\textwidth]{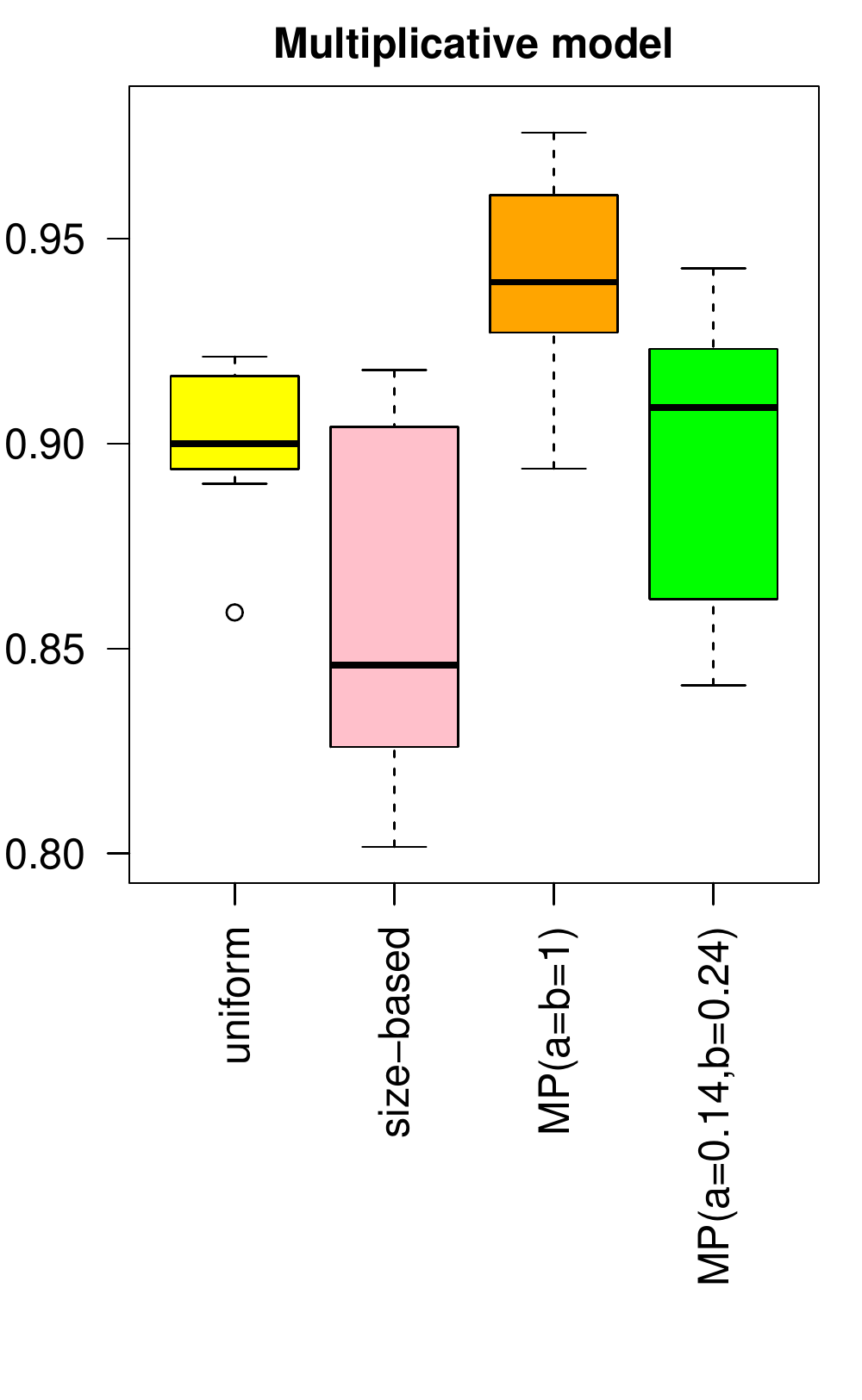}
\includegraphics[width=0.32\textwidth]{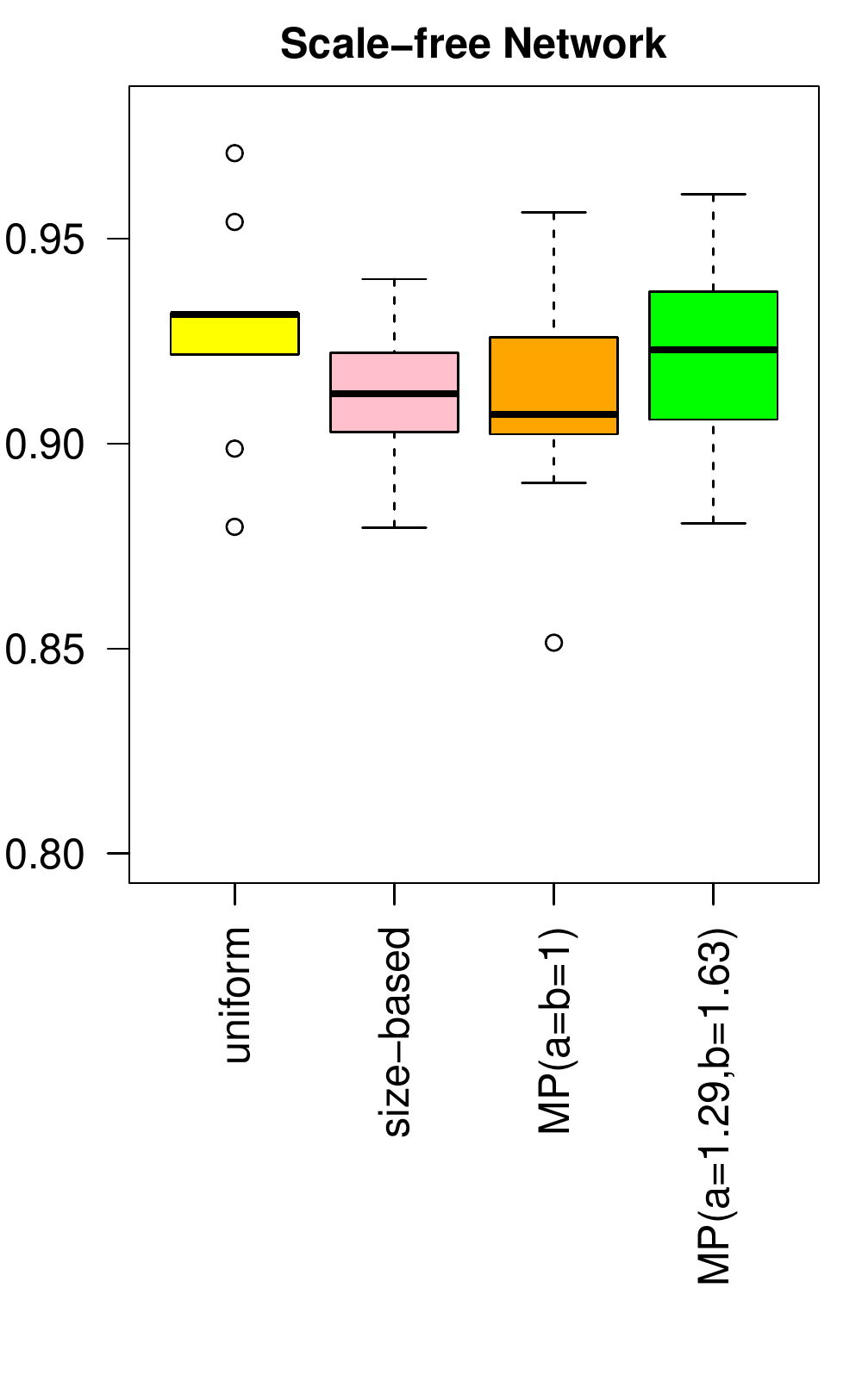}
\includegraphics[width=0.32\textwidth]{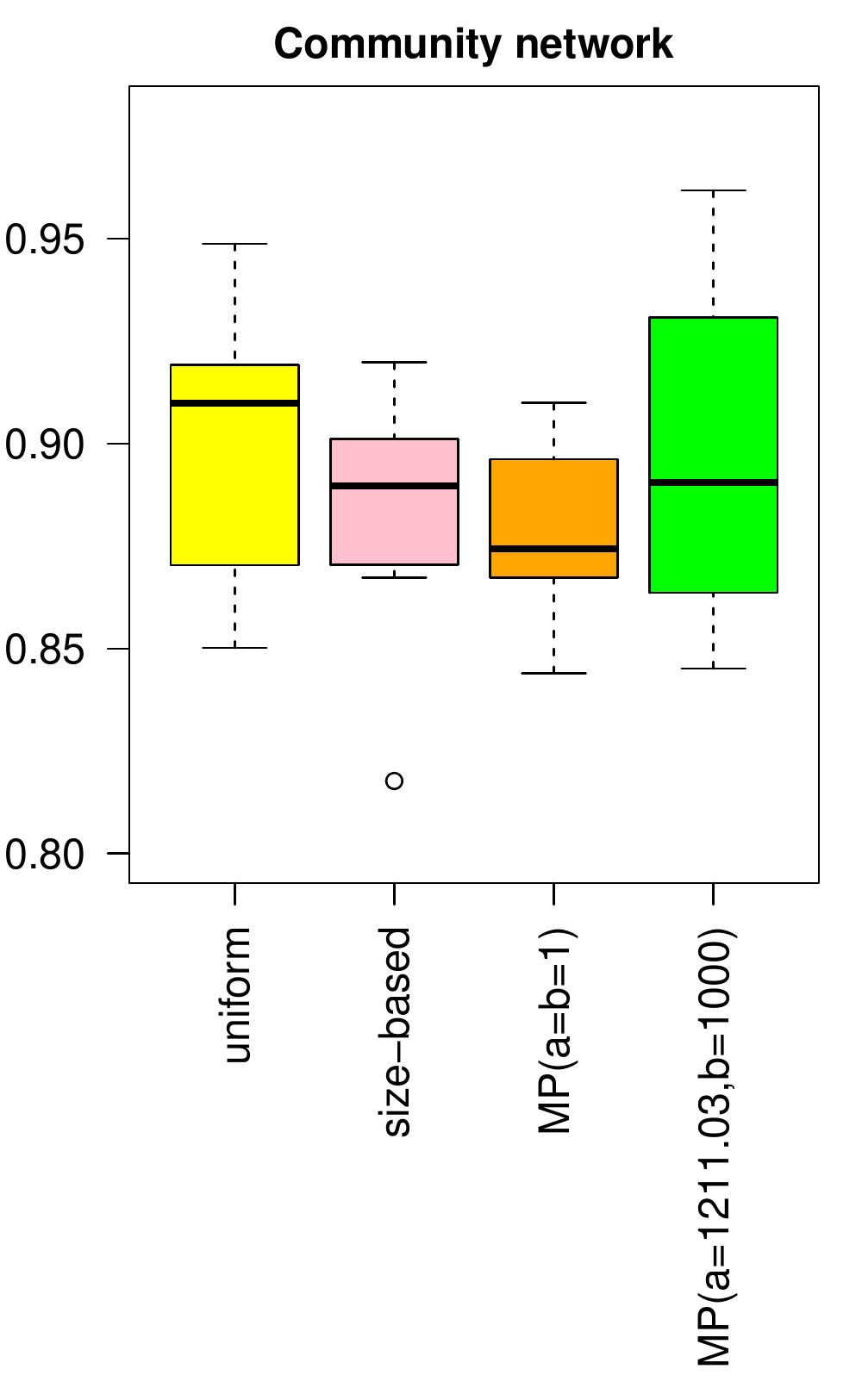}
\caption{Boxplots of AUC values for datasets simulated from different networks obtained using different priors. \label{AUCval}}
\end{figure}

Using the weighted samples, we compute the posterior probability of the occurrence of each edge and summarize the results using the area under the ROC curves (AUC). The boxplots of the AUC values are shown in Figure \ref{AUCval}. The multiplicative priors performed better than the uniform and especially the size-based prior for data simulated from the multiplicative model. For data simulated from the scale-free and community networks, the performance of the different priors are quite similar. For these networks, the multiplicative prior performed better if the hyperparameters were tuned to match the degree distribution of the true graph.

\begin{figure}[htb!]
\centering
\includegraphics[width=0.63\textwidth]{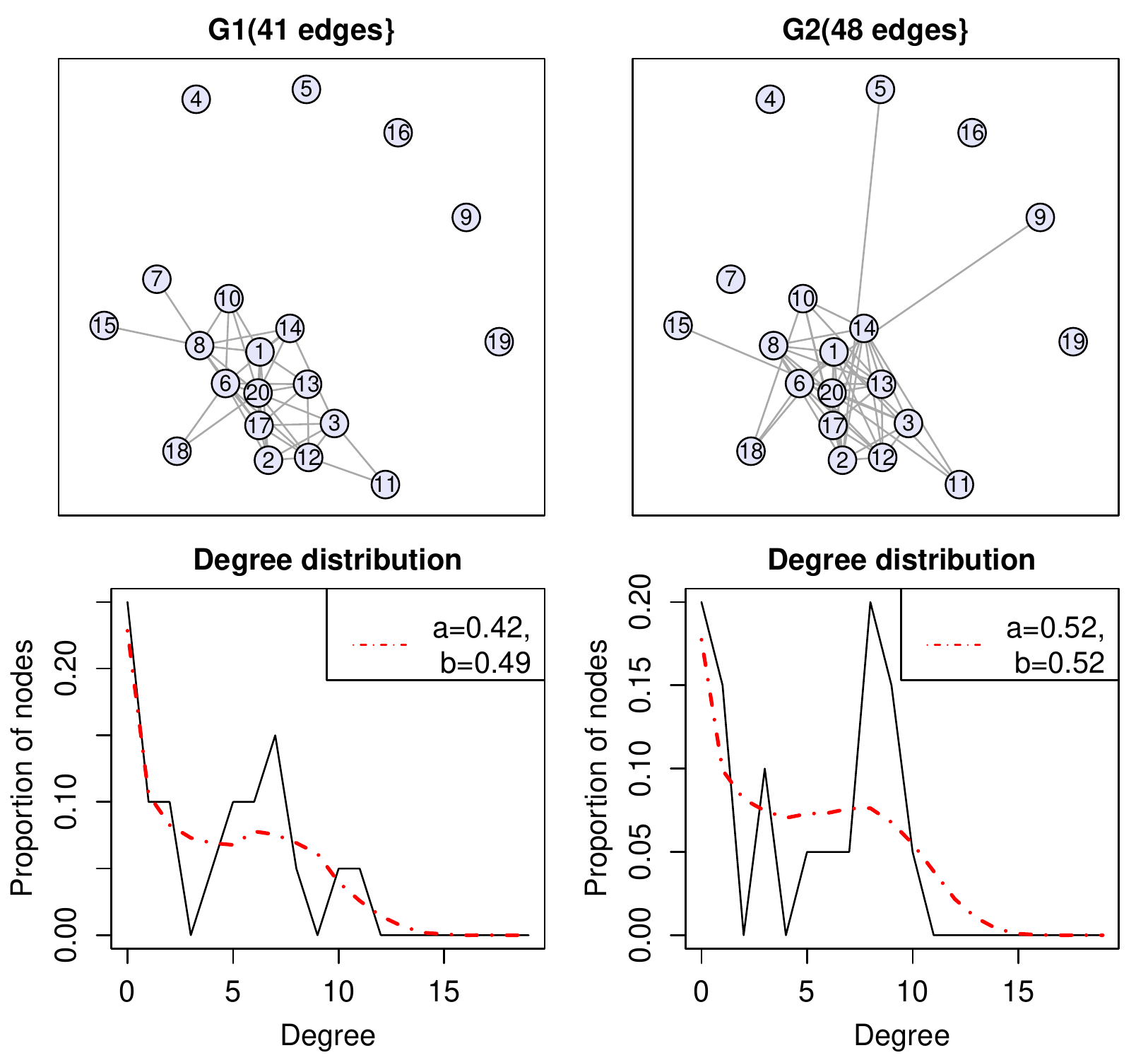}
\includegraphics[width=0.36\textwidth]{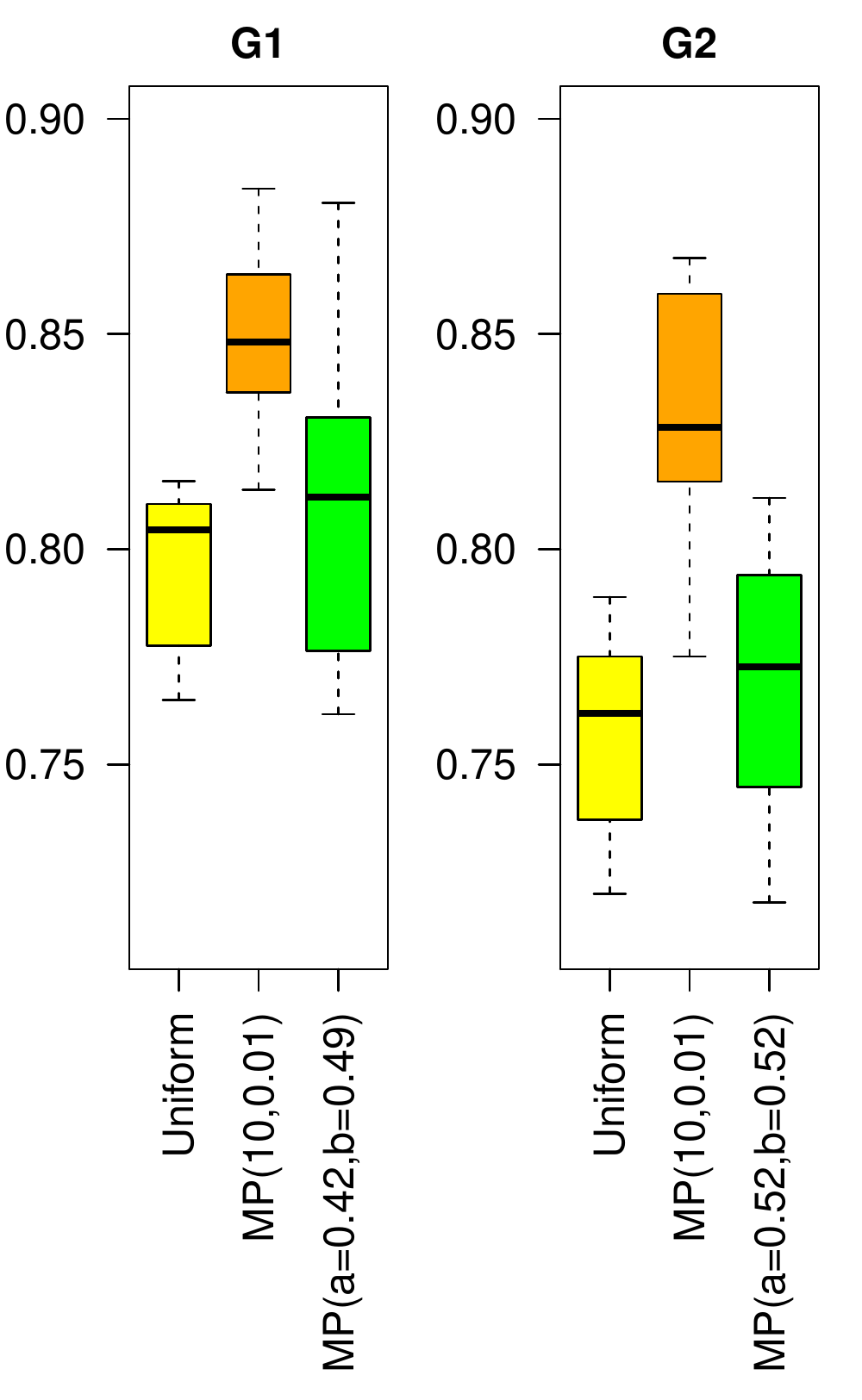}
\caption{Simulated networks ($K=2$), corresponding degree distributions and boxplots of AUC values obtained using different priors. \label{simeg14}}
\end{figure}

Next, we investigate the ability of the multiplicative prior to borrow information across graphs when the nodes have similar connectivity. We simulate 10 datasets each with $H=100$ observations, $p=20$ variables and $K=2$ groups. We assume that there are 50 observations in each group and set the covariates $x_1 = (1,0)$ and $x_2=(1,1)$. The underlying graphs are generated from the multiplicative model where the connectivities are simulated using the model in Section \ref{sec: mp_K_big} and precision matrices for each graph are constructed in the same manner as before. Setting $\sigma_1^2 = 10$ and $\sigma_2^2 = 0.01$, the connectivity of the nodes in $G_1$ may vary over a wide range (since $\beta_{i1}$ has a large variance) and the connectivity of each node in $G_1$ and $G_2$ are similar (since $\beta_{i2}$ is close to zero).  Figure \ref{simeg14} shows the simulated graphs and their degree distributions. 

We compare results obtained using (1) the uniform prior which assigns equal prior probability to each pair of graphs, (2) the joint multiplicative prior with $\sigma_1^2=10$, $\sigma_2^2=0.01$ and (3) where independent multiplicative priors are used for $G_1$ and $G_2$ with hyperparameters chosen to match the degree distributions of the true graphs. Using Algorithm 1 with the same setting as before, the average CPU time for the joint prior ($K=2$) case is $(12.4 \pm 0.5)$ hours and for the independent multiplicative priors case ($K=1$) is $(6.5 \pm 0.4)$ hours. The results are summarized using boxplots of the AUC values as shown in Figure \ref{simeg14}.
The joint multiplicative prior performs better than the uniform prior and the case of independent multiplicative priors indicating the ability of the multiplicative prior to encourage similarity in connectivity of nodes across graphs.

In the Supplementary Material, we also provide details of a small experiment which shows the significant improvement that SMC provides over standard MCMC. In particular, SMC has a higher acceptance rate and achieves higher average log target density for the same number of MCMC steps.

\subsection{Application to urinary metabolic data}
We analyze urinary metabolic data for $H=127$ individuals acquired using $^1$H NMR spectroscopy (see \cite{Ellis2012} for details). These individuals live close to a lead and zinc smelter at Avonmouth in Bristol, UK, which produces large quantities of airborne cadmium (Cd). Here, we investigate the correlation structure of $p=22$ urinary metabolites listed in Table \ref{metab} in response to cadmium exposure through GGMs. This dataset has also been studied by \cite{Salamanca2014} using Bayesian hierarchical models. We perform two analyses. In the first case, we consider the individuals as a homogeneous group. In the second case, we divide the individuals into two groups; $S_1$ (a control group with level of exposure to cadmium lower than or equal to the median) and $S_2$ (a diseased group with level of cadmium higher than the median). In each case, we first use the {\tt R} package {\tt GeneNet} \citep {Schaefer2015} to obtain fast shrinkage estimators of partial correlation in the network. The degree distributions obtained (see Supplementary Material Figure \ref{F-genenetdegdistn}) can be used as a basis for determining appropriate hyperparameters for the multiplicative model. The observations of each variable are first normalized to have zero mean and standard deviation of one in each group. For the SMC sampler, we set the number of samples $N =500$, and the number of edges flipped at each iteration in the MCMC step $M=5$. The sequence $\{\phi_t\}$ is set as (0.005, 0.01, \dots, 1) with $T=200$. The CPU time taken on average for each experiment is ($24.7 \pm 3.0$) hours for $K=1$ and $(48.0 \pm 7.5)$ hours for $K=2$.

\begin{table}[ht]
	\centering \ra{1.1} \tabcolsep=0.15cm
	\begin{footnotesize}
	\resizebox{\columnwidth}{!}{\begin{tabular}{@{}ll|cccc|cccc@{}}
			\hline
			& & \multicolumn{4}{c}{Degree} \vline &  \multicolumn{4}{c}{Betweenness} \\ 
			Metabolites & Abbrev & M(1, 1) & M(0.1, 0.1) & SB & UF & M(1, 1) & M(0.1, 0.1) & SB & UF \\ 
			\hline
			Trimethylamine oxide & TMAO & \textbf{4.52} & \textbf{4.62} & \textbf{4.20} & 5.33 &\textbf{ 0.15} & \textbf{0.13} &\textbf{ 0.20} & 0.08 \\ 
			P-cresol-sulphate & PCS & 4.33 & 3.81 & 3.72 & \textbf{6.39} & 0.10 & 0.07 & 0.11 & 0.09 \\ 
			Succinate & Suc & 3.95 & 4.07 & 3.79 & 5.71 & 0.10 & 0.10 & 0.17 & 0.07 \\ 
			Dimethylamine & DMA & 3.81 & 4.05 & 3.47 & 5.68 & 0.10 & 0.08 & 0.12 & 0.07 \\ 
			Creatinine & Creat & 3.54 & 3.38 & 3.09 & 4.93 & 0.08 & 0.05 & 0.09 & 0.06 \\ 
			4-deoxyerythronic acid & 4-DEA & 3.52 & 1.56 & 2.63 & 5.23 & 0.09 & 0.02 & 0.11 & 0.08 \\ 
			Pyruvate & Pyr & 3.21 & 3.19 & 3.01 & 5.41 & 0.05 & 0.05 & 0.06 & 0.06 \\ 
			Citrate & Cit & 3.15 & 2.58 & 2.42 & 4.84 & 0.09 & 0.07 & 0.12 & 0.05 \\ 
			3-hydroxyisovalerate & 3-HV & 2.85 & 2.87 & 2.74 & 4.58 & 0.10 & 0.08 & 0.17 & 0.06 \\ 
			Glycine & Gly & 2.70 & 2.79 & 2.41 & 4.42 & 0.07 & 0.06 & 0.10 & 0.05 \\ 
			Urea & Urea & 2.62 & 3.50 & 2.09 & 6.17 & 0.07 & 0.07 & 0.06 & \textbf{0.09} \\ 
			Alanine & Ala & 2.48 & 2.01 & 2.38 & 5.45 & 0.06 & 0.03 & 0.10 & 0.07 \\ 
			Phenylacetylglutamine & PAG & 2.24 & 2.12 & 2.16 & 4.55 & 0.02 & 0.02 & 0.05 & 0.04 \\ 
			Acetate & AcO & 1.93 & 0.29 & 1.32 & 5.13 & 0.04 & 0.00 & 0.04 & 0.05 \\ 
			Hippurate & Hip & 1.55 & 2.98 & 1.76 & 4.86 & 0.02 & 0.05 & 0.04 & 0.06 \\ 
			Dimethylgycine & DMG & 1.49 & 1.88 & 1.44 & 4.70 & 0.02 & 0.03 & 0.04 & 0.06 \\ 
			Trimethylamine & TMA & 1.28 & 1.18 & 1.60 & 3.28 & 0.01 & 0.01 & 0.04 & 0.03 \\ 
			Lactate & Lac & 1.08 & 0.62 & 0.96 & 4.56 & 0.02 & 0.01 & 0.02 & 0.04 \\ 
			Proline-betaine & PB & 0.64 & 0.06 & 0.84 & 2.49 & 0.00 & 0.00 & 0.01 & 0.01 \\ 
			N-methyl-nicotinic acid & NMNA & 0.56 & 0.09 & 1.06 & 3.06 & 0.01 & 0.00 & 0.01 & 0.02 \\ 
			Formate & For & 0.36 & 0.03 & 0.43 & 2.85 & 0.00 & 0.00 & 0.01 & 0.02 \\ 
			Creatine & Crea & 0.12 & 0.01 & 0.28 & 1.94 & 0.00 & 0.00 & 0.00 & 0.01 \\ 
			\hline
		\end{tabular}}
	\end{footnotesize}
	\caption{List of 22 urinary metabolites and their abbreviations. Columns 3--6 and columns 7--10 show the weighted mean degree and betweenness respectively, under the multiplicative model with $a=b=1$ (M(1, 1)) and $a=b=0.1$ (M(0.1, 0.1)), the size-based prior (SB) and the uniform prior(UF). The highest value in each column (3--10) is highlighted in bold.}
	\label{metab}
\end{table}

\subsection{Case: $K=1$}
\begin{figure}[tb]
	\centering
	\includegraphics[width=\textwidth]{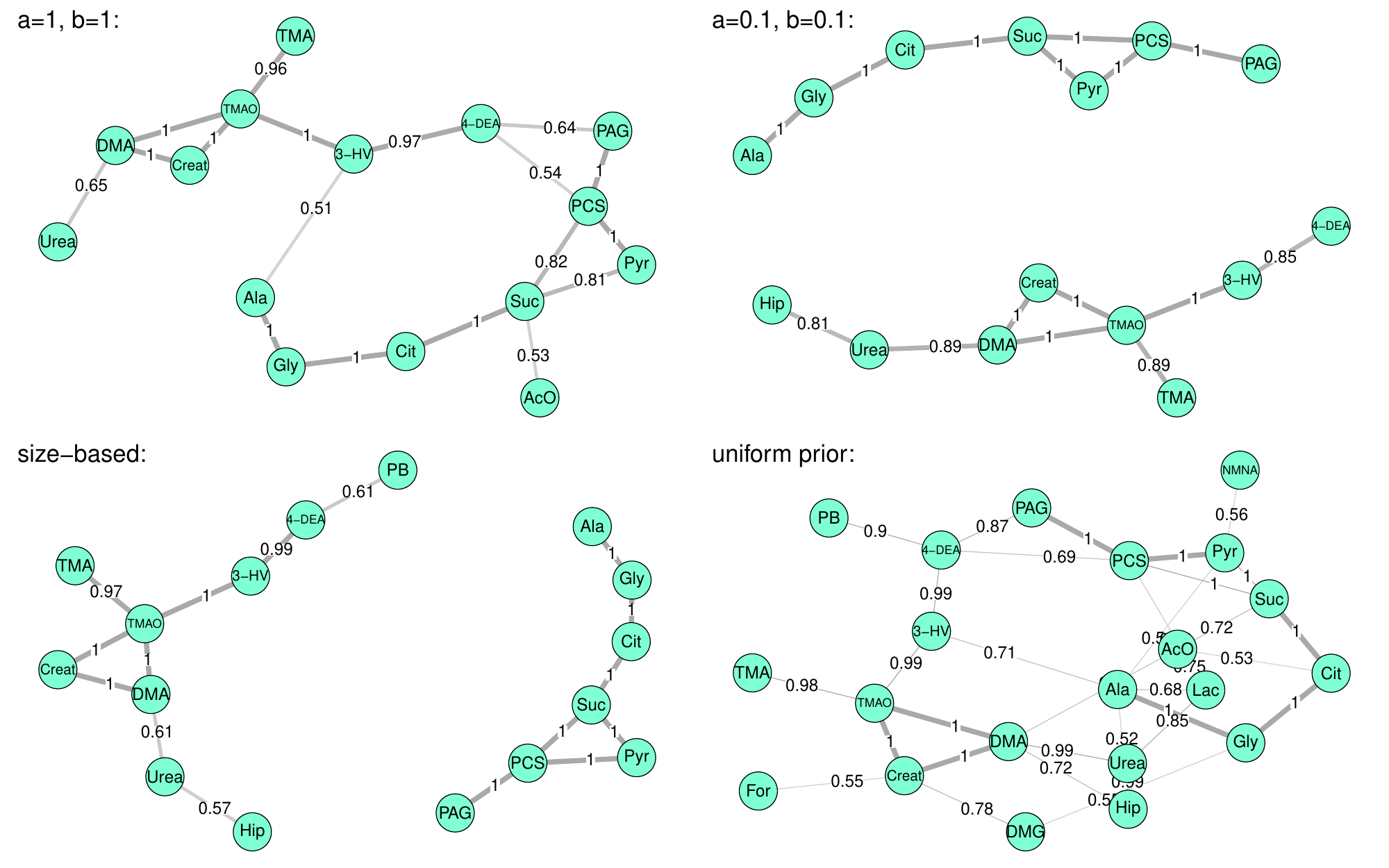}
	\caption{Graphs corresponding to different priors. Only edges with posterior weights greater than 0.5 are shown.}
	\label{F-networkall}
\end{figure}

In this section, we study the correlation structure of the metabolites treating the individuals as one homogeneous group. We compare the performance of four priors on the graphical space, namely, the multiplicative model with $a=b=1$ and $a=b=0.1$, the size-based prior and the uniform prior. We fit a GGM to the data using Algorithm 1, obtaining $N=500$ weighted samples from the posterior distribution in each case. Using these weighted samples, we compute the posterior probability of occurrence of each edge. Figure \ref{F-networkall} shows the graphs obtained under each prior. Only edges with posterior probability greater than 0.5 and associated nodes are shown and the width of each edge is proportional to its posterior probability. Graphs showing the full set of nodes and all possible edges are given in the Supplementary Material Figure \ref{F-Gall}. The graphs obtained under the multiplicative model and the size-based prior have a high degree of similarity and are much sparser than that of the uniform prior. 

Table \ref{metab} shows the weighted mean degree and betweenness centrality measures for each metabolite. The metabolites have been sorted in terms of weighted mean degree in decreasing order according to M(1,1), the multiplicative model with $a=b=1$.  Under the multiplicative model and size-based prior, TMAO has the highest degree as well as betweenness. Under the uniform prior, PCS has the highest degree with Urea close behind; these two metabolites also have the highest betweenness. 

\begin{figure}[tb]
	\centering
	\includegraphics[width=0.975\textwidth]{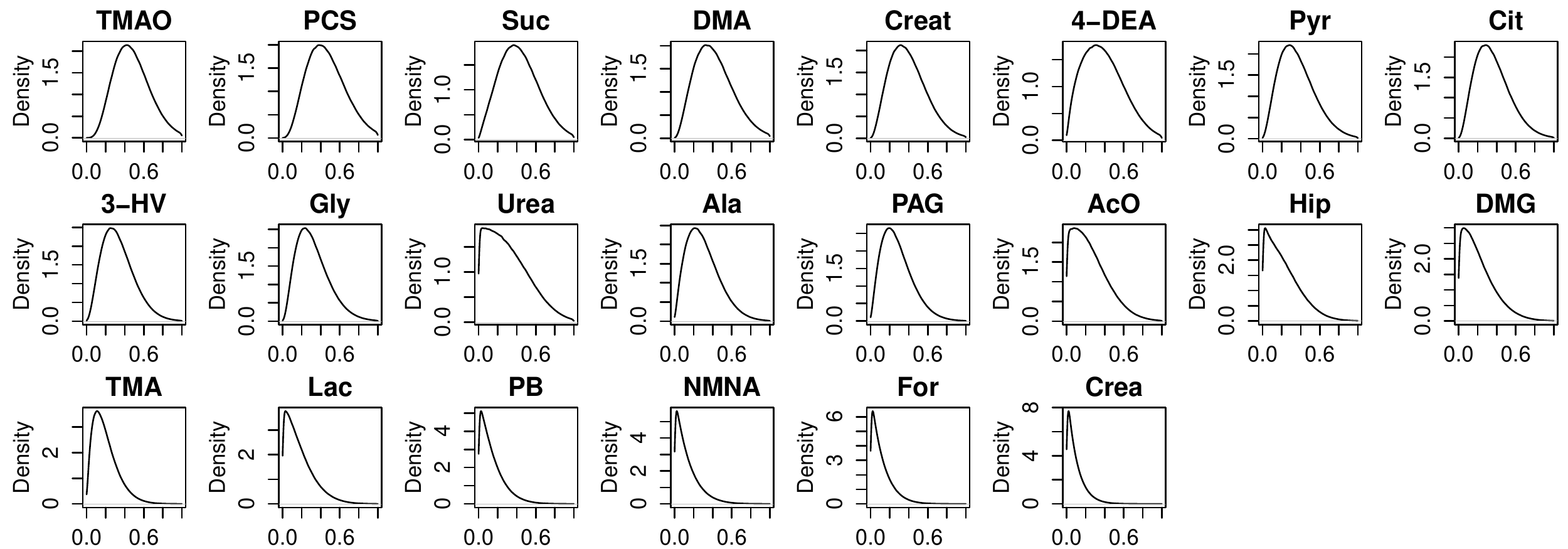} 
	\includegraphics[width=0.975\textwidth]{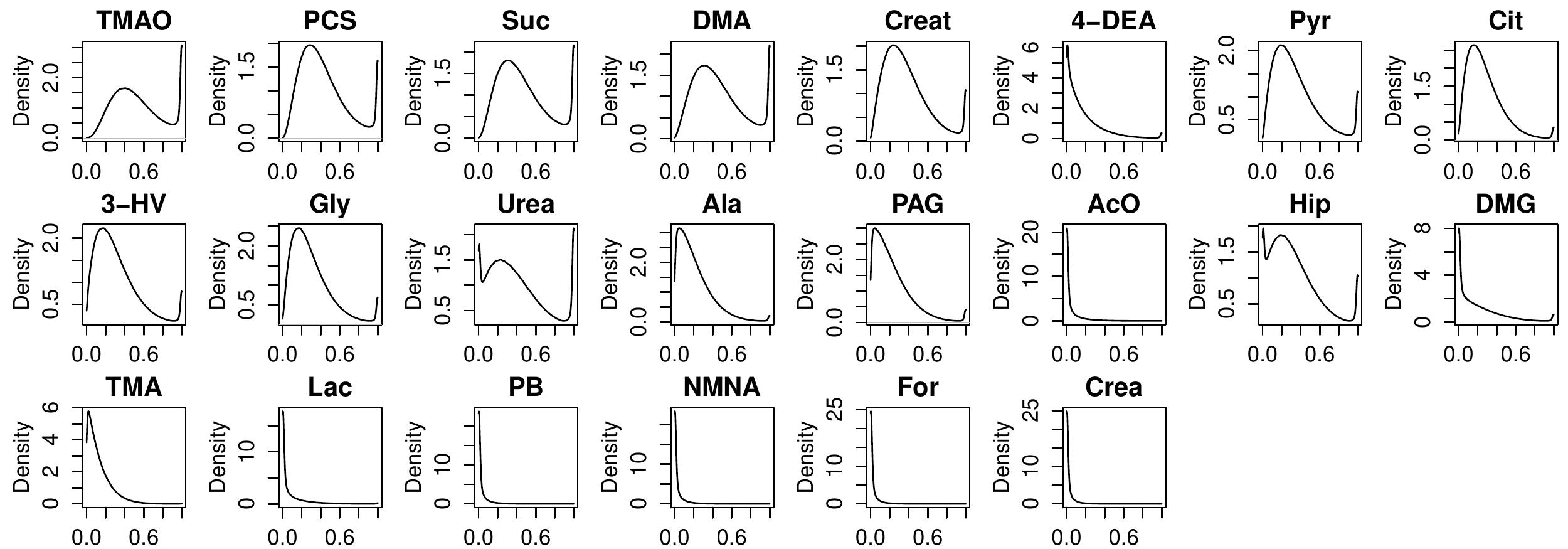} 
	\caption{Posterior distribution of the connectivity ($\pi_i$) of different metabolites under the multiplicative model with $a=b=1$ (first 3 rows) and $a=b=0.1$ (last 3 rows).}
	\label{F-Gconnect}
\end{figure}

For the multiplicative model, we can also obtain uncertainty measures of the tendency of each node to form connections with other nodes. Figure \ref{F-Gconnect} shows the posterior distributions of the connectivities $\pi_i$ of each metabolite obtained via simulations. It appears that the multiplicative model with $a=b=0.1$ is too strong and places too much prior weight on values of $\pi_i$ at the extremes of 0 and 1. The multiplicative model with $a=b=1$ provides a better fit. The mean and 95\% credible interval of the connectivity of each metabolite, and the mean covariance matrix corresponding to the multiplicative model with $a=b=1$ are given in the Supplementary Material Tables \ref{Gcon} and \ref{CovK1}.

\subsection{Case: $K=2$}
Next, we investigate the difference in correlation structure of the urinary metabolites between the two groups of individuals $S_1$ (with level of exposure to cadmium lower than or equal to the median) and $S_2$ (level of exposure higher than the median).  We consider the covariates $x_k$ for the $k$th graph to include an intercept and an indicator for level of exposure to cadmium (1 if above the median and 0 otherwise) so that $x_1 = (1,0)$ and $x_2=(1,1)$. The difference in graphical structure between $G_1$ and $G_2$ due to exposure to urinary cadmium is of interest. We fit a GGM with $K=2$ to the data using the SMC algorithm under four priors. The first three are the multiplicative model with $\sigma_1^2=\sigma_2^2=1$, $\sigma_1^2=1$ and $\sigma_2^2=10$ and $\sigma_1^2=\sigma_2^2=10$, and the last is the uniform prior. From Figure \ref{F-degpair} and the preliminary degree distributions obtained using GeneNet (see Supplementary Material Figure \ref{F-genenetdegdistn}), taking $\sigma_1^2=\sigma_2^2=1$ seems appropriate but we wish to investigate if there is any benefit to be gained by allowing the structure of $G_2$ to vary more significantly from that of $G_1$ by taking $\sigma_2^2$ to be 10 and whether a prior which assumes the tendencies to connect are closer to the extremes of 0 and 1 is more appropriate ($\sigma_1^2=\sigma_2^2=10$).

Using Algorithm 1, we obtain weighted samples from the posterior distribution under each of the four priors. The ESS and acceptance rate in the SMC sampler are monitored at each iteration and these plots are given in the Supplementary Material Figure \ref{F-ESSA} for the multiplicative prior with $\sigma_1^2=\sigma_2^2=1$. Typically, the ESS decreases as the algorithm proceeds until it falls below the threshold, $N_{\text{threshold}} = N/3$, and it bounces back after resampling is performed. Due to bridging of target densities using tempering, the acceptance rate is usually high at the beginning when the temperature $\phi_t$ is close to zero and proposals have a high probability of being accepted [see \eqref{E-MCMCaccept}]. As the temperature increases, the samples becomes more concentrated in the regions of high posterior probability and the acceptance rate falls.   

\begin{figure}[tb!]
\centering
\includegraphics[width=0.9\textwidth]{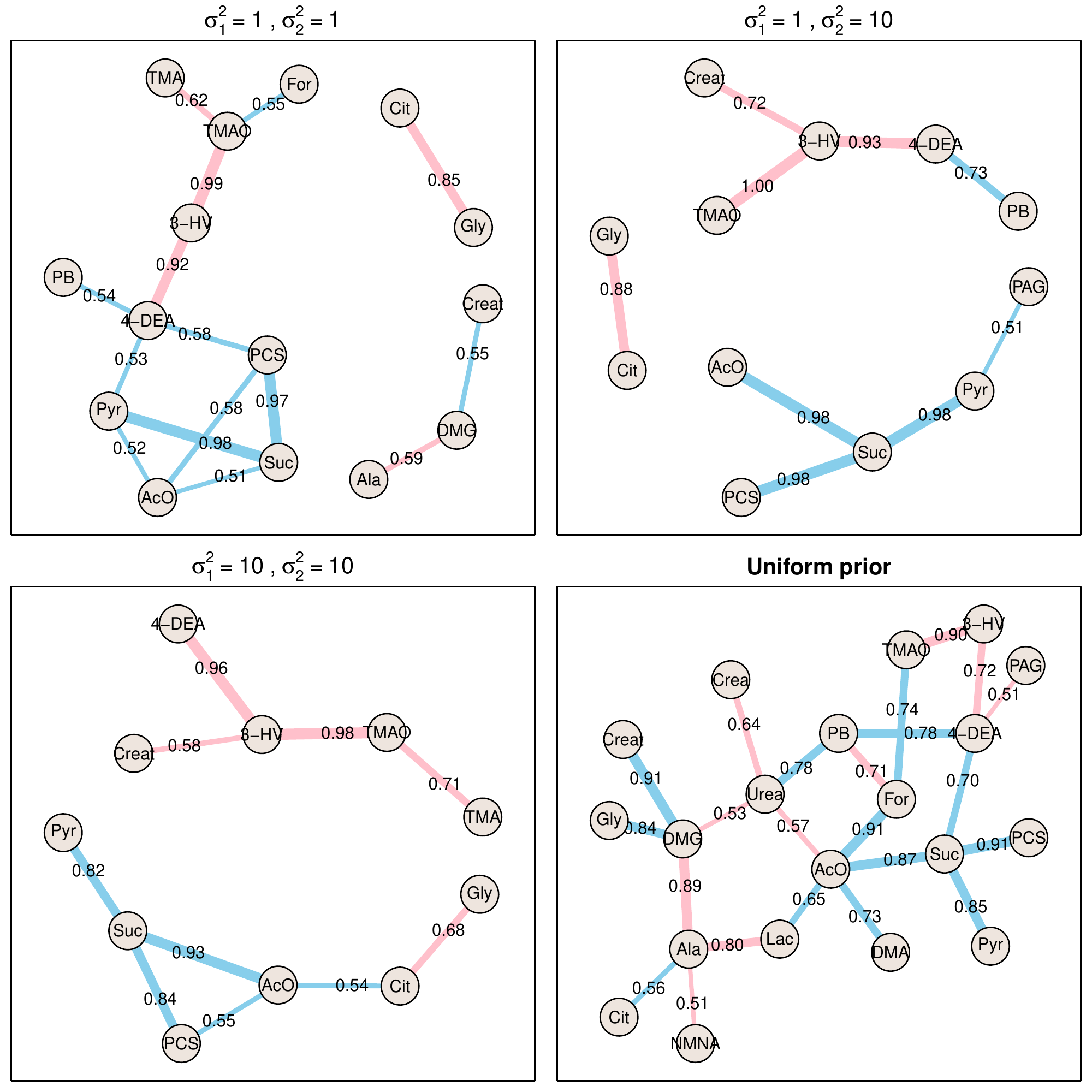} 
\caption{Differential network corresponding to the different priors. Edges in blue are likely to appear in $G_1$ but not in $G_2$ and pink edges are likely to appear in $G_2$ but not in $G_1$. The labels indicate the estimate of $|\rho_{ij}^1 - \rho_{ij}^2|$ for each edge $(i,j)$.}
\label{difnet}
\end{figure}
To compare the differences in edges between $G_1$ and $G_2$, we construct differential networks which display only edges likely to appear in one graph but not the other. Differential networks serve as powerful tools for exploring the changes in correlation structures across different conditions and have been considered widely in recent research. For instance, \cite{Valcarcel2011} define an edge as differential if the partial correlations estimated via linear shrinkage estimators differ significantly between two graphs while \cite{Peterson2015} and \cite{Mitra2016} consider the posterior probability of an edge differing across conditions. Here we adopt another definition which enables us to differentiate more easily between the edges which are more likely to appear in $G_1$ than $G_2$ and vice versa. Let $\rho_{ij}^k$ denote the posterior marginal probability of inclusion of the edge $(i,j)$ in $G_k$ for $k=1,2$. We estimate $\rho_{ij}^k$ as the proportion of SMC samples for which the edge $(i,j)$ was included in $G_k$ and consider an edge to be differential if $|\rho_{ij}^1 - \rho_{ij}^2| > \kappa$ for some $0 <\kappa <1$. Figure \ref{difnet} shows the differential network corresponding to the different priors for $\kappa=0.5$. The estimates of $\rho_{ij}^1$ and $\rho_{ij}^2$ for the edges in the differential networks are given in Table \ref{G2diff} in the Supplementary Material. Due to space limitations, we have also included further detailed results in the Supplementary Material. These include weighted graphs obtained from Algorithm 1 under different priors (Figures \ref{G2all} and \ref{network2all}), posterior distributions (Figures \ref{post211}, \ref{post2110} and \ref{post21010}), betweenness centrality measures, weighted means and 95\% credible intervals of the connectivities ($\pi_{i,k}$) and regression coefficients ($\beta_{iq}$) of each metabolite (Tables \ref{G2tab1}, \ref{G2tab2} and \ref{G2tab3}) and the mean covariance matrices corresponding to the multiplicative prior with $\sigma_1^2=\sigma_2^2=1$ (Tables \ref{Cov21} and \ref{Cov22}).

The full network in the $K=1$ case and the differential network in the $K=2$ case both show similar topological characteristics corresponding to sub-graphs of metabolites. For the case of $K=2$, the different prior hyperparameters only lead to different levels of shrinkage, but there is a high degree of similarity in terms of biological interpretation. For example, both Figures \ref{F-networkall} and \ref{difnet} show three different sub-graphs linking metabolites with shared metabolic origin. First, a group of organic acids including succinate, pyruvate, acetate and para-cresol sulphate (PCS) are connected, sometimes also with phenylacetylglutamine (PAG). Several of these metabolites (PCS and PAG) are known to be of gut bacterial origin, and Cd stress is known to modulate gut microbiota populations in mice \citep{Liu2014}. Increased acetate is a known consequence of renal damage, which could be linked to high Cd levels in this population. The second group contains trimethylamine (TMA) and its oxidation product trimethylamine-N-oxide (TMAO), both part of choline metabolism, plus 3-HV and 4-DEA which are products of amino acid catabolism. Choline is an essential nutrient and is metabolised primarily in the liver. Due to its long biological half-life, Cd accumulates in human tissues, especially the liver and kidney, so this observation may point towards a possible mechanistic connection. Moreover, in their original study of this data set, \cite{Ellis2012} reported a positive correlation between urinary Cd and both 4-DEA and 3-HV, though this relationship did not survive correction for age and sex. The third group links citrate and glycine, closely associated via malate and glyoxylate in central carbon metabolism (the network of metabolic reactions essential to life). A strong correlation between Cd and citrate was found by \cite{Ellis2012}, while \cite{Valcarcel2011} found a significant deregulation of the dependency network associated with dimethylglycine, a biproduct of the synthesis of glycine from choline. Thus, it is plausible that several of the metabolites found in the networks of Figures \ref{F-networkall} and \ref{difnet} are involved in pathways disregulated due to Cd exposure. However, metabolite associations derive from a variety of factors and many may be indirect, and possibly non-biochemical in origin, e.g. change in expression of membrane transporters. Thus interpretation of dependency networks, such as those generated here, is difficult. Nonetheless, they give us a novel view of the data not exposed in conventional analyses, and may serve to help generate new hypotheses to be investigated by future biochemical experiments.

\section{Conclusion}
This article proposes using the multiplicative or Chung-Lu random graph model as a prior on the graphical space of GGMs, where the probability of inclusion of each edge is a product of the connectivities of the end nodes. This model can be used to encourage sparsity or particular degree structures, when such prior knowledge is available, say from a database or based on expert opinion. A Bayesian approach is adopted and priors are further placed on the connectivity of the nodes. We study the degree and clustering properties of the multiplicative prior and note that this prior is able to accommodate a wider range of degree structures than the Erd\H{o}s-R\'{e}nyi model. For example, we can use it to encourage shrinkage towards the extremes of 0 and 1, or degree distributions that are right-skewed by varying the hyperparameters,  We illustrate how this prior can be applied to both single and multiple GGMs and a SMC sampler is developed for posterior inference. We find the performance of this sampler to be stable and consistent in our experiments and it can also be parallelized easily. The multiplicative prior also yields rich posterior inference, enabling a study of the connectivity of each node and how the propensity to connect varies across different experimental conditions in the case of multiple GGMs. This allows deeper exploration into the structure of dependency networks and may aid in the formulation of new scientific hypothesis and in opening further lines of investigations.

\section*{Acknowledgments}
Linda Tan is supported by the National University of Singapore Overseas Postdoctoral Fellowship. Ajay Jasra is supported by AcRF Tier 1 grant R-155-000-156-112. T.M.D.E. acknowledges support from the EU PhenoMeNal project (Horizon 2020, 654241). We thank the referees and the associate editor for their comments which have helped to greatly improve the manuscript. 

\begin{supplement}[id=suppA]
	\sname{Supplement}
	\stitle{“Bayesian inference for multiple Gaussian graphical models with application to metabolic association networks}
	\runtitle{Multiple Gaussian graphical models”}
	\slink[doi]{COMPLETED BY THE TYPESETTER}
	\sdatatype{.pdf}
	\sdescription{We provide additional material to support the results in this paper. This include Matlab code, further discussions, detailed derivations and further results on the application to urinary metabolic data.}
\end{supplement}

\newpage

\begin{frontmatter}

\title{Supplement to ``Bayesian inference for multiple Gaussian graphical models with application to metabolic association networks"}
\runtitle{Supplement}

\begin{aug}
	\author{\fnms{Linda S. L.}  \snm{Tan}\thanksref{m1}\ead[label=e1]{st2924@columbia.edu}},
	\author{\fnms{Ajay} \snm{Jasra}\thanksref{m1}\ead[label=e2]{staja@nus.edu.sg}}
	\author{\fnms{Maria}
		\snm{De Iorio}\thanksref{m2}\ead[label=e3]{m.deiorio@ucl.ac.uk}}
	\and
	\author{\fnms{Timothy M. D.}  \snm{Ebbels}\thanksref{m3}
		\ead[label=e4]{t.ebbels@imperial.ac.ukg}}
	
	\runauthor{Tan et al.}
	
	\affiliation{National University of Singapore\thanksmark{m1}, University College London\thanksmark{m2} and Imperial College London\thanksmark{m3}}
	
\end{aug}

\end{frontmatter}

\setcounter{section}{0} \renewcommand{\thesection}{S\arabic{section}}
\setcounter{figure}{0} \renewcommand{\thefigure}{S\arabic{figure}}
\setcounter{table}{0} \renewcommand{\thetable}{S\arabic{table}}

\section{Properties of multiplicative model} 

\subsection{Derivations}
This section provides the proofs of properties P1 - P9 given in Section \ref{S-Properties} of the manuscript.

\begin{itemize} 
	\setlength\itemsep{1em}
\item Proof of \ref{prob_given_pii}:
The probability of a random node $j$ being connected to a node $i$ with connectivity $\pi_i$ is given by $\int_0^1 \pi_i \pi_j p(\pi_j) \, \text{d} \pi_j = \pi_i \mu$.
	
	\item Proof of \ref{av_deg_given_pii}: From \ref{prob_given_pii}, $A_{ij}|\pi_i \sim \text{Bernoulli}(\pi_i\mu)$ for any $j \neq i$. Since $D_i =\sum_{j \neq i} A_{ij}$, $D_i| \pi_i \sim \text{Binomial}(p-1, \pi_i \mu)$. Therefore $\text{E}(D_i|\pi_i) =(p-1)\mu \pi_i $. We also have $E(D_i^2|\pi_i) = (p-1)\mu \pi_i + (p-1)(p-2) \mu^2 \pi_i^2$.
	
	\item Proof of \ref{pgf}:
	\begin{equation*}
	\begin{aligned}
	\text{E}(z^{D_i}| \pi_i) &= \sum_{d=0}^{p-1} Pr(D_i=d| \pi_i) z^d \\
	& =\sum_{d=0}^{p-1} {p-1 \choose d} (\mu \pi_iz)^d (1-\mu\pi)^{p-1-d} \\
	& = (1-\mu\pi_i +\mu\pi_i z)^{p-1}.
	\end{aligned}
	\end{equation*}
	Therefore $\text{E}(z^{D_i}) = \text{E}\{ \text{E}(z^{D_i}| \pi_i)\}=\int_0^1 (1 - \mu\pi_i  + \mu \pi_i z )^{p-1} p(\pi_i) \, \text{d} \pi_i.$
	\begin{equation*}
	\begin{aligned}
	G_{D_i}^{(k)}(z) &= \int_0^1 p(\pi_i) \frac{\partial^k}{\partial^k z}(1 - \mu\pi_i  + \mu \pi_i z )^{p-1}  \, \text{d} \pi_i \\
	& = \int_0^1 p(\pi_i) \frac{(p-1)!}{(p-1-k)!}(\mu \pi_i)^k (1 - \mu\pi_i  + \mu \pi_i z )^{p-1}  \, \text{d} \pi_i.
	\end{aligned}
	\end{equation*}
	Hence $G_{D_i}^{(k)}(1) =\frac{(p-1)!}{(p-1-k)!} \mu^k \int_0^1 p(\pi_i)  \pi_i^k  \, \text{d} \pi_i = \frac{(p-1)!B(a+k,b)}{(p-1-k)!B(a,b)} \mu^k$.
	
	\item Proof of \ref{av_deg}: We have
	\begin{equation*}
	\begin{aligned}
	\text{E}(D_i) &=\text{E}\{ \text{E}(D_i|\pi_i) \} = (p-1) \mu^2. \\
	\text{E}(D_i^2) &=\text{E}\{ \text{E}(D_i^2|\pi_i) \} = (p-1) \mu^2 + (p-1)(p-2) \mu^2 (\mu^2 + \sigma^2). \\
	\end{aligned}
	\end{equation*}
	Hence, $\text{Var}(D_i) = \text{E}(D_i^2) - \text{E}(D_i^2) = (p-1)\mu^2 \{1 - \mu^2 + (p-2) \sigma^2\}$.
	
	\item Proof of \ref{Rdegdistn}: Since $P(D_i=d|\pi_i) = {p-1 \choose d} (\mu \pi_i)^d (1-\mu\pi)^{p-1-d} $, we have $P(D_i=d) = \int_0^1 P(D_i=d|\pi_i)p(\pi_i) \text{d} \pi_i$.
	
	\item Proof of \ref{Rdisp}:
	The dispersion index of the degree distribution can be computed as $\frac{\text{Var}(D_i)}{\text{E}(D_i)} = 1-\mu^2+(p-2)\sigma^2$. Substituting $\mu = \frac{a}{a+b}$ and $\sigma^2 = \frac{ab}{(a+b)^2 (a+b+1)}$, we get the result.
	
	\item Proof of \ref{Rskew}:
	\begin{equation*}
	\begin{aligned}
	\text{E}(D_i^3|\pi_i) &= (p-1)(p-2)(p-3)(\mu \pi_i)^3 \\
	& \quad + 3(p-1)(p-2)(\mu\pi_i)^2 + (p-1) \mu\pi_i. \\
	\text{Hence } \text{E}(D_i^3) &=  (p-1)(p-2)(p-3)\mu^3 E(\pi_i^3) \\
	& \quad  + 3(p-1)(p-2)\mu^2( \mu^2+\sigma^2) + (p-1) \mu^2,
	\end{aligned}
	\end{equation*}
	where $\text{E}(\pi_i^3) = \frac{(a+2)(a+1)a}{(a+b+2)(a+b+1)(a+b)}$. 
	
	\item Proof of \ref{Rdegneighbour_connect}: The average degree of node $j$ given that it is connected to a node with connectivity $\pi_i$ is given by  
	\begin{equation*}
	\begin{aligned}
	& \sum_{d=1}^{p-1} d \, \text{P}(D_j=d| A_{ij}=1, \pi_i) \\
	& = \int \sum_{d=1}^{p-1} d \, \text{P}(D_j=d| A_{ij}=1, \pi_i, \pi_j) \text{P}(\pi_j|A_{ij} =1, \pi_i) \, d\pi_j \\
	& = \int \sum_{d=1}^{p-1} d {p-2 \choose d-1}(\pi_j \mu)^{d-1} (1-\pi_j \mu)^{p-2-(d-1)} \\
	& \quad \times   \frac{\text{P}(A_{ij} =1| \pi_i, \pi_j)p(\pi_i) p(\pi_j)}{\text{P}(A_{ij} =1| \pi_j)p(\pi_i)}  \, d\pi_j  \\
	& = \int \sum_{x=0}^{p-2} (x+1) {p-2 \choose x}(\pi_j \mu)^x (1-\pi_j \mu)^{p-2-x}  \frac{\pi_i \pi_j }{\pi_i \mu} p(\pi_j) \, d\pi_j \\
	& =  \int \{(p-2)\pi_j \mu + 1\}  \frac{\pi_j }{\mu} p(\pi_j) \, d\pi_j \\
	& = (p-2) (\mu^2 + \sigma^2)  + 1.
	\end{aligned}
	\end{equation*}
	Therefore the average degree of a neighbour of node $i$ is independent of its connectivity $\pi_i$.
	Similarly, the average degree of node $j$ given that it is connected to a node with degree $k$ is given by
	\begin{equation} \label{eq}
	\begin{aligned}
	& \sum_{d=1}^{p-1} d \, \text{P}(D_j=d| A_{ij}=1, D_i=k) \\
	& = \int \sum_{d=1}^{p-1} d \, \text{P}(D_j=d| A_{ij}=1, D_i=k,\pi_i) \text{P}(\pi_i| A_{ij}=1, D_i=k) \, d\pi_i \\
	& = \int \sum_{d=1}^{p-1} d \, \text{P}(D_j=d| A_{ij}=1,\pi_i) \frac{\text{P}(\pi_i, A_{ij}=1, D_i=k)}{\text{P}(A_{ij}=1, D_i=k)} \, d\pi_i.
	\end{aligned}
	\end{equation}
	From above, we have $\sum_{d=1}^{p-1} d \, \text{P}(D_j=d| A_{ij}=1,\pi_i) =  (p-2) (\mu^2 + \sigma^2)  + 1$ and 
	\begin{equation*}
	\begin{aligned}
	\frac{\text{P}(\pi_i, A_{ij}=1, D_i=k)}{\text{P}(A_{ij}=1, D_i=k)} &= \frac{\text{P}(D_i=k| \pi_i, A_{ij}=1) \text{P}(A_{ij}=1| \pi_i)p(\pi_i)}{\int \text{P}(D_i=k| \pi_i, A_{ij}=1) \text{P}(A_{ij}=1| \pi_i)p(\pi_i) d\pi_i} \\
	&= \frac{ {p-2 \choose k-1} (\pi\mu)^k (1-\pi\mu)^{p-1-k} \, p(\pi_i)}{\int {p-2 \choose k-1} (\pi\mu)^k (1-\pi\mu)^{p-1-k} \, p(\pi_i) d\pi_i} \\
	&= \frac{ \text{P}(D_i=k| \pi_i)\, p(\pi_i)}{\text{P}(D_i=k)}. \\
	\end{aligned}
	\end{equation*}
	Therefore \eqref{eq} simplifies to $(p-2) (\mu^2 + \sigma^2)  + 1$, which is again independent of the degree of $i$.
	
	\item Proof of \ref{PClust}: The global clustering coefficient can be written as 
	\begin{equation*}
	\begin{aligned}
		\frac{ \int (\pi_i \pi_k) (\pi_k \pi_j)(\pi_j \pi_i)p(\pi_i) p(\pi_j) p(\pi_k) \; d\pi_i d\pi_j d\pi_k}{\int (\pi_i \pi_k) (\pi_k \pi_j)p(\pi_i) p(\pi_j) p(\pi_k) \; d\pi_i d\pi_j d\pi_k} &= \frac{(\sigma^2 + \mu^2)^3}{(\sigma^2 + \mu^2)^2 \mu}  \\
		&= \frac{a+1}{a+b+1}.
	\end{aligned}
	\end{equation*}
\end{itemize}

\subsection{Dispersion index and skewness}
\begin{figure}
	\centering
	\includegraphics[width=0.35\linewidth]{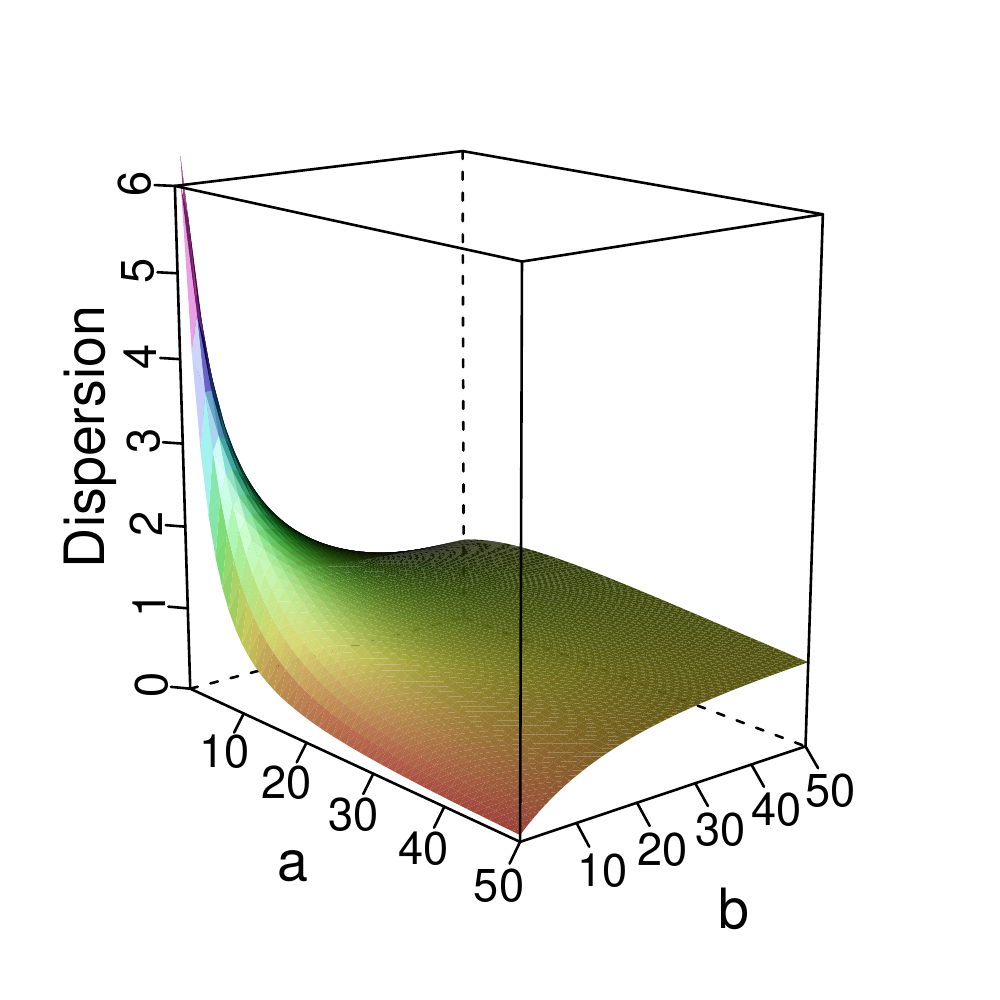}
	\includegraphics[width=0.34\linewidth]{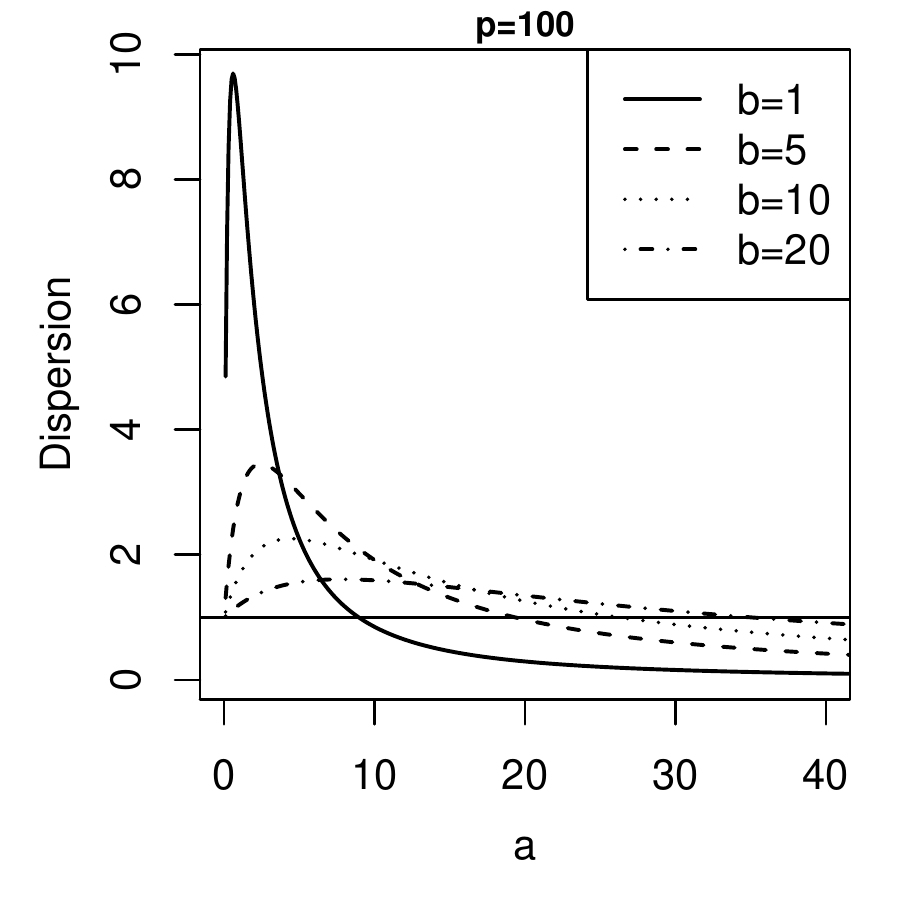}
	\caption{\label{disp}Left: Dispersion index as a function of hyperparameters $a$ and $b$ in the Beta prior for $p=100$. Right: Cross-sectional plots for $b=1,5,10,20$.}
\end{figure}

\begin{figure}[h]
	\centering
	\includegraphics[width=0.35\linewidth]{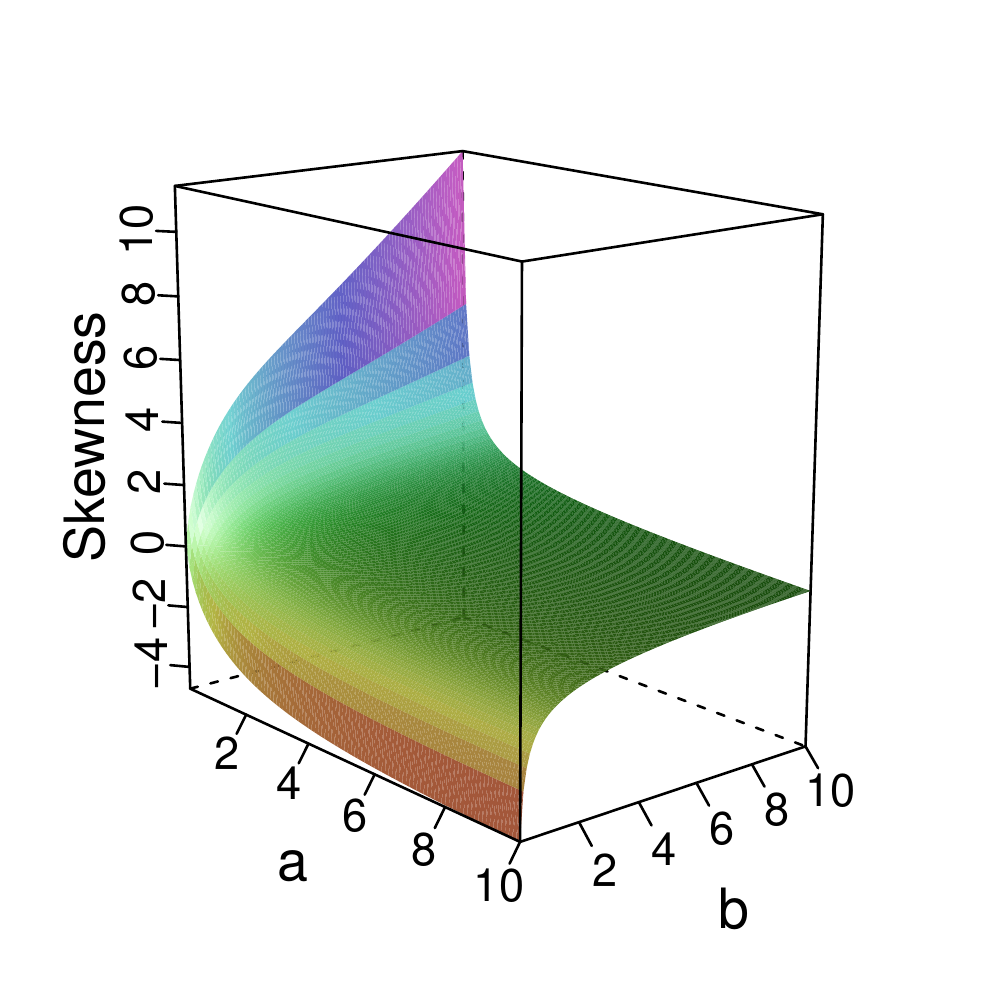}
	\includegraphics[width=0.34\linewidth]{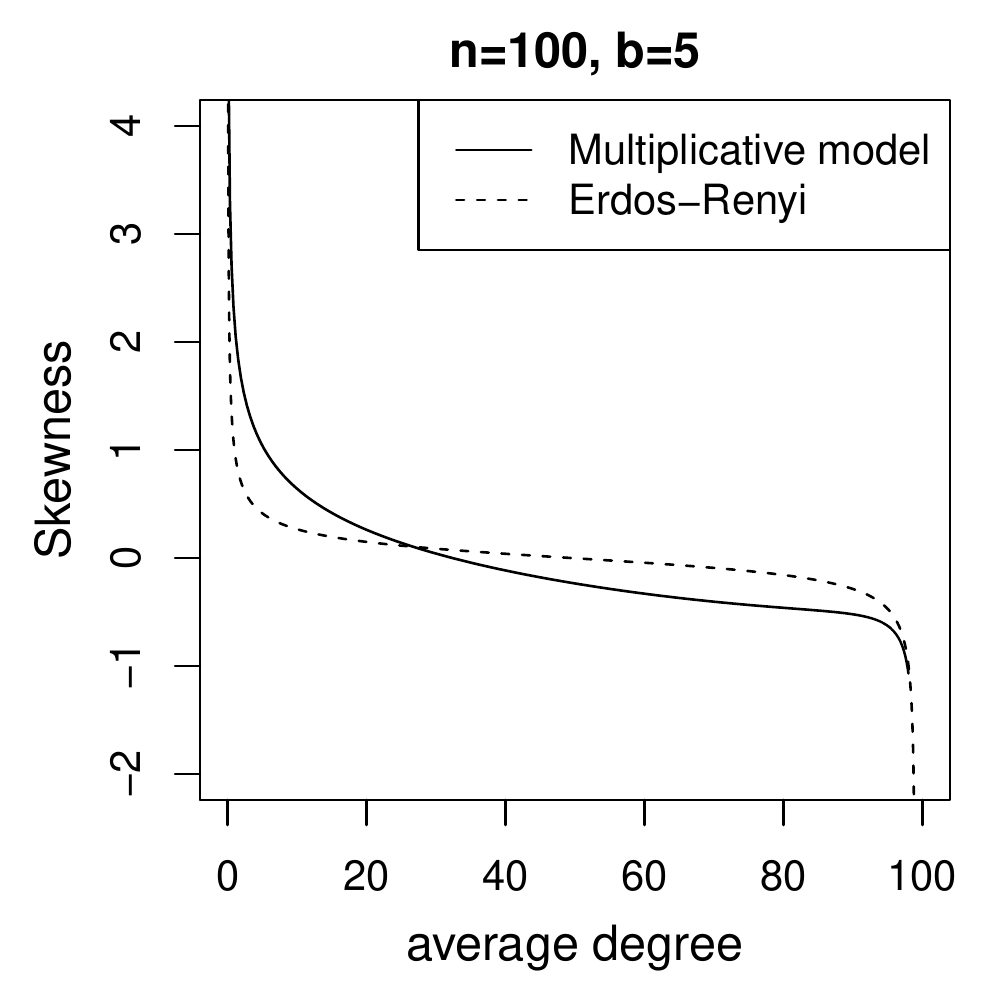}
	\caption{\label{skew}Left: Skewness as a function of hyperparameters $a$ and $b$ in the Beta prior for $p=100$. Right: Comparison of skewness of multiplicative prior with the Erd\H{o}s-R\'{e}nyi model for the same mean degree.}
\end{figure}

Figures \ref{disp} and \ref{skew} show plots of the dispersion index and skewness as a function of $a$ and $b$ for $p=100$. The left plots in Figures \ref{disp} and \ref{skew} show how the dispersion index and skewness vary across a wide range of hyperparameter values. The right plot in Figure \ref{disp} shows some cross-sectional plots of the skewness while the right plot in Figure \ref{skew} compares the skewness of the multiplicative prior with the Erd\H{o}s-R\'{e}nyi model for the same mean degree when $p=100$ and $b=5$. The multiplicatve model tends to be more positively skewed for small values of $a$ and less so for large values of $a$ as compared to the Erd\H{o}s-R\'{e}nyi model. 

\section{Illustration of connectivity of nodes when $K=2$}
\begin{figure}[htb!]
	\centering \includegraphics[width=0.4\linewidth]{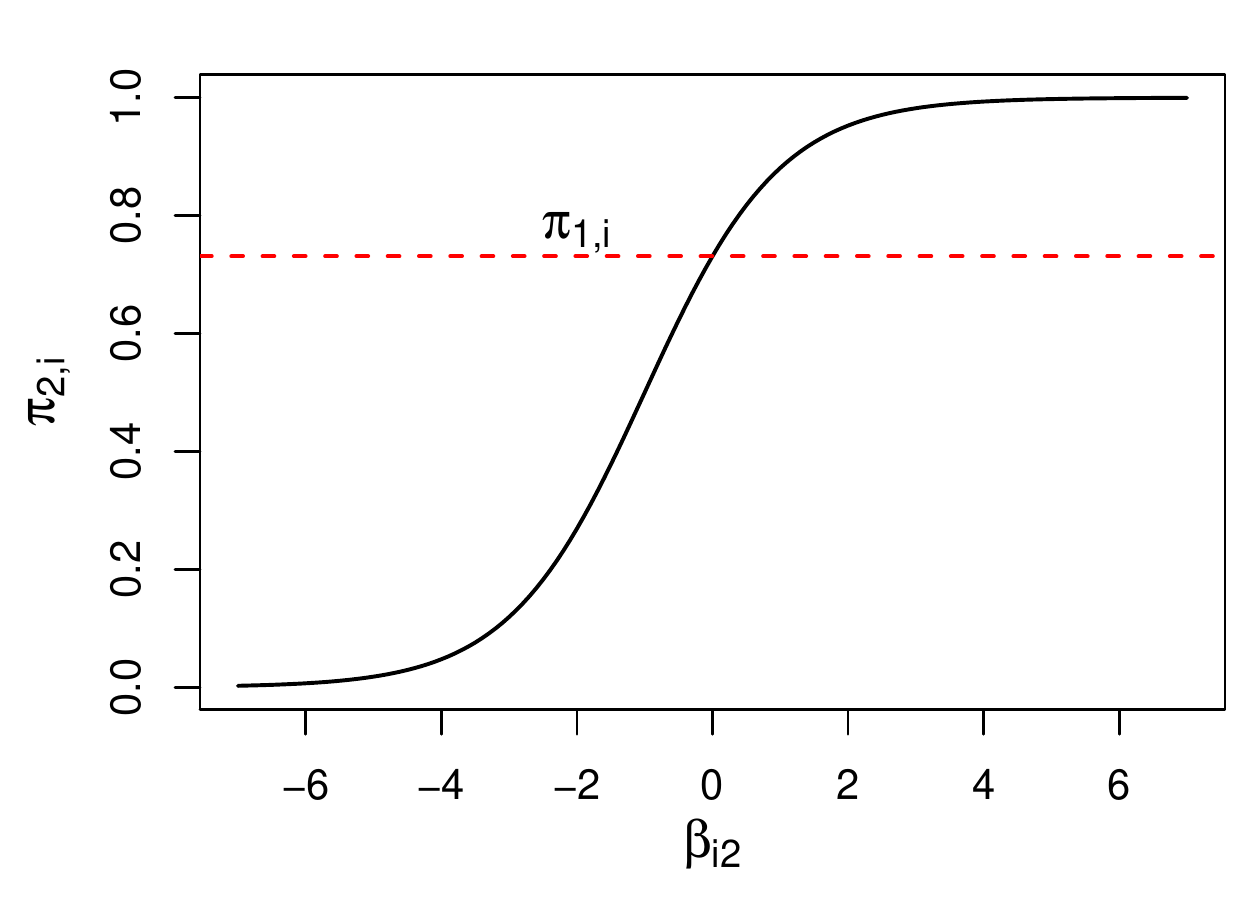}
	\caption{\label{F-betas} Connectivity of node $i$ in $G_2$ ($\pi_{2,i}$) as a function of $\beta_{i2}$ (black line), with $\beta_{i1}$ fixed at 1. Red dotted line marks the value of $\pi_{1,i}$.}
\end{figure}

In our proposed prior for multiple GGMs, the connectivity of node $i$ is $\pi_{1,i} = \{1+\exp(-\beta_{i1})\}^{-1}$ in $G_1$ and $\pi_{2,i} = \{1+\exp(-\beta_{i1} - \beta_{i2})\}^{-1}$ in $G_2$ when $K=2$, $x_1 = (1,0)$ and $x_2=(1,1)$. The relationship between the connectivity of a node and its regression coefficients is illustrated in Figure \ref{F-betas}.

\section{Laplace approximation} \label{A-Laplace}
The gradients and Hessians required in the Laplace approximation for computing prior probabilities of graphs described in Section 4.1 are given below. They can be derived using vector differential calculus and a useful reference is \cite{Magnus1988}. For simplicity, we have omitted dependence of $f$ on other variables in its expression below. 

For $K=1$, we take 
\begin{equation*}
\begin{aligned}
f(u) & =  \sum_i \left\{ (a+d_i) u_i - (a+d_i+1) \log[1+\exp(u_i)] + (b-1) \log (1-\pi_i) \right\}  \\
& \quad + \sum_{i<j} (1-A_{ij}) \log (1- \pi_i \pi_j ) -p \log B(a,b).
\end{aligned}
\end{equation*}
We have $\frac{d\pi_i}{d u_i} = \frac{\pi_i}{1+\exp(u_i)}$, $\frac{d^2\pi_i}{du_i^2} =  \frac{d\pi_i}{du_i} \frac{1-\exp(u_i)}{1+\exp(u_i)}$ and
\begin{itemize}
	\item $\frac{\partial f}{\partial u_i} = a+d_i - (a+d_i+1)\pi_i -(b-1)\frac{d\pi_i}{d u_i}/(1-\pi_i) - \sum_{j \neq i} (1-A_{ij})\pi_j \frac{d\pi_i}{d u_i}/(1-\pi_i \pi_j)$, 
	\item $\frac{\partial^2f}{\partial u_i \partial u_j} = - (1-A_{ij}) \frac{d\pi_j}{d u_j} \frac{d\pi_i}{d u_i}/ (1-\pi_i \pi_j)^2$,
	\item $\frac{\partial^2f}{\partial u_i^2} = - (a+d_i+1)\frac{d\pi_i}{d u_i} -(b-1)\left\{ \frac{d^2\pi_i}{d u_i^2}/(1-\pi_i) + (\frac{d\pi_i}{d u_i})^2 / (1-\pi_i)^2 \right\}$ \\ 
	\hspace*{10mm} $ - \sum_{j \neq i} (1-A_{ij}) \left\{ \pi_j \frac{d^2\pi_i}{d u_i^2} / (1-\pi_i \pi_j) +\pi_j^2( \frac{d\pi_i}{d u_i})^2 / (1-\pi_i \pi_j)^2 \right\}$.
\end{itemize}

For $K>1$, we take 
\begin{multline*}
f(\beta) = \sum_{k=1}^K\left\{ \sum_{i=1}^p d_{k,i} \log \pi_{k,i} + \sum_{i <j}(1-A_{k,ij}) \log(1 - \pi_{k,i} \pi_{k,j} )   \right\} \\
- \sum_{i=1}^p \sum_{q=1}^Q \left\{ \frac{1}{2} \log(2\pi \sigma_q^2) + \frac{\beta_{iq}^2}{2\sigma_q^2} \right\}.
\end{multline*}
We have $ \frac{\partial \pi_{k,i} }{\partial \beta_i} = \frac{ \pi_{k,i} x_k}{1+\exp(\beta_i^Tx_k)}$, $\frac{\partial^2 \pi_{k,i}}{\partial \beta_i \partial \beta_i^T} = \pi_{k,i} \frac{1 - \exp(\beta_i^Tx_k)} {\{1+\exp(\beta_i^Tx_k)\}^2} x_kx_k^T$ and
\begin{itemize}
	\item $\frac{\partial f}{\partial \beta_i}= \sum_{k=1}^K \left\{ \frac{d_{k,i}}{1+\exp(\beta_i^Tx_k)}  - \sum_{j \neq i} \frac{ (1-A_{k,ij}) \pi_{k,i} \pi_{k,j}} {(1-\pi_{k,i} \pi_{k,j}) \{1+ \exp(\beta_i^Tx_k)\}}  \right\}x_k \\ - \text{diag}(\frac{1}{\sigma^2})\beta_i$,
	\item $\frac{\partial^2 f}{\partial \beta_i \partial \beta_j^T} = - \sum_{k=1}^K (1-A_{k,ij})(1-\pi_{k,i} \pi_{k,j})^{-2} \; \frac{\partial\pi_{k,i} }{\partial \beta_i} \left( \frac{\partial \pi_{k,j}}{\partial \beta_j} \right)^T$,
	\item $\frac{\partial^2 f}{\partial \beta_i \beta_i^T} = - \sum_{k=1}^K\Big\{ \frac{d_{k,i}\pi_{k,i}}{1+\exp(\beta_i^Tx_k)} \\
	+\sum_{j \neq i}  \frac{(1-A_{k,ij}) \pi_{k,i} \pi_{k,j} \{1 - \exp(\beta_i^Tx_k) (1-\pi_{k,i} \pi_{k,j})\}} {(1-\pi_{k,i} \pi_{k,j})^2 \{1+\exp(\beta_i^Tx_k)\}^2 }  \Big\}x_k x_k^T  -  \text{diag}(\frac{1}{\sigma^2})$,
\end{itemize}
where $\frac{1}{\sigma^2} = (\frac{1}{\sigma_1^2}, \dots, \frac{1}{\sigma_Q^2})$ is evaluated element-wise.

\section{Simulation} \label{supsec: simulation}
\begin{figure}[htb!]
\centering
\includegraphics[width=\textwidth]{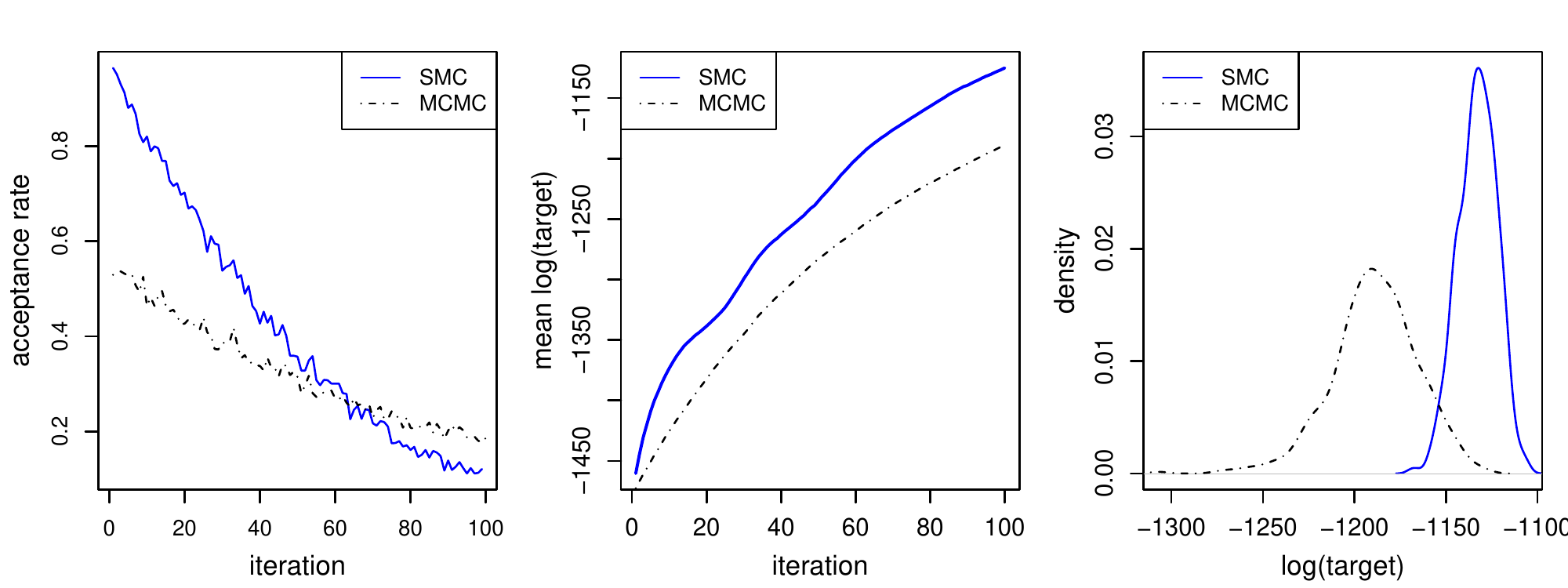}
\caption{Acceptance rate (left), mean log(target) (center) at each iteration and density of the log(target) of 500 samples (right).  \label{SMC_MCMC}}
\end{figure}

In this simulation, we compare the performance of SMC with standard MCMC for a dataset simulated from the multiplicative model in Section \ref{sec:simulations}. Using the multiplicative prior MP(1,1), we run standard MCMC (where the temperature $\phi$ is fixed at 1 throughout) for 100 iterations and set $N=500$ and $M=3$. We compare results obtained with that of Algorithm 1 for the same number of iterations.  Figure \ref{SMC_MCMC} shows the acceptance rate (left) and mean log(target) (center) at each iteration and the distribution of the log(target) of the 500 samples at the end of each algorithm. It is clear that the SMC algorithm has better mixing and achieves higher log target density than standard MCMC.

\section{Urinary metabolic data}
\begin{figure}[H]
\centering
\includegraphics[width=0.88\textwidth]{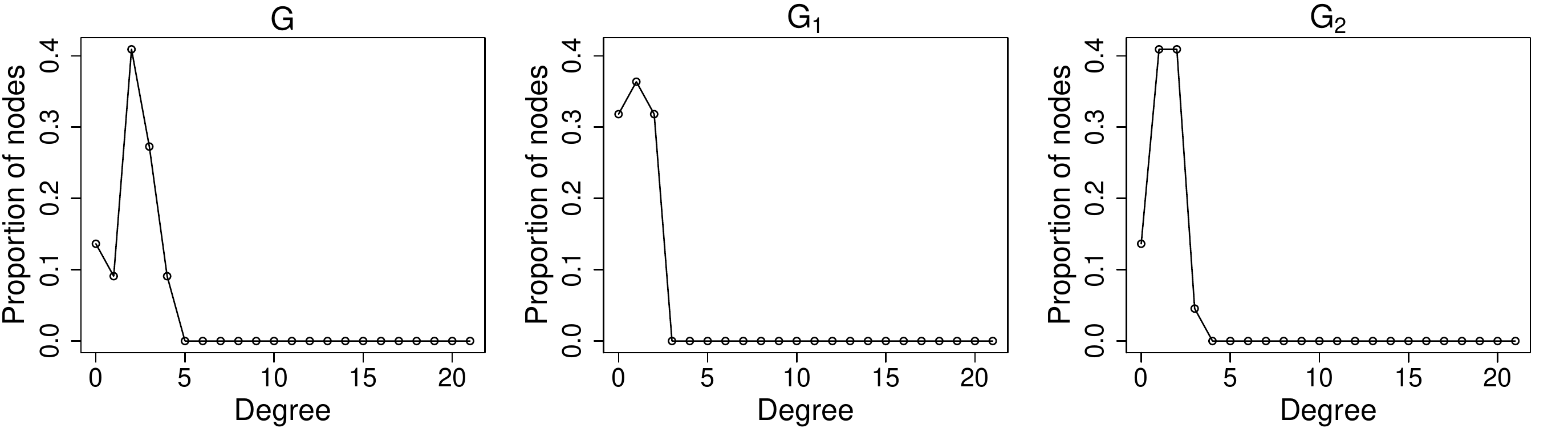}
\caption{\label{F-genenetdegdistn} Degree distributions estimated using GeneNet for the case where the individuals are treated as one heterogeneous group (left) and where they are divided into two groups $S_1$ (middle) and $S_2$ (right).}
\end{figure}

Figure \ref{F-genenetdegdistn} shows the degree distributions estimated using GeneNet.

\subsection{Case: $K=1$}
\begin{figure}[p]
	\centering
	\includegraphics[width=0.7\textwidth]{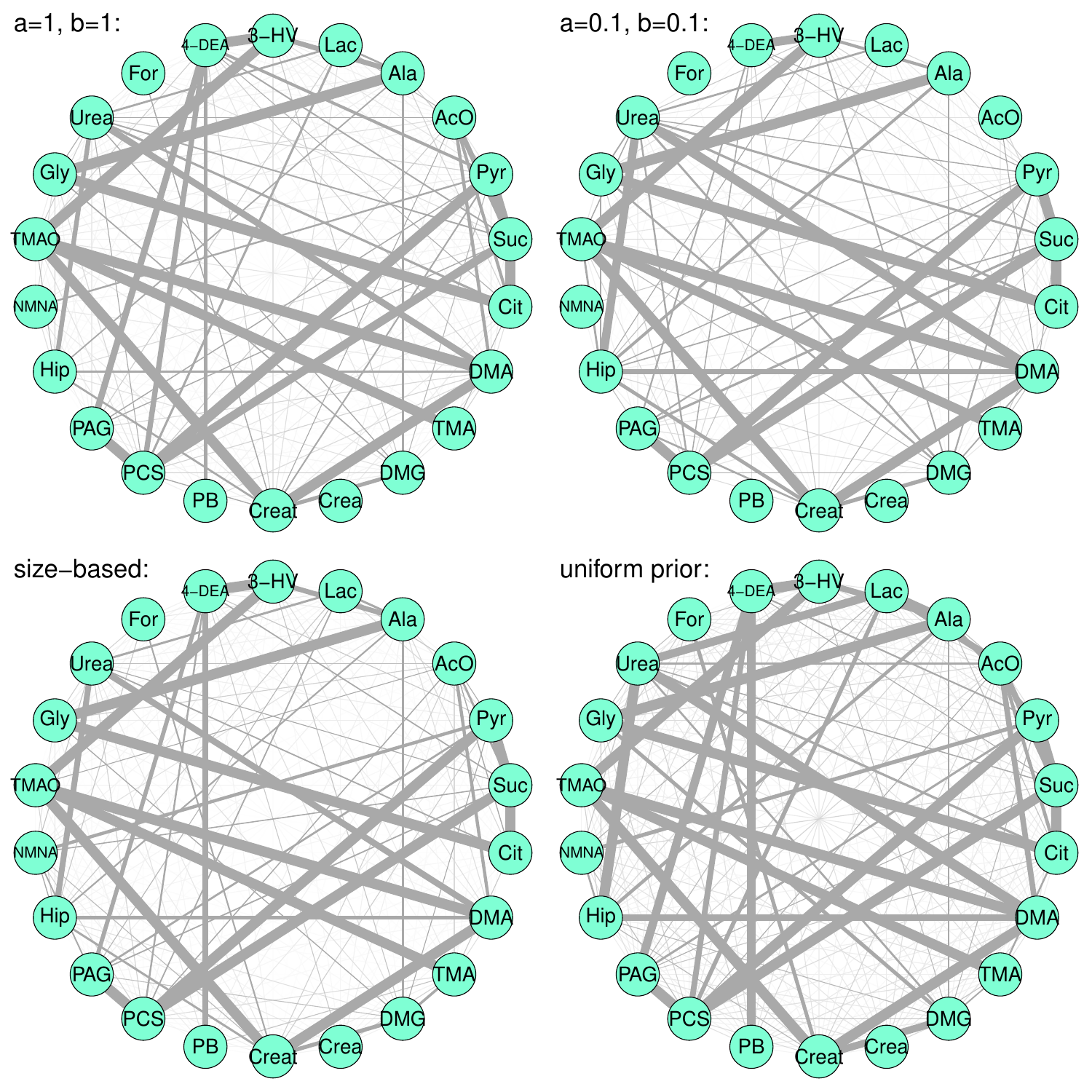}
	\caption{Graphs corresponding to different priors. Width of edges are proportional to their posterior weights.}
	\label{F-Gall}
\end{figure}

Figure \ref{F-Gall} shows the graphs obtained from Algorithm 1 under each prior. The width of every edge is proportional to its posterior weight. Table \ref{Gcon} shows the mean and 95\% credible interval of the connectivity of each metabolite. Table \ref{CovK1} shows the mean precision matrix corresponding to the multiplicative prior with $a=b=1$.

\begin{table}[htb!]
	\centering \ra{1.05} 
	\begin{footnotesize}
		\begin{tabular}{@{}l|cc|cc@{}}
			\hline
			& \multicolumn{2}{c}{ M(1, 1)} \vline&  \multicolumn{2}{c}{M(0.1, 0.1)} \\ 
			Abbrev & $\bar{\pi}_i$ & CI ($\pi_i$) & $\bar{\pi}_i$ & CI ($\pi_i$)  \\ 
			\hline
			TMAO & 0.48 & (0.15, 0.84) & 0.54 & (0.00, 0.90) \\ 
			PCS & 0.46 & (0.11, 0.84) & 0.43 & (0.06, 0.96) \\ 
			Suc & 0.43 & (0.06, 0.81) & 0.46 & (0.07, 0.95) \\ 
			DMA & 0.42 & (0.07, 0.80) & 0.46 & (0.07, 0.96) \\ 
			Creat & 0.39 & (0.07, 0.76) & 0.38 & (0.06, 0.99) \\ 
			4-DEA & 0.39 & (0.02, 0.79) & 0.18 & (0.00, 0.14) \\ 
			Pyr & 0.36 & (0.05, 0.73) & 0.36 & (0.05, 0.99) \\ 
			Cit & 0.36 & (0.05, 0.71) & 0.28 & (0.00, 0.65) \\ 
			3-HV & 0.33 & (0.05, 0.66) & 0.32 & (0.00, 0.81) \\ 
			Gly & 0.32 & (0.04, 0.66) & 0.31 & (0.00, 0.76) \\ 
			Urea & 0.31 & (0.00, 0.75) & 0.40 & (0.01, 1.00) \\ 
			Ala & 0.30 & (0.02, 0.63) & 0.22 & (0.00, 0.62) \\ 
			PAG & 0.28 & (0.02, 0.60) & 0.23 & (0.00, 0.68) \\ 
			AcO & 0.25 & (0.00, 0.63) & 0.05 & (0.00, 0.05) \\ 
			Hip & 0.22 & (0.00, 0.60) & 0.33 & (0.04, 1.00) \\ 
			DMG & 0.22 & (0.00, 0.58) & 0.21 & (0.00, 0.19) \\ 
			TMA & 0.20 & (0.00, 0.46) & 0.13 & (0.00, 0.11) \\ 
			Lac & 0.18 & (0.00, 0.52) & 0.08 & (0.00, 0.07) \\ 
			PB & 0.14 & (0.00, 0.47) & 0.02 & (0.00, 0.03) \\ 
			NMNA & 0.14 & (0.00, 0.48) & 0.03 & (0.00, 0.03) \\ 
			For & 0.12 & (0.00, 0.44) & 0.02 & (0.00, 0.03) \\ 
			Crea & 0.10 & (0.00, 0.43) & 0.02 & (0.00, 0.03) \\ 
			\hline
		\end{tabular}
	\end{footnotesize}
	\caption{Mean and 95\% credible interval of the connectivity ($\pi_i$) of each metabolite under the multiplicative prior with $a=b=1$ (M(1, 1)) and $a=b=0.1$ (M(0.1, 0.1)). }
	\label{Gcon}
\end{table}

\begin{table}[htb!]
	\centering \tabcolsep=0.1cm \ra{1.08}
	\resizebox{\columnwidth}{!}{\begin{tabular}{@{}lrrrrrrrrrrrrrrrrrrrrrr@{}}
			\hline
			& TMAO & PCS & Suc & DMA & Creat & 4-DEA & Pyr & Cit & 3-HV & Gly & Urea & Ala & PAG & AcO & Hip & DMG & TMA & Lac & PB & NMNA & For & Crea \\ 
			\hline
			TMAO & 2.64 & 0.00 & 0.00 & -1.86 & 0.82 & 0.00 & 0.00 & 0.00 & -0.52 & -0.00 & -0.01 & -0.00 & -0.00 & 0.00 & 0.02 & -0.01 & -0.43 & -0.00 & -0.00 & -0.00 & -0.00 & -0.00 \\ 
			PCS & . & 5.93 & -0.82 & 0.00 & 0.02 & -0.36 & -3.62 & -0.00 & 0.00 & 0.01 & -0.02 & -0.00 & -2.11 & -0.02 & 0.00 & 0.01 & -0.00 & 0.06 & 0.00 & 0.01 & 0.00 & 0.00 \\ 
			Suc & . & . & 1.53 & 0.00 & -0.00 & -0.04 & 0.67 & -0.57 & -0.00 & -0.00 & 0.05 & -0.00 & 0.05 & -0.19 & -0.00 & 0.00 & 0.00 & 0.00 & -0.00 & 0.00 & -0.01 & -0.00 \\ 
			DMA & . & . & . & 3.02 & -1.30 & -0.01 & -0.00 & -0.00 & -0.00 & -0.00 & 0.24 & -0.00 & -0.00 & 0.10 & 0.09 & -0.01 & -0.02 & 0.00 & -0.00 & 0.00 & 0.00 & -0.00 \\ 
			Creat & . & . & . & . & 1.71 & -0.01 & 0.02 & 0.00 & -0.00 & -0.00 & -0.00 & 0.00 & 0.00 & 0.02 & 0.06 & -0.12 & -0.00 & 0.00 & 0.00 & 0.01 & 0.03 & 0.00 \\ 
			4-DEA & . & . & . & . & . & 1.34 & 0.15 & 0.00 & -0.41 & -0.00 & 0.00 & -0.00 & 0.36 & -0.00 & 0.00 & -0.00 & -0.00 & -0.00 & -0.13 & 0.00 & 0.00 & -0.00 \\ 
			Pyr & . & . & . & . & . & . & 4.23 & -0.00 & 0.00 & -0.00 & -0.01 & -0.02 & -0.08 & 0.01 & 0.00 & 0.02 & -0.00 & 0.01 & 0.00 & 0.06 & 0.00 & 0.00 \\ 
			Cit & . & . & . & . & . & . & . & 1.54 & -0.00 & -0.52 & 0.07 & -0.01 & -0.00 & 0.11 & 0.00 & 0.00 & 0.02 & 0.00 & 0.00 & 0.00 & -0.00 & 0.00 \\ 
			3-HV & . & . & . & . & . & . & . & . & 1.46 & -0.00 & 0.00 & -0.15 & 0.00 & 0.00 & 0.00 & -0.02 & -0.00 & -0.00 & -0.00 & -0.00 & 0.00 & 0.00 \\ 
			Gly & . & . & . & . & . & . & . & . & . & 1.56 & 0.03 & -0.60 & 0.00 & 0.01 & 0.02 & -0.05 & -0.00 & -0.00 & -0.00 & 0.00 & -0.00 & 0.00 \\ 
			Urea & . & . & . & . & . & . & . & . & . & . & 1.21 & 0.00 & -0.00 & -0.04 & 0.17 & -0.03 & -0.00 & 0.06 & 0.00 & 0.00 & 0.00 & -0.00 \\ 
			Ala & . & . & . & . & . & . & . & . & . & . & . & 1.41 & 0.00 & 0.00 & 0.05 & -0.07 & -0.00 & -0.06 & 0.00 & 0.00 & -0.00 & -0.00 \\ 
			PAG & . & . & . & . & . & . & . & . & . & . & . & . & 2.83 & 0.00 & -0.00 & 0.00 & 0.00 & 0.02 & 0.00 & 0.00 & 0.00 & -0.00 \\ 
			AcO & . & . & . & . & . & . & . & . & . & . & . & . & . & 1.14 & 0.00 & 0.00 & 0.00 & -0.02 & 0.00 & 0.00 & -0.00 & -0.00 \\ 
			Hip & . & . & . & . & . & . & . & . & . & . & . & . & . & . & 1.13 & 0.01 & 0.00 & 0.00 & -0.00 & -0.01 & -0.00 & 0.00 \\ 
			DMG & . & . & . & . & . & . & . & . & . & . & . & . & . & . & . & 1.11 & -0.01 & -0.00 & -0.00 & 0.00 & -0.00 & -0.00 \\ 
			TMA & . & . & . & . & . & . & . & . & . & . & . & . & . & . & . & . & 1.21 & 0.00 & 0.00 & -0.00 & -0.00 & 0.00 \\ 
			Lac & . & . & . & . & . & . & . & . & . & . & . & . & . & . & . & . & . & 1.08 & -0.00 & 0.00 & -0.00 & -0.00 \\ 
			PB & . & . & . & . & . & . & . & . & . & . & . & . & . & . & . & . & . & . & 1.07 & 0.00 & -0.00 & 0.00 \\ 
			NMNA & . & . & . & . & . & . & . & . & . & . & . & . & . & . & . & . & . & . & . & 1.05 & -0.00 & 0.00 \\ 
			For & . & . & . & . & . & . & . & . & . & . & . & . & . & . & . & . & . & . & . & . & 1.04 & 0.00 \\ 
			Crea & . & . & . & . & . & . & . & . & . & . & . & . & . & . & . & . & . & . & . & . & . & 1.03 \\ 
			\hline
		\end{tabular}}
		\caption{Mean precision matrix $\Omega$ correponding to $K=1$ and multiplicative prior with $a=b=1$. \label{CovK1}}
	\end{table}

\subsection{Case: $K=2$}
\begin{figure}[htb!]
		\centering
		\includegraphics[width=0.8\textwidth]{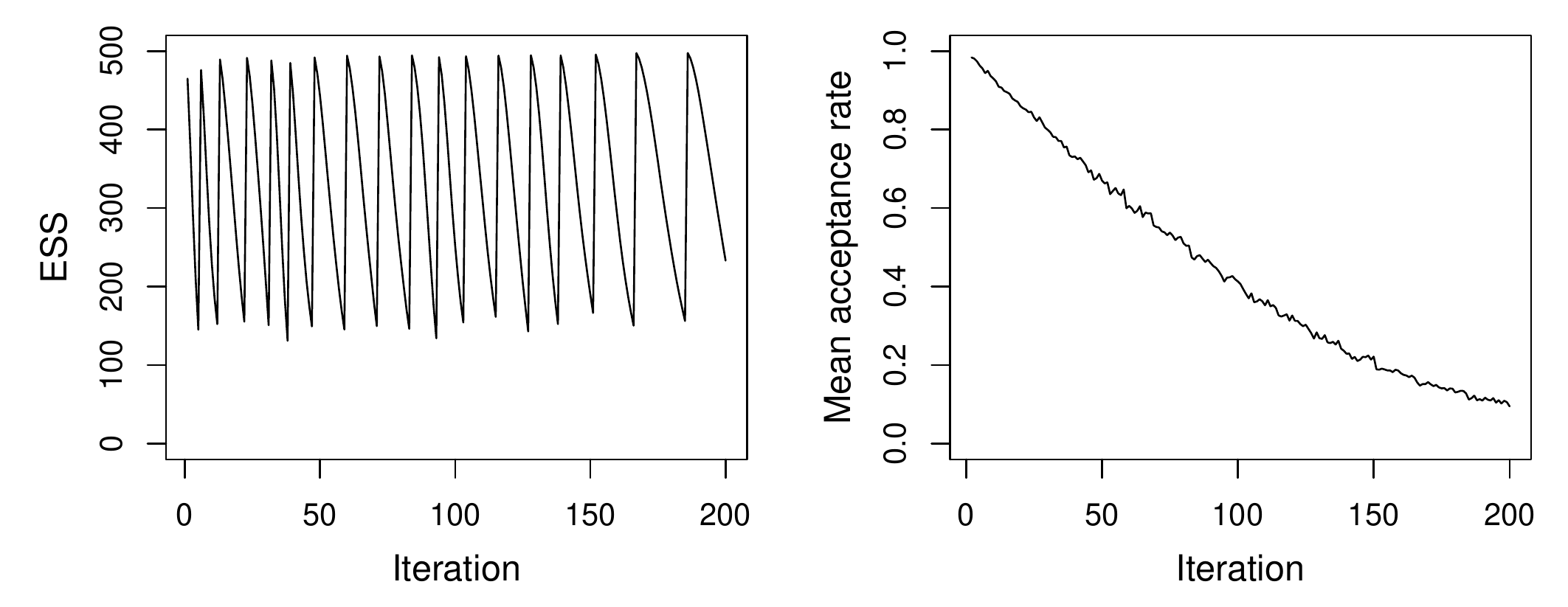}
		\caption{\label{F-ESSA} Typical plot of ESS (left) and mean acceptance rate (right) of SMC algorithm. This plot is obtained from fitting the urinary metabolic data using $K=2$ for the multiplicative prior with $\sigma_1^2=\sigma_2^2=1$.}
	\end{figure}

\begin{figure}[htb!]
		\centering
		\includegraphics[width=0.65\textwidth]{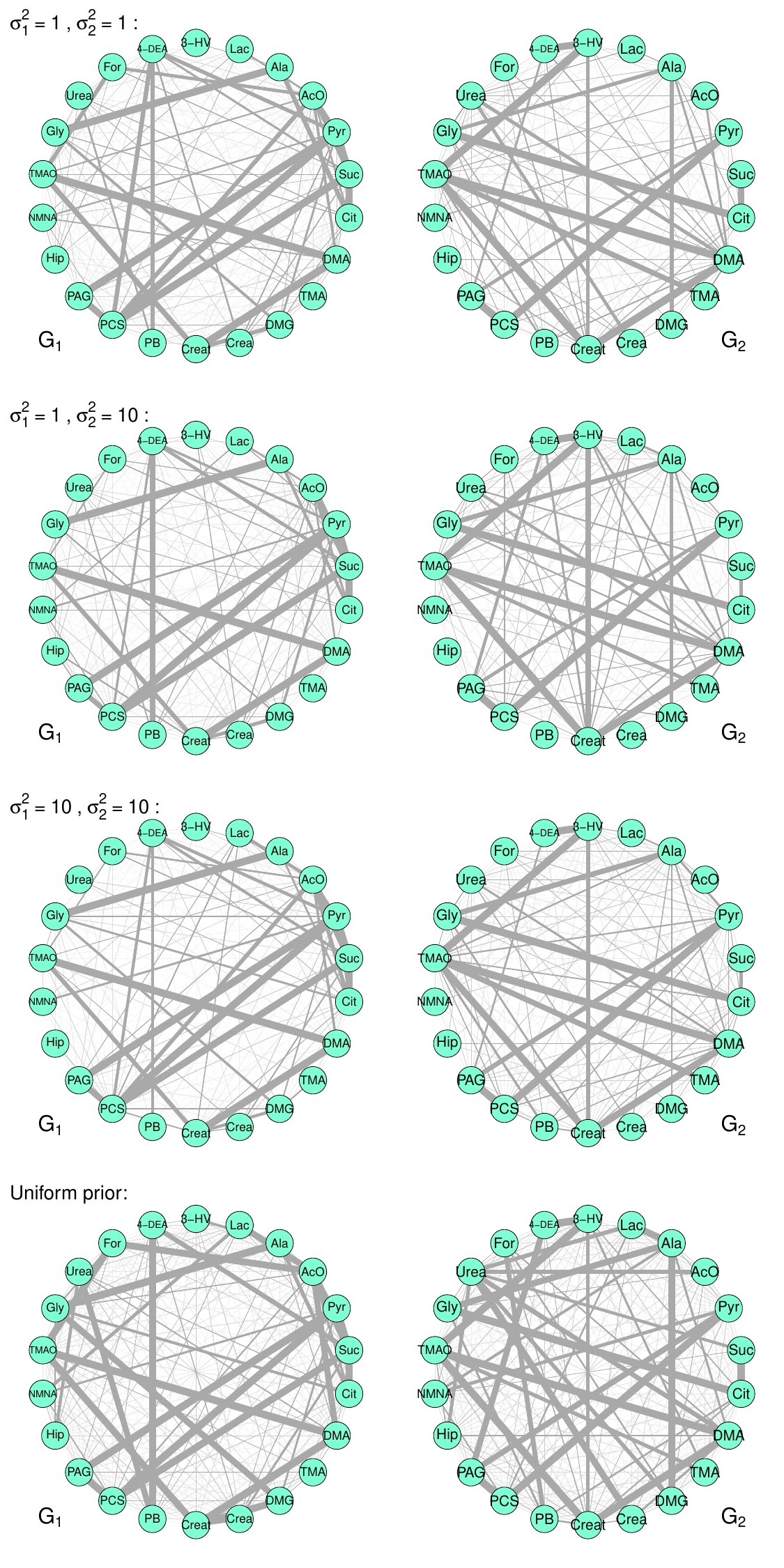} 
		\caption{Graphs of $G_1$ and $G_2$ corresponding to different priors. Width of edges are proportional to their posterior weights.}
		\label{G2all}
	\end{figure}

\begin{figure}[htb!]
		\centering
		\includegraphics[width=\textwidth]{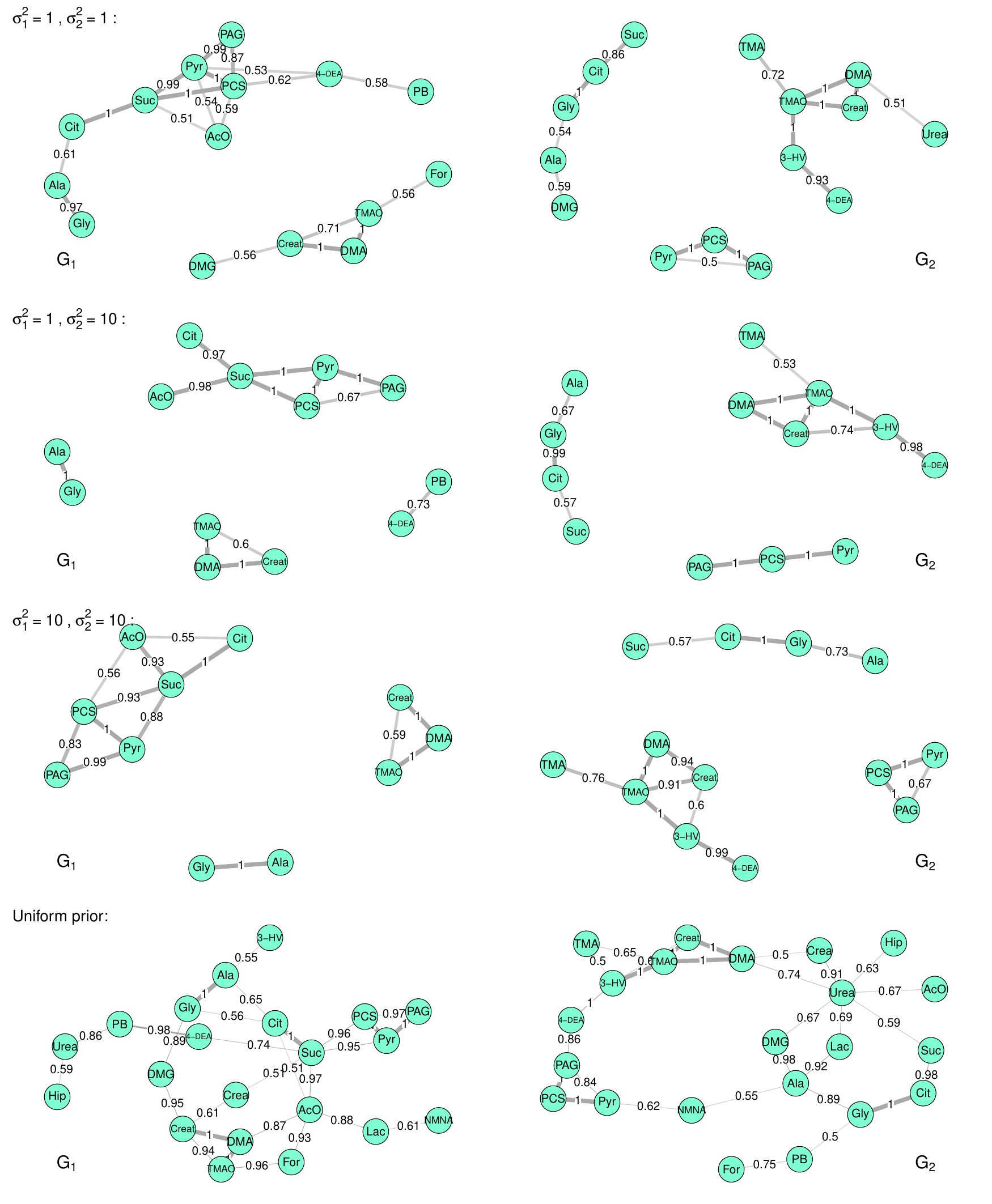} 
		\caption{Plots of $G_1$ and $G_2$ showing edges with weights greater than 0.5 corresponding to different priors.}
		\label{network2all}
	\end{figure}

\begin{figure}[htb!]
		\centering
		\includegraphics[width=\textwidth]{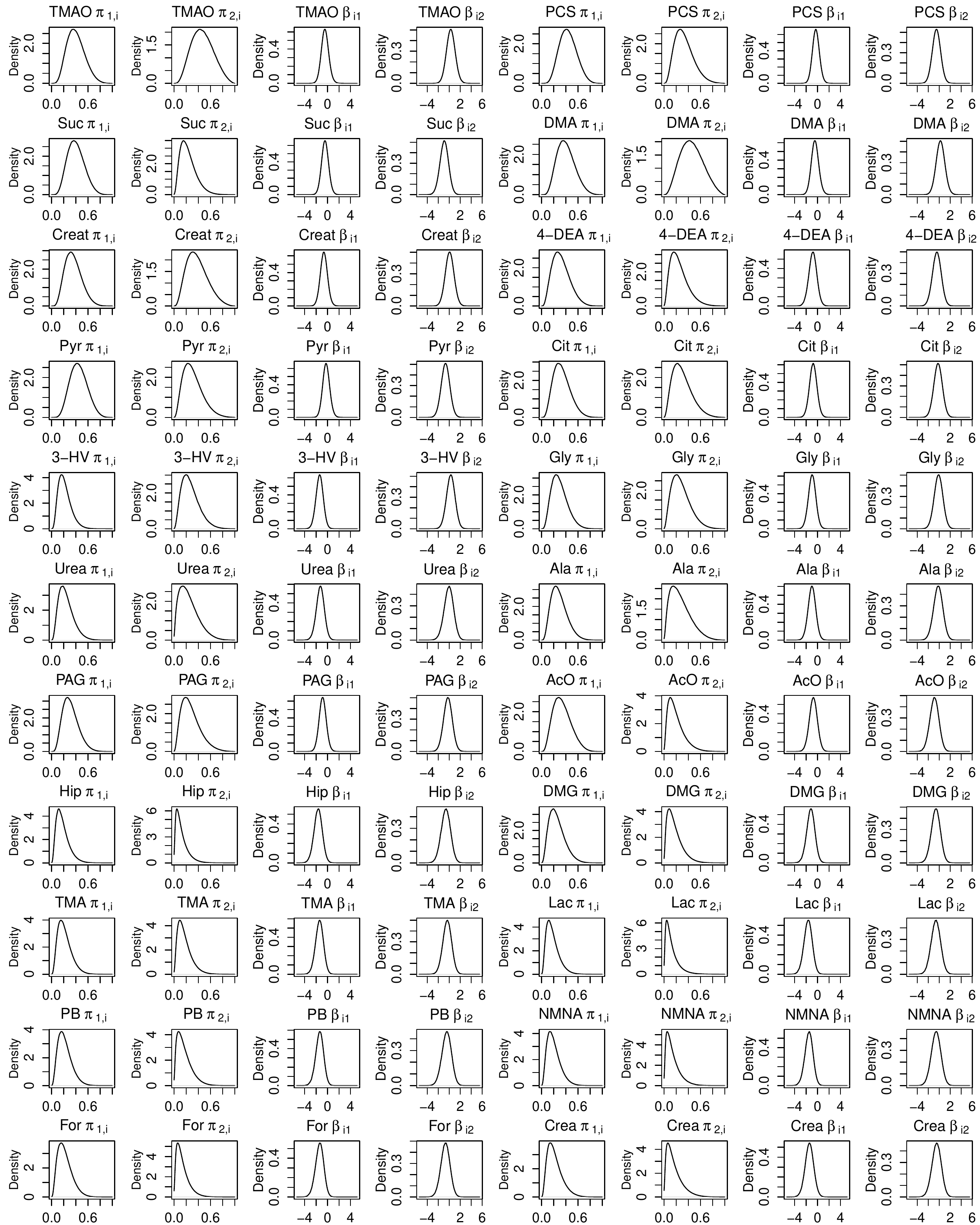} 
		\caption{Posterior distributions of the connectivity $\pi_{k,i}$ and the regression coefficients $\beta_{iq}$ for each metabolite for $k=1,2$ and $q=1,2$ corresponding to the multiplicative prior with $\sigma_1^2=\sigma_2^2=1$.}
		\label{post211}
	\end{figure}

\begin{figure}[htb!]
		\centering
		\includegraphics[width=\textwidth]{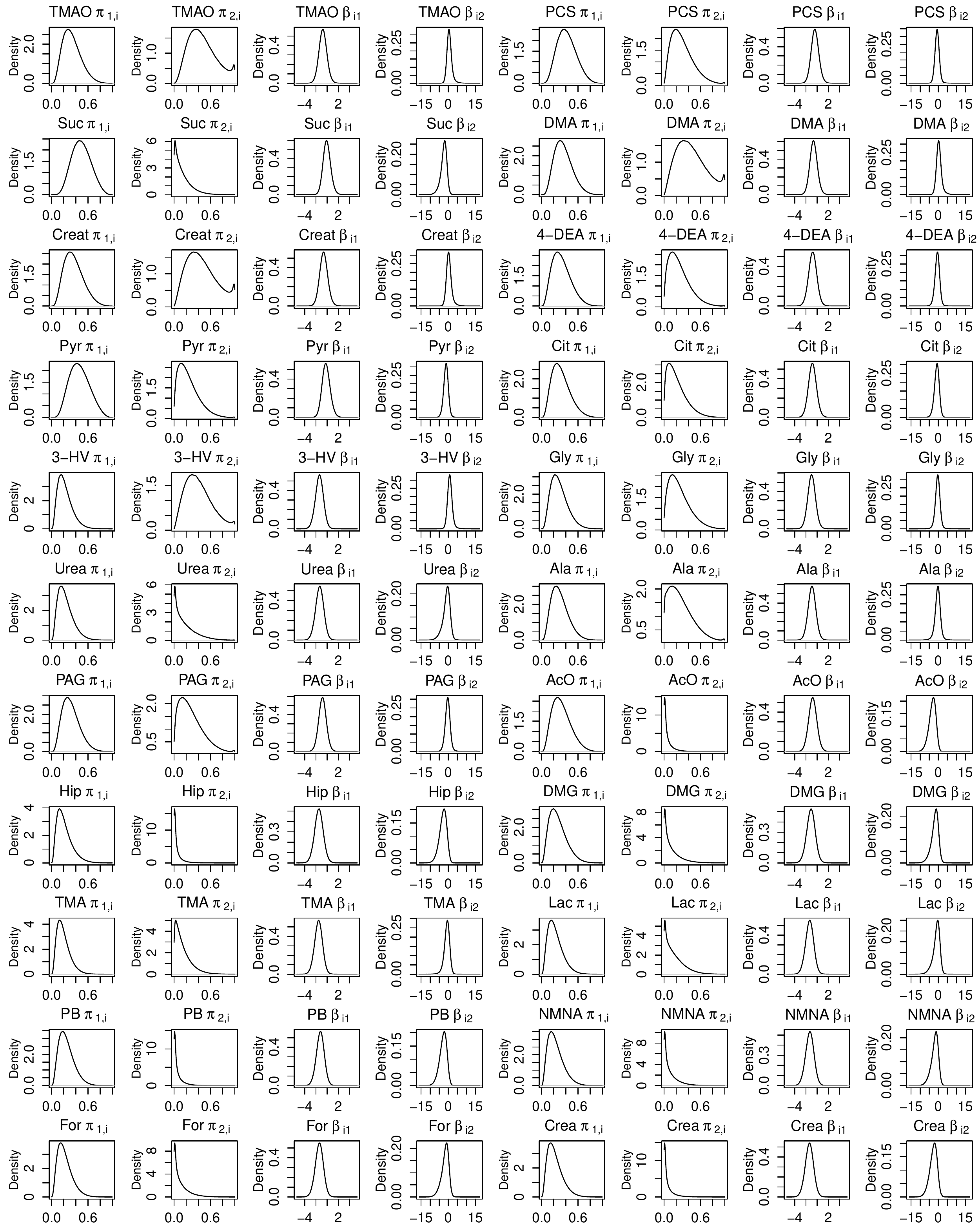} 
		\caption{Posterior distributions of the connectivity $\pi_{k,i}$ and the regression coefficients $\beta_{iq}$ for each metabolite for $k=1,2$ and $q=1,2$ corresponding to the multiplicative prior with $\sigma_1^2=1$ and $\sigma_2^2=10$.}
		\label{post2110}
	\end{figure}

\begin{figure}[htb!]
		\centering
		\includegraphics[width=\textwidth]{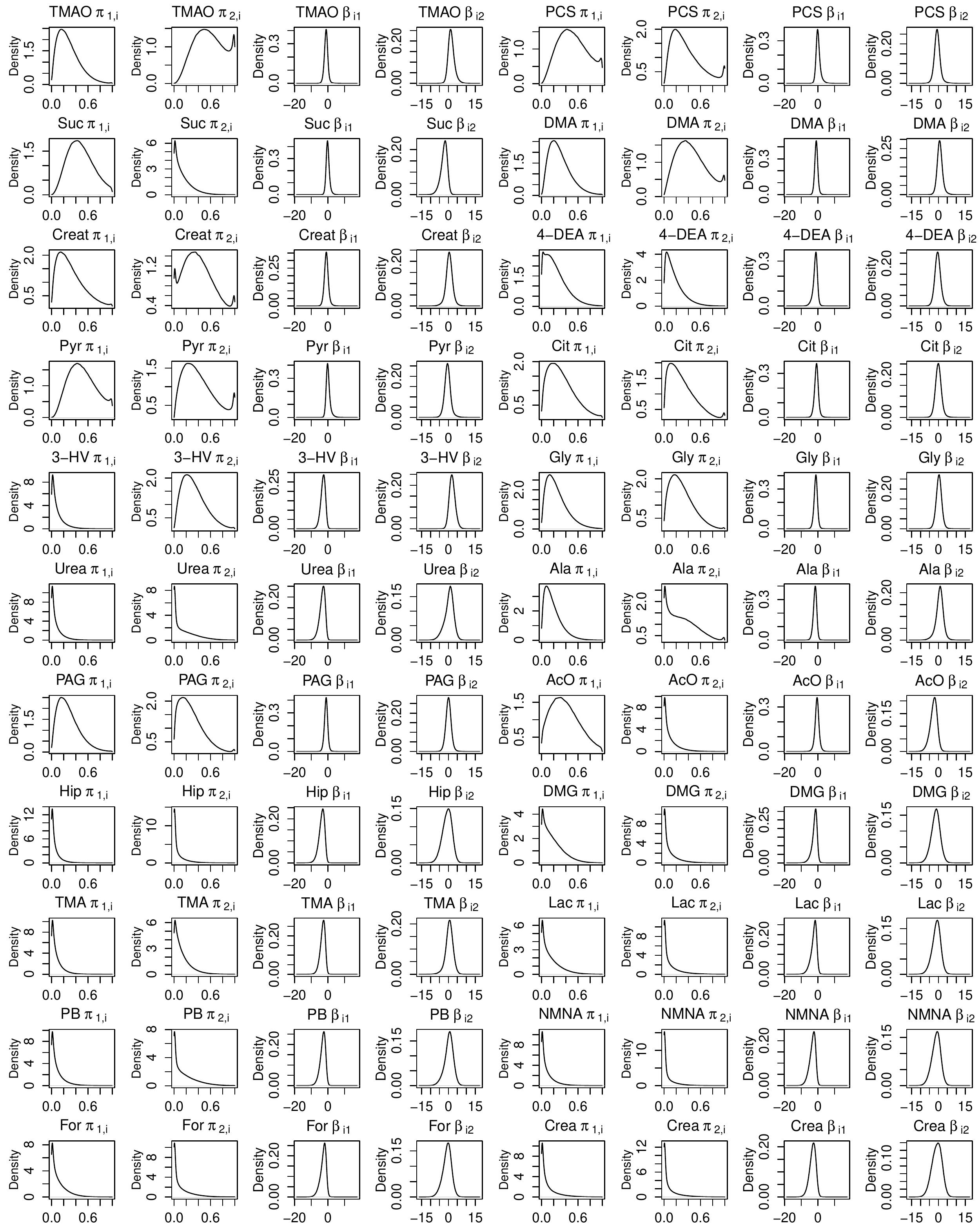} 
		\caption{Posterior distributions of the connectivity $\pi_{k,i}$ and the regression coefficients $\beta_{iq}$ for each metabolite for $k=1,2$ and $q=1,2$ corresponding to the multiplicative prior with $\sigma_1^2=\sigma_2^2=10$.}
		\label{post21010}
	\end{figure}

Figure \ref{F-ESSA} shows a plot of the ESS and mean acceptance rate in the MCMC steps at each iteration for the multiplicative prior with $\sigma_1^2=\sigma_2^2=1$. Figure \ref{G2all} shows the graphs of $G_1$ and $G_2$ corresponding to different priors. The width of the edges are proportional to their posterior weights. Figure \ref{network2all} shows these graphs but displaying only edges with posterior weights greater than 0.5 and associated nodes. Table \ref{G2diff} shows a list of the differential edges which are likely to appear in $G_1$ but not in $G_2$ and vice versa under each prior. Tables \ref{G2tab1}, \ref{G2tab2} and \ref{G2tab3} show the mean and 95\% credible interval of (1) the connectivity ($\pi_{ik}$) and (2) the regression coefficients ($\beta_{iq}$) of each metabolite for $k=1,2$, $q=1,2$, and the weighted mean betweenness centrality measure in each graph under the multplicative priors. Figures \ref{post211}, \ref{post2110} and \ref{post21010} show the posterior distributions of the connectivity and regression coefficients of each metabolite under the multplicative priors. Tables \ref{Cov21} and \ref{Cov22} show the mean precision matrix of $G_1$ and $G_2$ corresponding to the multiplicative prior with $\sigma_1^2 =\sigma_2^2=1$.
	
\begin{table}[htb!]
\centering \ra{1.1} \tabcolsep=0.16cm
\begin{footnotesize}
\resizebox{\columnwidth}{!}{
\begin{tabular}{@{}l|lccc|lccc@{}}
\hline
& \multicolumn{4}{c}{In $G_1$ not in $G_2$}\vline & \multicolumn{4}{c}{In $G_2$ not in $G_1$} \\
& Edge & $\rho_{ij}^1$ & $\rho_{ij}^2$ & $|\rho_{ij}^1 -\rho_{ij}^2|$ & Edge & $\rho_{ij}^1$ & $\rho_{ij}^2$ & $|\rho_{ij}^1 -\rho_{ij}^2|$ \\ \hline
\multirow{ 10}{*}{$\sigma_1^2=\sigma_2^2=1$:} &
Pyr  --  Suc & 0.99 & 0.01 & 0.98 & 3-HV  --  TMAO & 0.01 & 1.00 & 0.99 \\ 
   & Suc  --  PCS & 1.00 & 0.03 & 0.97 & 3-HV  --  4-DEA & 0.00 & 0.93 & 0.92 \\ 
   & PCS  --  4-DEA & 0.62 & 0.03 & 0.58 & Cit  --  Gly & 0.15 & 1.00 & 0.85 \\ 
   & AcO  --  PCS & 0.59 & 0.01 & 0.58 & TMA  --  TMAO & 0.10 & 0.72 & 0.62 \\ 
   & DMG  --  Creat & 0.56 & 0.01 & 0.55 & Ala  --  DMG & 0.01 & 0.59 & 0.59 \\ 
   & TMAO  --  For & 0.56 & 0.01 & 0.55 &  &  &  &  \\ 
   & PB  --  4-DEA & 0.58 & 0.03 & 0.54 &  &  &  &  \\ 
   & Pyr  --  4-DEA & 0.53 & 0.01 & 0.53 &  &  &  &  \\ 
   & AcO  --  Pyr & 0.54 & 0.02 & 0.52 &  &  &  &  \\ 
   & AcO  --  Suc & 0.51 & 0.00 & 0.51 &  &  &  &  \\ [2mm]
\hline\multirow{ 5}{*}{$\sigma_1^2=1$, $\sigma_2^2=10$:} 
 & Suc  --  PCS & 1.00 & 0.01 & 0.98 & 3-HV  --  TMAO & 0.00 & 1.00 & 1.00 \\ 
   & AcO  --  Suc & 0.98 & 0.00 & 0.98 & 3-HV  --  4-DEA & 0.06 & 0.98 & 0.93 \\ 
   & Pyr  --  Suc & 1.00 & 0.02 & 0.98 & Cit  --  Gly & 0.11 & 0.99 & 0.88 \\ 
   & PB  --  4-DEA & 0.73 & 0.00 & 0.73 & 3-HV  --  Creat & 0.01 & 0.74 & 0.72 \\ 
   & Pyr  --  PAG & 1.00 & 0.49 & 0.51 &  &  &  &  \\ [2mm]
\hline \multirow{5}{*}{$\sigma_1^2=\sigma_2^2=10$:}  
& AcO  --  Suc & 0.93 & 0.00 & 0.93 & 3-HV  --  TMAO & 0.02 & 1.00 & 0.98 \\ 
   & Suc  --  PCS & 0.93 & 0.09 & 0.84 & 3-HV  --  4-DEA & 0.03 & 0.99 & 0.96 \\ 
   & Pyr  --  Suc & 0.88 & 0.05 & 0.82 & TMA  --  TMAO & 0.06 & 0.76 & 0.71 \\ 
   & AcO  --  PCS & 0.56 & 0.01 & 0.55 & Cit  --  Gly & 0.32 & 1.00 & 0.68 \\ 
   & AcO  --  Cit & 0.55 & 0.01 & 0.54 & 3-HV  --  Creat & 0.01 & 0.60 & 0.58 \\ [2mm]
\hline	\multirow{10}{*}{Uniform prior:} 
 & Suc  --  PCS & 0.96 & 0.05 & 0.91 & 3-HV  --  TMAO & 0.10 & 1.00 & 0.90 \\ 
   & AcO  --  For & 0.93 & 0.02 & 0.91 & Ala  --  DMG & 0.09 & 0.98 & 0.89 \\ 
   & DMG  --  Creat & 0.95 & 0.04 & 0.91 & Lac  --  Ala & 0.12 & 0.92 & 0.80 \\ 
   & AcO  --  Suc & 0.97 & 0.10 & 0.87 & 3-HV  --  4-DEA & 0.28 & 1.00 & 0.72 \\ 
   & Pyr  --  Suc & 0.95 & 0.10 & 0.85 & PB  --  For & 0.04 & 0.75 & 0.71 \\ 
   & DMG  --  Gly & 0.89 & 0.05 & 0.84 & Crea  --  Urea & 0.27 & 0.91 & 0.64 \\ 
   & PB  --  Urea & 0.86 & 0.07 & 0.78 & AcO  --  Urea & 0.10 & 0.67 & 0.57 \\ 
   & PB  --  4-DEA & 0.98 & 0.20 & 0.78 & DMG  --  Urea & 0.14 & 0.67 & 0.53 \\ 
   & TMAO  --  For & 0.96 & 0.22 & 0.74 & Ala  --  NMNA & 0.04 & 0.55 & 0.51 \\ 
   & AcO  --  DMA & 0.87 & 0.14 & 0.73 & PAG  --  4-DEA & 0.36 & 0.86 & 0.51 \\ 
   & Suc  --  4-DEA & 0.74 & 0.04 & 0.70 &  &  &  &  \\ 
   & Lac  --  AcO & 0.88 & 0.24 & 0.65 &  &  &  &  \\ 
   & Ala  --  Cit & 0.65 & 0.09 & 0.56 &  &  &  &  \\ \hline
\end{tabular}}
\end{footnotesize}
\caption{List of differential edges which are likely to appear in $G_1$ but not in $G_2$ (left) and in $G_2$ but not in $G_1$ (right) corresponding to the different priors. $\rho_{ij}^1$ and $\rho_{ij}^2$ represent the posterior probability of the edge $(i,j)$ being present in $G_1$ and $G_2$ respectively.\label{G2diff}}
\end{table}

	\begin{table}[htb!]
		\centering \ra{1.15} \tabcolsep=0.2cm
		\begin{footnotesize}
	\resizebox{\columnwidth}{!}{\begin{tabular}{@{}ccccccccccc@{}}
				\hline
				Node $i$ & $\bar{\beta}_{i1}$ & CI ($\beta_{i1}$) & $\bar{\beta}_{i2}$ & CI ($\beta_{i2}$) & $\bar{\pi}_{1,i}$ & CI ($\pi_{1,i}$) &  $\bar{\pi}_{2,i}$ & CI($\pi_{2,i}$) & $\bar{B}_{1,i}$ & $\bar{B}_{2,i}$    \\ \hline
				TMAO & -0.44 & (-1.69, 0.81) & 0.30 & (-1.27, 1.87) & 0.40 & (0.14, 0.68) & 0.47 & (0.14, 0.81) & 0.07 & 0.18 \\ 
				PCS & -0.28 & (-1.48, 0.94) & -0.50 & (-1.96, 0.99) & 0.44 & (0.18, 0.71) & 0.33 & (0.07, 0.63) & 0.10 & 0.07 \\ 
				Suc & -0.43 & (-1.61, 0.75) & -0.91 & (-2.41, 0.60) & 0.40 & (0.15, 0.67) & 0.23 & (0.03, 0.49) & 0.18 & 0.02 \\ 
				DMA & -0.43 & (-1.69, 0.84) & 0.23 & (-1.33, 1.81) & 0.40 & (0.14, 0.68) & 0.46 & (0.12, 0.80) & 0.08 & 0.19 \\ 
				Creat & -0.63 & (-1.88, 0.61) & 0.08 & (-1.50, 1.66) & 0.36 & (0.12, 0.62) & 0.38 & (0.08, 0.71) & 0.06 & 0.12 \\ 
				4-DEA & -0.83 & (-2.23, 0.55) & -0.43 & (-2.02, 1.16) & 0.32 & (0.07, 0.60) & 0.25 & (0.02, 0.53) & 0.08 & 0.04 \\ 
				Pyr & -0.23 & (-1.44, 1.00) & -0.67 & (-2.24, 0.90) & 0.45 & (0.18, 0.72) & 0.31 & (0.05, 0.62) & 0.13 & 0.05 \\ 
				Cit & -0.76 & (-2.04, 0.52) & -0.21 & (-1.72, 1.31) & 0.33 & (0.09, 0.60) & 0.30 & (0.04, 0.61) & 0.09 & 0.11 \\ 
				3-HV & -1.39 & (-2.65, -0.14) & 0.28 & (-1.22, 1.79) & 0.22 & (0.05, 0.42) & 0.27 & (0.04, 0.55) & 0.00 & 0.08 \\ 
				Gly & -0.95 & (-2.23, 0.32) & -0.12 & (-1.70, 1.46) & 0.30 & (0.07, 0.55) & 0.28 & (0.03, 0.57) & 0.05 & 0.11 \\ 
				Urea & -1.27 & (-2.67, 0.11) & -0.07 & (-1.76, 1.59) & 0.24 & (0.04, 0.48) & 0.24 & (0.01, 0.55) & 0.02 & 0.07 \\ 
				Ala & -0.98 & (-2.29, 0.32) & -0.19 & (-1.85, 1.46) & 0.29 & (0.07, 0.54) & 0.27 & (0.02, 0.59) & 0.05 & 0.10 \\ 
				PAG & -0.85 & (-2.09, 0.38) & -0.24 & (-1.83, 1.33) & 0.31 & (0.09, 0.56) & 0.28 & (0.03, 0.58) & 0.02 & 0.06 \\ 
				AcO & -0.74 & (-2.13, 0.65) & -0.89 & (-2.52, 0.75) & 0.34 & (0.09, 0.62) & 0.20 & (0.01, 0.46) & 0.12 & 0.02 \\ 
				Hip & -1.65 & (-3.08, -0.24) & -0.62 & (-2.28, 1.05) & 0.18 & (0.02, 0.39) & 0.12 & (0.00, 0.32) & 0.01 & 0.01 \\ 
				DMG & -1.19 & (-2.62, 0.21) & -0.62 & (-2.26, 0.99) & 0.25 & (0.04, 0.50) & 0.17 & (0.00, 0.41) & 0.06 & 0.01 \\ 
				TMA & -1.42 & (-2.79, -0.06) & -0.37 & (-1.96, 1.20) & 0.22 & (0.04, 0.43) & 0.17 & (0.01, 0.39) & 0.02 & 0.02 \\ 
				Lac & -1.66 & (-3.08, -0.24) & -0.61 & (-2.28, 1.05) & 0.18 & (0.02, 0.39) & 0.13 & (0.00, 0.35) & 0.01 & 0.01 \\ 
				PB & -1.39 & (-2.77, -0.03) & -0.47 & (-2.18, 1.18) & 0.22 & (0.04, 0.44) & 0.17 & (0.00, 0.41) & 0.02 & 0.01 \\ 
				NMNA & -1.52 & (-2.91, -0.14) & -0.58 & (-2.28, 1.09) & 0.20 & (0.03, 0.41) & 0.14 & (0.00, 0.37) & 0.01 & 0.01 \\ 
				For & -1.40 & (-2.85, -0.01) & -0.70 & (-2.33, 0.92) & 0.22 & (0.03, 0.44) & 0.14 & (0.00, 0.34) & 0.03 & 0.01 \\ 
				Crea & -1.42 & (-2.87, 0.03) & -0.48 & (-2.18, 1.16) & 0.22 & (0.03, 0.46) & 0.17 & (0.00, 0.42) & 0.02 & 0.02 \\ 
				\hline
			\end{tabular}}
		\end{footnotesize}
		\caption{Weighted mean ($\bar{\pi}_{k,i}$) and 95\% credible interval (CI) of $\pi_{k,i}$, weighted mean ($\bar{\beta}_{iq}$) and 95\% credible interval (CI) of $\beta_{iq}$, and weighted mean betweenness centrality ($\bar{B}_{k,i}$) for each node $i$ for $k=1,2$ and $q=1,2$ corresponding to the multiplicative prior with  $\sigma_1^2=\sigma_2^2=1$.}
		\label{G2tab1}
	\end{table}

	\begin{table}[htb!]
		\centering \ra{1.15} \tabcolsep=0.2cm
		\begin{footnotesize}
	\resizebox{\columnwidth}{!}{\begin{tabular}{@{}ccccccccccc@{}}
				\hline
				Node $i$ & $\bar{\beta}_{i1}$ & CI ($\beta_{i1}$) & $\bar{\beta}_{i2}$ & CI ($\beta_{i2}$) & $\bar{\pi}_{1,i}$ & CI ($\pi_{1,i}$) &  $\bar{\pi}_{2,i}$ & CI($\pi_{2,i}$) & $\bar{B}_{1,i}$ & $\bar{B}_{2,i}$    \\ \hline
				TMAO & -0.76 & (-2.16, 0.64) & 0.71 & (-1.96, 3.51) & 0.34 & (0.08, 0.62) & 0.47 & (0.07, 0.92) & 0.04 & 0.10 \\ 
				PCS & -0.38 & (-1.73, 1.00) & -0.55 & (-3.01, 1.86) & 0.41 & (0.14, 0.71) & 0.31 & (0.02, 0.69) & 0.08 & 0.07 \\ 
				Suc & -0.04 & (-1.37, 1.34) & -2.46 & (-6.03, 0.61) & 0.49 & (0.20, 0.79) & 0.14 & (0.00, 0.11) & 0.24 & 0.02 \\ 
				DMA & -0.64 & (-1.97, 0.70) & 0.48 & (-2.35, 3.31) & 0.36 & (0.11, 0.64) & 0.45 & (0.04, 0.92) & 0.06 & 0.12 \\ 
				Creat & -0.61 & (-2.03, 0.81) & 0.48 & (-2.28, 3.37) & 0.37 & (0.09, 0.67) & 0.46 & (0.06, 0.94) & 0.08 & 0.09 \\ 
				4-DEA & -0.80 & (-2.24, 0.64) & -0.50 & (-3.14, 1.99) & 0.33 & (0.07, 0.62) & 0.26 & (0.00, 0.61) & 0.06 & 0.05 \\ 
				Pyr & -0.17 & (-1.56, 1.28) & -1.20 & (-3.80, 1.33) & 0.46 & (0.16, 0.77) & 0.25 & (0.00, 0.61) & 0.13 & 0.03 \\ 
				Cit & -0.83 & (-2.24, 0.59) & -0.78 & (-3.53, 1.86) & 0.32 & (0.07, 0.61) & 0.22 & (0.00, 0.56) & 0.07 & 0.05 \\ 
				3-HV & -1.39 & (-2.82, 0.02) & 0.98 & (-1.69, 3.70) & 0.22 & (0.03, 0.45) & 0.41 & (0.04, 0.84) & 0.00 & 0.09 \\ 
				Gly & -0.99 & (-2.37, 0.38) & -0.27 & (-3.07, 2.39) & 0.29 & (0.06, 0.56) & 0.27 & (0.00, 0.64) & 0.04 & 0.08 \\ 
				Urea & -1.35 & (-2.82, 0.08) & -0.98 & (-5.71, 2.78) & 0.23 & (0.03, 0.47) & 0.18 & (0.00, 0.15) & 0.02 & 0.04 \\ 
				Ala & -0.93 & (-2.31, 0.45) & -0.27 & (-3.73, 2.98) & 0.30 & (0.07, 0.57) & 0.29 & (0.00, 0.76) & 0.05 & 0.09 \\ 
				PAG & -0.84 & (-2.20, 0.53) & -0.26 & (-3.21, 2.52) & 0.32 & (0.08, 0.59) & 0.30 & (0.00, 0.74) & 0.03 & 0.06 \\ 
				AcO & -0.79 & (-2.24, 0.66) & -3.25 & (-7.22, 0.28) & 0.33 & (0.07, 0.63) & 0.05 & (0.00, 0.05) & 0.09 & 0.00 \\ 
				Hip & -1.47 & (-3.01, 0.08) & -2.89 & (-6.96, 0.81) & 0.21 & (0.03, 0.46) & 0.04 & (0.00, 0.04) & 0.02 & 0.00 \\ 
				DMG & -1.12 & (-2.63, 0.36) & -1.84 & (-5.84, 1.60) & 0.27 & (0.04, 0.54) & 0.12 & (0.00, 0.10) & 0.05 & 0.01 \\ 
				TMA & -1.55 & (-2.97, -0.17) & -0.74 & (-3.86, 2.19) & 0.20 & (0.03, 0.41) & 0.15 & (0.00, 0.52) & 0.00 & 0.02 \\ 
				Lac & -1.35 & (-2.80, 0.06) & -1.01 & (-4.92, 2.26) & 0.23 & (0.03, 0.47) & 0.16 & (0.00, 0.14) & 0.01 & 0.03 \\ 
				PB & -1.23 & (-2.65, 0.17) & -2.82 & (-6.89, 0.87) & 0.25 & (0.04, 0.49) & 0.06 & (0.00, 0.05) & 0.02 & 0.00 \\ 
				NMNA & -1.33 & (-2.86, 0.17) & -1.96 & (-6.03, 1.53) & 0.23 & (0.03, 0.48) & 0.10 & (0.00, 0.08) & 0.01 & 0.01 \\ 
				For & -1.41 & (-2.84, 0.02) & -1.74 & (-5.84, 1.73) & 0.22 & (0.03, 0.45) & 0.11 & (0.00, 0.09) & 0.01 & 0.01 \\ 
				Crea & -1.40 & (-2.86, 0.04) & -2.68 & (-6.82, 1.07) & 0.22 & (0.03, 0.46) & 0.06 & (0.00, 0.05) & 0.02 & 0.00 \\ 
				\hline
			\end{tabular}}
		\end{footnotesize}
		\caption{Weighted mean ($\bar{\pi}_{k,i}$) and 95\% credible interval (CI) of $\pi_{k,i}$, weighted mean ($\bar{\beta}_{iq}$) and 95\% credible interval (CI) of $\beta_{iq}$, and weighted mean betweenness centrality ($\bar{B}_{k,i}$) for each node $i$ for $k=1,2$ and $q=1,2$ corresponding to the multiplicative prior with  $\sigma_1^2=1$, $\sigma_2^2=10$.}
		\label{G2tab2}
	\end{table}

\begin{table}[htb!]
	\centering \ra{1.15} \tabcolsep=0.2cm
	\begin{footnotesize}
	\resizebox{\columnwidth}{!}{\begin{tabular}{@{}ccccccccccc@{}}
			\hline
				Node $i$ & $\bar{\beta}_{i1}$ & CI ($\beta_{i1}$) & $\bar{\beta}_{i2}$ & CI ($\beta_{i2}$) & $\bar{\pi}_{1,i}$ & CI ($\pi_{1,i}$) &  $\bar{\pi}_{2,i}$ & CI($\pi_{2,i}$) & $\bar{B}_{1,i}$ & $\bar{B}_{2,i}$    \\ \hline
				TMAO & -1.09 & (-3.33, 1.02) & 1.63 & (-1.59, 5.13) & 0.29 & (0.01, 0.66) & 0.58 & (0.00, 0.90) & 0.03 & 0.14 \\ 
				PCS & 0.16 & (-2.26, 3.02) & -0.69 & (-4.25, 2.86) & 0.52 & (0.15, 0.99) & 0.38 & (0.00, 0.90) & 0.10 & 0.05 \\ 
				Suc & -0.07 & (-2.12, 2.09) & -2.48 & (-6.31, 0.87) & 0.48 & (0.11, 0.89) & 0.14 & (0.00, 0.12) & 0.12 & 0.01 \\ 
				DMA & -0.97 & (-2.90, 0.88) & 0.82 & (-2.19, 3.86) & 0.30 & (0.02, 0.66) & 0.45 & (0.03, 0.92) & 0.05 & 0.09 \\ 
				Creat & -0.91 & (-3.55, 1.74) & 0.43 & (-3.52, 4.39) & 0.33 & (0.00, 0.78) & 0.42 & (0.05, 0.99) & 0.07 & 0.06 \\ 
				4-DEA & -1.59 & (-4.62, 0.95) & -0.38 & (-3.72, 3.00) & 0.23 & (0.00, 0.63) & 0.18 & (0.00, 0.53) & 0.03 & 0.01 \\ 
				Pyr & 0.12 & (-2.19, 2.74) & -0.45 & (-4.18, 3.20) & 0.51 & (0.14, 0.96) & 0.42 & (0.02, 0.94) & 0.10 & 0.07 \\ 
				Cit & -0.88 & (-3.55, 1.66) & -0.01 & (-3.32, 3.26) & 0.33 & (0.00, 0.76) & 0.33 & (0.00, 0.85) & 0.08 & 0.07 \\ 
				3-HV & -2.95 & (-5.83, -0.33) & 2.14 & (-1.12, 5.59) & 0.09 & (0.00, 0.07) & 0.34 & (0.01, 0.75) & 0.00 & 0.08 \\ 
				Gly & -1.30 & (-3.55, 0.81) & 0.30 & (-2.92, 3.40) & 0.26 & (0.00, 0.60) & 0.31 & (0.00, 0.72) & 0.06 & 0.07 \\ 
				Urea & -3.41 & (-7.04, -0.33) & 0.37 & (-5.25, 5.06) & 0.08 & (0.00, 0.06) & 0.17 & (0.00, 0.17) & 0.00 & 0.03 \\ 
				Ala & -1.75 & (-3.90, 0.24) & 0.69 & (-3.78, 4.73) & 0.19 & (0.00, 0.48) & 0.35 & (0.00, 0.32) & 0.02 & 0.09 \\ 
				PAG & -1.12 & (-3.19, 0.88) & 0.05 & (-3.05, 3.13) & 0.28 & (0.01, 0.64) & 0.30 & (0.00, 0.76) & 0.02 & 0.04 \\ 
				AcO & -0.55 & (-3.26, 2.09) & -2.70 & (-6.71, 0.94) & 0.40 & (0.00, 0.83) & 0.10 & (0.00, 0.08) & 0.10 & 0.00 \\ 
				Hip & -3.75 & (-7.40, -0.62) & -0.75 & (-6.31, 4.33) & 0.06 & (0.00, 0.05) & 0.07 & (0.00, 0.06) & 0.00 & 0.00 \\ 
				DMG & -2.04 & (-5.47, 0.67) & -1.55 & (-6.78, 3.40) & 0.19 & (0.00, 0.16) & 0.09 & (0.00, 0.08) & 0.05 & 0.01 \\ 
				TMA & -3.19 & (-6.19, -0.55) & 0.51 & (-3.85, 4.73) & 0.08 & (0.00, 0.06) & 0.13 & (0.00, 0.11) & 0.00 & 0.01 \\ 
				Lac & -2.51 & (-6.33, 0.52) & -1.19 & (-6.38, 3.46) & 0.15 & (0.00, 0.12) & 0.10 & (0.00, 0.08) & 0.01 & 0.01 \\ 
				PB & -3.14 & (-6.76, -0.12) & 0.26 & (-5.11, 5.06) & 0.09 & (0.00, 0.07) & 0.17 & (0.00, 0.15) & 0.00 & 0.02 \\ 
				NMNA & -3.30 & (-7.19, -0.12) & -1.31 & (-6.58, 3.46) & 0.09 & (0.00, 0.07) & 0.06 & (0.00, 0.06) & 0.00 & 0.00 \\ 
				For & -2.87 & (-6.69, 0.17) & -0.74 & (-6.05, 3.99) & 0.12 & (0.00, 0.09) & 0.12 & (0.00, 0.10) & 0.01 & 0.01 \\ 
				Crea & -3.18 & (-7.04, 0.31) & -0.89 & (-6.58, 4.46) & 0.11 & (0.00, 0.07) & 0.09 & (0.00, 0.07) & 0.01 & 0.01 \\ 
				\hline
			\end{tabular}}
		\end{footnotesize}
		\caption{Weighted mean ($\bar{\pi}_{k,i}$) and 95\% credible interval (CI) of $\pi_{k,i}$, weighted mean ($\bar{\beta}_{iq}$) and 95\% credible interval (CI) of $\beta_{iq}$, and weighted mean betweenness centrality ($\bar{B}_{k,i}$) for each node $i$ for $k=1,2$ and $q=1,2$ corresponding to the multiplicative prior with $\sigma_1^2=\sigma_2^2=10$.}
		\label{G2tab3}
	\end{table}

	\begin{table}[htb!]
		\centering \tabcolsep=0.1cm \ra{1.1}
		\resizebox{\columnwidth}{!}{\begin{tabular}{@{}lrrrrrrrrrrrrrrrrrrrrrr@{}}
				\hline
				& TMAO & PCS & Suc & DMA & Creat & 4-DEA & Pyr & Cit & 3-HV & Gly & Urea & Ala & PAG & AcO & Hip & DMG & TMA & Lac & PB & NMNA & For & Crea \\ 
				\hline
				TMAO & 1.74 & 0.00 & 0.01 & -1.09 & 0.44 & 0.01 & -0.00 & 0.00 & -0.00 & 0.00 & -0.00 & -0.00 & 0.01 & 0.06 & 0.00 & -0.00 & -0.03 & 0.00 & -0.00 & -0.00 & -0.24 & 0.00 \\ 
				PCS & . & 4.14 & -1.22 & 0.00 & 0.01 & -0.40 & -2.54 & 0.00 & 0.00 & 0.00 & -0.00 & 0.00 & -0.87 & -0.49 & 0.00 & 0.04 & 0.00 & 0.03 & 0.00 & 0.04 & -0.00 & 0.00 \\ 
				Suc & . & . & 2.19 & 0.01 & 0.00 & -0.13 & 1.06 & -0.72 & -0.00 & -0.00 & 0.00 & 0.00 & 0.02 & -0.35 & -0.00 & 0.00 & 0.01 & 0.00 & -0.00 & 0.01 & -0.05 & -0.07 \\ 
				DMA & . & . & . & 2.34 & -1.14 & -0.00 & 0.00 & -0.00 & 0.00 & 0.00 & 0.01 & 0.01 & 0.00 & 0.15 & 0.00 & -0.01 & -0.02 & 0.00 & -0.00 & 0.01 & -0.00 & 0.00 \\ 
				Creat & . & . & . & . & 1.80 & 0.00 & 0.00 & 0.00 & 0.00 & 0.01 & 0.00 & 0.00 & 0.00 & 0.01 & 0.01 & -0.25 & -0.00 & 0.00 & -0.00 & 0.01 & 0.00 & -0.02 \\ 
				4-DEA & . & . & . & . & . & 1.39 & 0.36 & -0.00 & -0.00 & -0.00 & 0.00 & 0.00 & 0.01 & -0.00 & -0.00 & 0.00 & -0.00 & 0.00 & -0.28 & -0.00 & -0.00 & -0.08 \\ 
				Pyr & . & . & . & . & . & . & 4.29 & -0.01 & 0.00 & -0.00 & -0.01 & -0.00 & -1.40 & 0.39 & 0.00 & 0.11 & 0.01 & 0.01 & -0.00 & 0.09 & 0.00 & 0.05 \\ 
				Cit & . & . & . & . & . & . & . & 1.56 & -0.00 & -0.05 & 0.04 & -0.27 & -0.00 & 0.12 & -0.00 & 0.02 & 0.01 & 0.00 & -0.00 & 0.00 & -0.00 & 0.03 \\ 
				3-HV & . & . & . & . & . & . & . & . & 1.06 & -0.00 & -0.00 & -0.03 & 0.00 & 0.00 & 0.00 & -0.01 & -0.00 & 0.00 & 0.00 & -0.00 & 0.00 & 0.00 \\ 
				Gly & . & . & . & . & . & . & . & . & . & 1.53 & 0.00 & -0.73 & -0.00 & 0.00 & 0.01 & -0.16 & -0.00 & -0.00 & 0.00 & 0.00 & -0.01 & 0.00 \\ 
				Urea & . & . & . & . & . & . & . & . & . & . & 1.13 & 0.00 & -0.00 & -0.00 & 0.02 & -0.00 & -0.02 & 0.01 & 0.11 & 0.00 & 0.00 & 0.01 \\ 
				Ala & . & . & . & . & . & . & . & . & . & . & . & 1.57 & 0.00 & 0.00 & 0.02 & -0.00 & -0.00 & -0.00 & 0.00 & -0.00 & -0.00 & 0.00 \\ 
				PAG & . & . & . & . & . & . & . & . & . & . & . & . & 2.74 & 0.00 & 0.00 & 0.00 & 0.01 & 0.01 & -0.00 & 0.01 & 0.00 & 0.00 \\ 
				AcO & . & . & . & . & . & . & . & . & . & . & . & . & . & 1.59 & -0.00 & 0.00 & -0.00 & -0.11 & 0.01 & 0.01 & -0.20 & -0.01 \\ 
				Hip & . & . & . & . & . & . & . & . & . & . & . & . & . & . & 1.08 & 0.00 & 0.00 & 0.00 & 0.00 & -0.02 & -0.02 & -0.00 \\ 
				DMG & . & . & . & . & . & . & . & . & . & . & . & . & . & . & . & 1.27 & -0.02 & 0.00 & 0.00 & -0.00 & 0.00 & -0.01 \\ 
				TMA & . & . & . & . & . & . & . & . & . & . & . & . & . & . & . & . & 1.09 & -0.00 & 0.01 & -0.00 & -0.00 & 0.01 \\ 
				Lac & . & . & . & . & . & . & . & . & . & . & . & . & . & . & . & . & . & 1.11 & 0.00 & 0.03 & 0.00 & 0.00 \\ 
				PB & . & . & . & . & . & . & . & . & . & . & . & . & . & . & . & . & . & . & 1.22 & 0.00 & 0.00 & 0.00 \\ 
				NMNA & . & . & . & . & . & . & . & . & . & . & . & . & . & . & . & . & . & . & . & 1.12 & -0.00 & -0.00 \\ 
				For & . & . & . & . & . & . & . & . & . & . & . & . & . & . & . & . & . & . & . & . & 1.23 & -0.00 \\ 
				Crea & . & . & . & . & . & . & . & . & . & . & . & . & . & . & . & . & . & . & . & . & . & 1.13 \\ 
				\hline
			\end{tabular}}
			\caption{Mean precision matrix $\Omega_1$ corresponding to $K=2$ and multiplicative prior with  $\sigma_1^2=\sigma_2^2=1$.  \label{Cov21}}
		\end{table}

		\begin{table}[htb!]
			\centering \tabcolsep=0.1cm \ra{1.1}
			\resizebox{\columnwidth}{!}{\begin{tabular}{@{}lrrrrrrrrrrrrrrrrrrrrrr@{}}
					& TMAO & PCS & Suc & DMA & Creat & 4-DEA & Pyr & Cit & 3-HV & Gly & Urea & Ala & PAG & AcO & Hip & DMG & TMA & Lac & PB & NMNA & For & Crea \\ 
					\hline
					TMAO & 5.23 & -0.01 & 0.00 & -3.27 & 1.38 & -0.00 & 0.02 & 0.00 & -1.85 & -0.01 & 0.01 & -0.02 & -0.01 & 0.01 & 0.01 & -0.00 & -0.45 & -0.00 & -0.18 & 0.01 & 0.00 & -0.03 \\ 
					PCS & . & 11.46 & -0.00 & -0.02 & 0.00 & 0.00 & -8.09 & -0.01 & 0.00 & 0.00 & -0.00 & -0.00 & -3.53 & 0.00 & -0.00 & 0.00 & -0.00 & 0.00 & 0.00 & -0.01 & -0.00 & 0.00 \\ 
					Suc & . & . & 1.30 & -0.00 & -0.03 & 0.00 & -0.00 & -0.48 & 0.00 & -0.01 & 0.04 & -0.00 & -0.00 & -0.00 & 0.00 & 0.00 & -0.00 & -0.00 & 0.00 & 0.00 & -0.00 & 0.00 \\ 
					DMA & . & . & . & 4.58 & -1.56 & -0.06 & -0.00 & -0.00 & -0.03 & -0.04 & 0.22 & -0.11 & -0.01 & 0.08 & 0.06 & -0.00 & -0.08 & -0.00 & 0.12 & -0.00 & 0.04 & -0.04 \\ 
					Creat & . & . & . & . & 1.86 & -0.05 & 0.00 & -0.06 & -0.33 & -0.03 & 0.00 & -0.00 & 0.01 & 0.01 & 0.04 & 0.00 & -0.00 & 0.00 & 0.03 & 0.00 & 0.06 & 0.00 \\ 
					4-DEA & . & . & . & . & . & 1.57 & 0.00 & -0.00 & -0.75 & -0.00 & 0.00 & -0.01 & 0.13 & 0.00 & 0.00 & -0.00 & 0.00 & -0.01 & -0.01 & 0.00 & 0.00 & 0.00 \\ 
					Pyr & . & . & . & . & . & . & 8.02 & -0.01 & 0.00 & -0.00 & -0.01 & -0.02 & 0.84 & -0.00 & -0.00 & 0.00 & -0.00 & 0.00 & 0.00 & 0.06 & 0.00 & 0.00 \\ 
					Cit & . & . & . & . & . & . & . & 1.83 & -0.00 & -0.87 & 0.00 & -0.04 & -0.01 & 0.01 & 0.00 & -0.02 & 0.00 & 0.00 & 0.04 & 0.00 & 0.00 & -0.00 \\ 
					3-HV & . & . & . & . & . & . & . & . & 3.03 & -0.01 & 0.02 & -0.01 & 0.00 & 0.01 & 0.00 & -0.00 & -0.05 & -0.00 & -0.00 & -0.03 & 0.00 & -0.00 \\ 
					Gly & . & . & . & . & . & . & . & . & . & 1.74 & 0.07 & -0.27 & 0.01 & 0.00 & 0.00 & -0.00 & 0.01 & -0.01 & -0.04 & 0.00 & -0.00 & -0.00 \\ 
					Urea & . & . & . & . & . & . & . & . & . & . & 1.33 & 0.03 & -0.00 & -0.04 & 0.02 & -0.04 & 0.00 & 0.05 & -0.00 & 0.03 & 0.00 & -0.21 \\ 
					Ala & . & . & . & . & . & . & . & . & . & . & . & 1.41 & 0.01 & 0.01 & 0.02 & -0.30 & -0.00 & -0.07 & 0.00 & 0.02 & -0.00 & -0.00 \\ 
					PAG & . & . & . & . & . & . & . & . & . & . & . & . & 3.43 & 0.00 & -0.00 & 0.00 & -0.00 & 0.00 & 0.01 & 0.01 & -0.00 & -0.00 \\ 
					AcO & . & . & . & . & . & . & . & . & . & . & . & . & . & 1.10 & 0.00 & 0.00 & 0.00 & -0.00 & -0.01 & 0.00 & 0.00 & -0.00 \\ 
					Hip & . & . & . & . & . & . & . & . & . & . & . & . & . & . & 1.10 & 0.00 & -0.00 & 0.00 & -0.00 & -0.00 & -0.00 & 0.00 \\ 
					DMG & . & . & . & . & . & . & . & . & . & . & . & . & . & . & . & 1.20 & -0.00 & -0.00 & -0.00 & 0.00 & -0.00 & -0.03 \\ 
					TMA & . & . & . & . & . & . & . & . & . & . & . & . & . & . & . & . & 1.34 & 0.00 & -0.00 & -0.00 & 0.00 & 0.00 \\ 
					Lac & . & . & . & . & . & . & . & . & . & . & . & . & . & . & . & . & . & 1.10 & -0.00 & 0.00 & -0.00 & -0.00 \\ 
					PB & . & . & . & . & . & . & . & . & . & . & . & . & . & . & . & . & . & . & 1.15 & 0.00 & -0.04 & -0.00 \\ 
					NMNA & . & . & . & . & . & . & . & . & . & . & . & . & . & . & . & . & . & . & . & 1.10 & -0.00 & 0.02 \\ 
					For & . & . & . & . & . & . & . & . & . & . & . & . & . & . & . & . & . & . & . & . & 1.10 & 0.01 \\ 
					Crea & . & . & . & . & . & . & . & . & . & . & . & . & . & . & . & . & . & . & . & . & . & 1.16 \\ 
					\hline
				\end{tabular}}
				\caption{Mean precision matrix $\Omega_2$ corresponding to $K=2$ and multiplicative prior with $\sigma_1^2=\sigma_2^2=1$. \label{Cov22}}
			\end{table}

\end{document}